\documentclass[11pt]{article}

\usepackage{graphicx,pstricks}
\usepackage{moreverb}
\usepackage{subfigure}
\usepackage{amsmath}
\usepackage{mathrsfs}
\usepackage{epsfig}
\usepackage{txfonts}
\usepackage{palatino}
\usepackage{epstopdf}
\usepackage{xfrac}
\usepackage{mdwlist}
\usepackage{morefloats}
\usepackage{cite}
\usepackage{multirow}
\usepackage{authblk}

\DeclareMathAlphabet{\mathsfsl}{OT1}{cmss}{m}{sl}
\newcommand{\fepx}{{\bfseries{\slshape{FEpX}}}}

\begin{document}
\title{Initiation and Propagation of Plastic Yielding\\ in Duplex Stainless Steel}
\author[1]{Andrew C. Poshadel}
\author[2]{Michael Gharghouri}
\author[1]{Paul R. Dawson}
\affil[1]{Sibley School of Mechanical and Aerospace Engineering, Cornell University, Ithaca, New York, USA}
\affil[2] {Canadian Nuclear Laboratories, Chalk River, Ontario, Canada}
\date{}
\maketitle

\begin{abstract}
The elastoplastic behavior of a two-phase stainless steel alloy is explored at 
the crystal scale for five levels of stress biaxiality. 
The crystal lattice (elastic) strains were measured with neutron diffraction using tubular samples subjected to different combinations of axial load and internal pressure to achieve a range of biaxial stress ratios. 
Finite element simulations were conducted on virtual polycrystals using loading histories that mimicked the experimental protocols.  
For this, two-phase microstructures were instantiated based on microscopy images of the grain and phase topologies and on crystallographic orientation distributions from neutron diffraction.
Detailed comparisons were made between the measured  and computed lattice strains for several crystal reflections in both phases for scattering vectors in the axial, radial and hoop directions that confirm the model's ability  to accurate predict the evolving local stress states.
The strength-to-stiffness parameter for multiaxial stress states reported in \cite{pos_daw_multiaxial-y2e} was applied to explain the initiation of yielding across 
the polycrystalline  samples across the five levels of stress biaxiality.  
Finally, building off the multiaxial strength-to-stiffness, the propagation of yielding over the volume of a polycrystal was estimated in terms of an equation that provides the local stress necessary to initiate  within crystals in terms of the macroscopic stress.

\end{abstract}
\clearpage

\section{Introduction}
\label{sec:introduction}
Metallic alloys used in structural components commonly are polyphase and polycrystalline solids.  
The phases often exhibit contrasting values of stiffness and strength.  
The constituent crystals of the phases exhibit mechanical anisotropy stemming from their crystalline structures.   
These attributes of the material structure dictate that the mechanical response 
to loading is spatially heterogeneous  at the scale of the phases and the crystals within the phases,
even within domains that are homogeneous at the macroscopic scale. 
One aspect of the mechanical response of great interest in the performance of
a structural component is the intensity of the stress,  and in particular, if it is sufficient to
induce plastic deformations.  
The heterogeneity of the stress that develops as a consequence of a material's microstructure complicates addressing this issue because the complexity of
interactions between phases and crystals as they react to imposed loads.   

In this paper, the initiation and propagation of plastic yielding in a duplex stainless steel is examined  in 
the context of its dual phase, polycrystalline microstructure.  
In particular, our interest is in quantifying the onset of yielding throughout the microstructure
as functions of the type of phase and the orientation of the crystallographic lattice of grains within the phases. 
We employ a recently developed metric for multiaxial strength-to-stiffness to rank crystals in terms of
the relative order in which they will yield as the intensity of the loading increases.  
For the loading, biaxial stress states range from uniaxial tension to balanced biaxial tension.

The paper is organized with the following structure.
First, we discuss the study material.
The duplex stainless steel, LDX-2101, was chosen because it has relatively equal phase volume fractions and  comparable strengths and stiffnesses for the two phases. 
The phases differ in their crystal structures (FCC versus BCC), their morphologies and topologies, and in their single-crystal mechanical properties (elastic and plastic).  
Next, we present a summary of the experimental program.
Using neutron diffraction with {\it in situ} loading,  crystal lattice (elastic) strains were measured subjected to different combinations of axial load and internal pressure to achieve a range of biaxial stress ratios. 
Following description of the experiments,  we summarize the modeling program.
Finite element simulations of virtual polycrystals embodying the two-phase microstructure were conducted using loading histories that mimicked the experimental protocols. 
We then present detailed comparisons  between the measured  and computed lattice strains for several crystal reflections in both phases for scattering vectors in the axial, radial and hoop directions.  
These comparisons confirm the model's ability  to accurate predict the evolving local stress states.
With confidence in the model's predictive capability gained through comparisons to measured responses,
we then turn to the challenge of predicting the initiation of yielding over the microstructure.  
The strength-to-stiffness parameter for multiaxial stress states is demonstrated to explain the initiation of yielding over the volume of the polycrystalline  samples for the full range of stress biaxiality.  
Lastly, using the multiaxial strength-to-stiffness framework,  an equation that provides the local stress necessary to initiate  within crystals in terms of the macroscopic stress is shown to capture the behavior simulated with model.

\clearpage

\section{Description of the Duplex Steel }
\label{sec:material}
LDX-2101 is a dual phase austenitic-ferritic stainless steel manufactured by Outokumpu. Austenite, or $\gamma$-phase, has a face-centered cubic (FCC) crystal structure. Ferrite, or $\alpha$-phase, has a body-centered cubic (BCC) structure. The name LDX-2101 refers to the material's dual phase nature and chemical composition, presented in Table~\ref{tab:chem_comp}. LDX stands for lean duplex, a dual phase steel low in alloying elements nickel and molybdenum. The number 2101 denotes 21\% chromium and 1\% nickel composition by weight.

The material in this study started as continuously cast billet, which was then continuously hot rolled into 38~mm (1.5~in) round bar. At the start of the rolling process, the temperature was around 1450~K, and at the end of the rolling process, the temperature was between 1310~K and 1340~K. The bar was water-quenched after rolling. Specimens were then machined from the round bar.

\begin{table}[h]
	\centering
	\caption{Chemical composition of LDX-2101 by weight. Remaining balance is iron}
	\begin{tabular} {c c c c c}		
		Cr & Ni & Mn & Mo & N \\
		21.0\%-22.0\% & 1.35\%-1.70\% & 4.00\%-6.00\% & 0.10\%-0.80\% & 0.20\%-0.25\%  \\ \hline
		C & Si & P & Cu & S \\
		$\leqslant$0.040\% & $\leqslant$1.00\% & $\leqslant$0.040\% & 0.10\%-0.80\% & $\leqslant$0.030\%
	\end{tabular}	
	\label{tab:chem_comp}
\end{table}

\begin{figure}[h]
\centering
\subfigure[Axial]{\includegraphics[scale = 0.4]{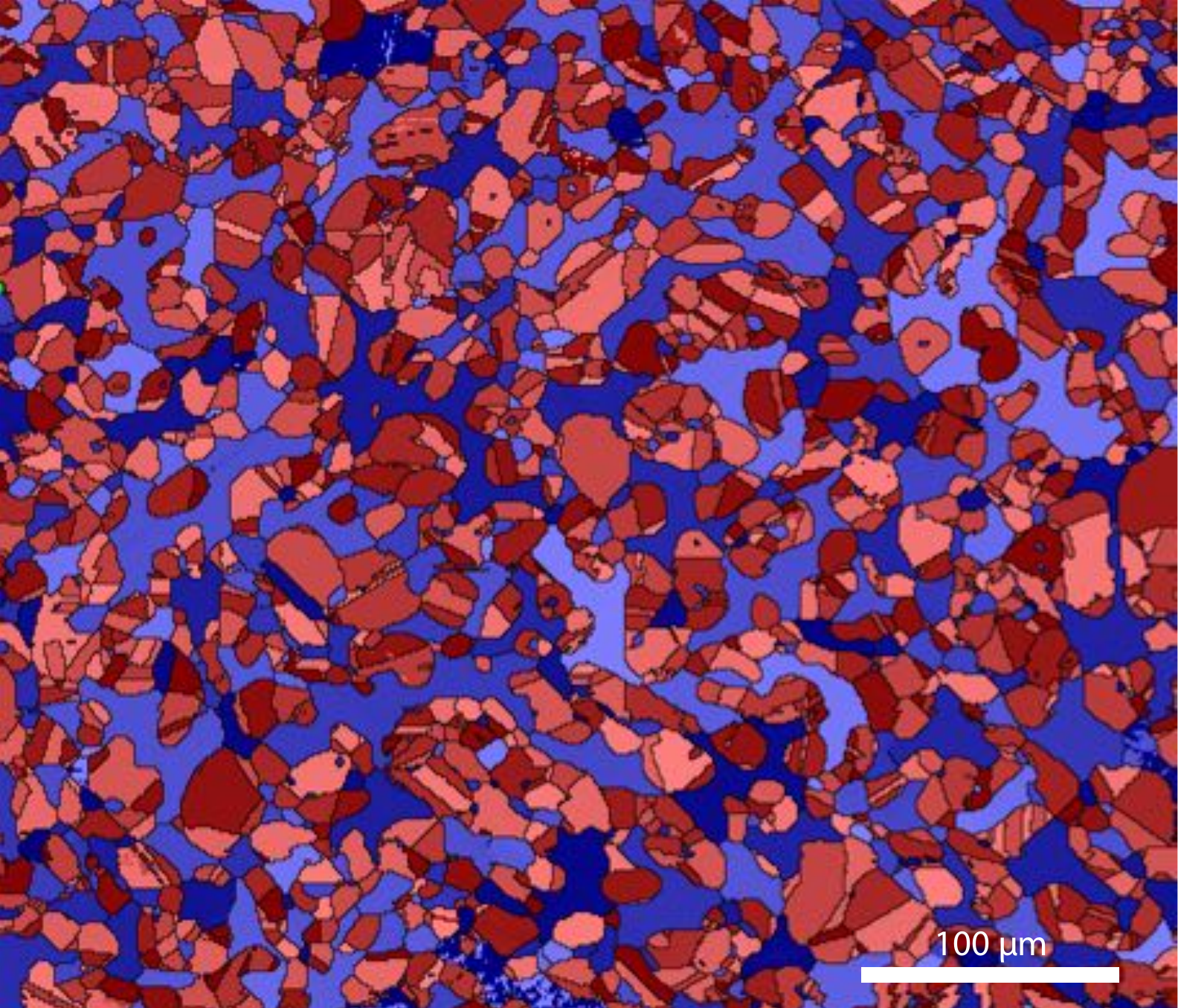}\label{fig:microstruct_axial}}
\subfigure[Transverse]{\includegraphics[scale = 0.4]{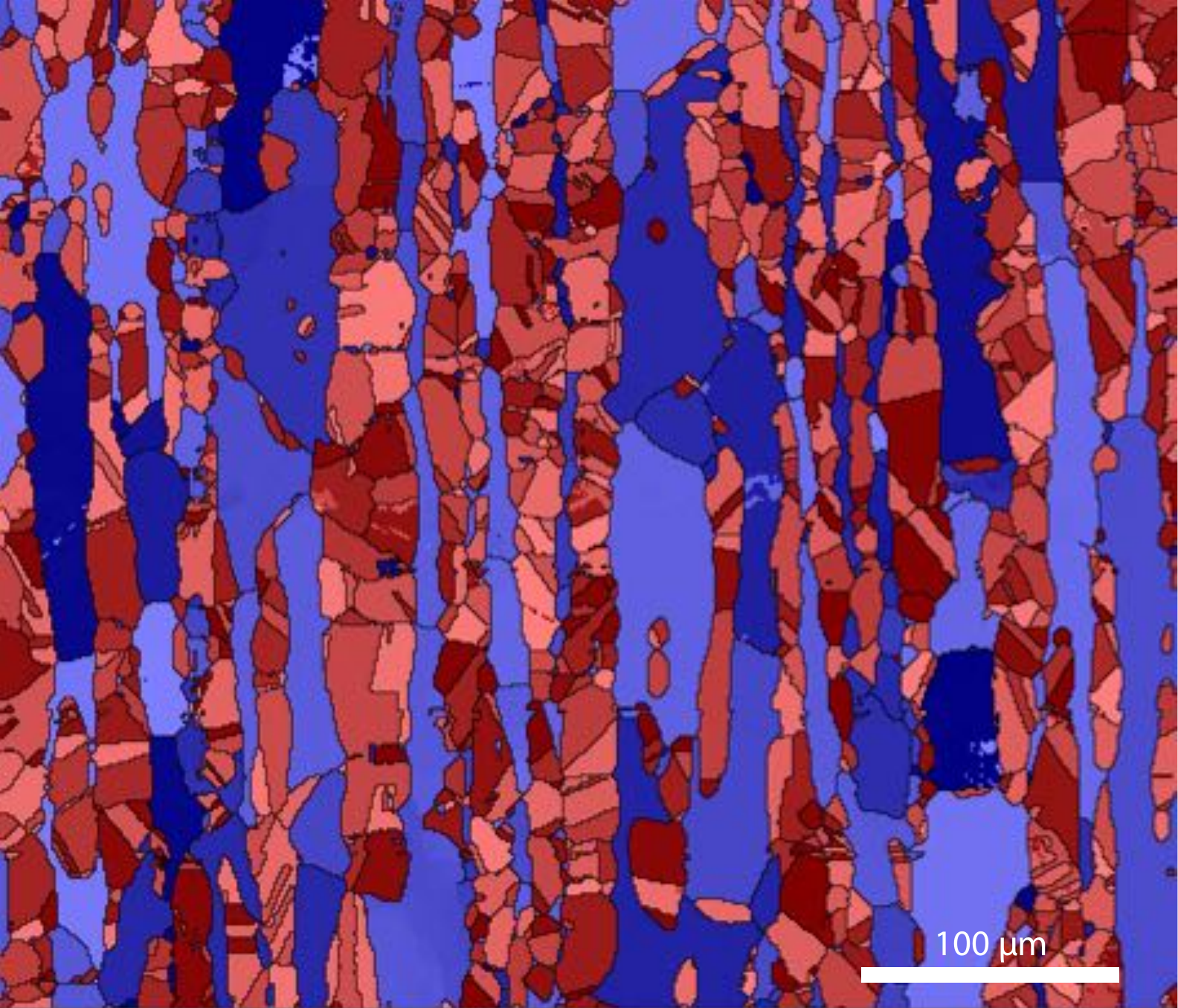}\label{fig:microstruct_transverse}}
\caption{False-color EBSD images of LDX-2101 microstructure. Austenite is colored red and ferrite is colored blue. Grains are shaded according to crystallographic orientation.}\label{fig:microstruct}
\end{figure}

\begin{figure}[h]
\centering
\subfigure[Austenite]{\includegraphics[trim = 0in 0in 0in 0.4in, clip]{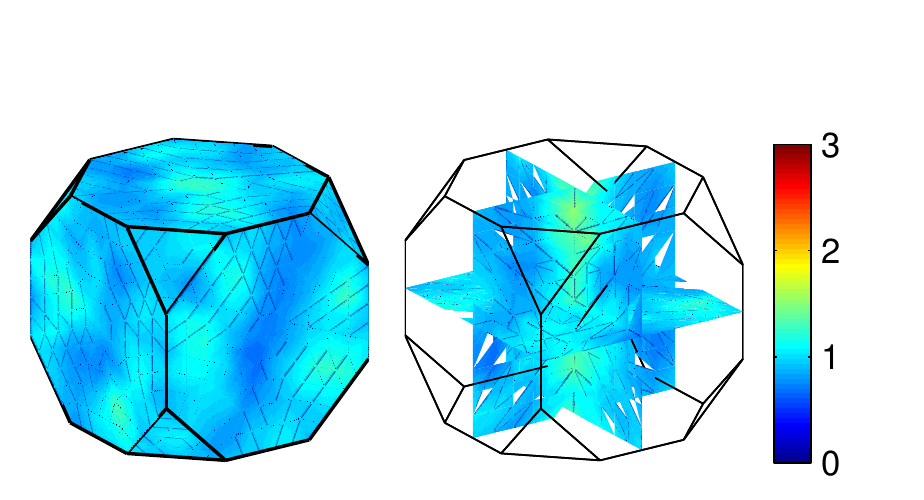}\label{fig:ODF-FCC}}
\subfigure[Ferrite]{\includegraphics[trim = 0in 0in 0in 0.4in, clip]{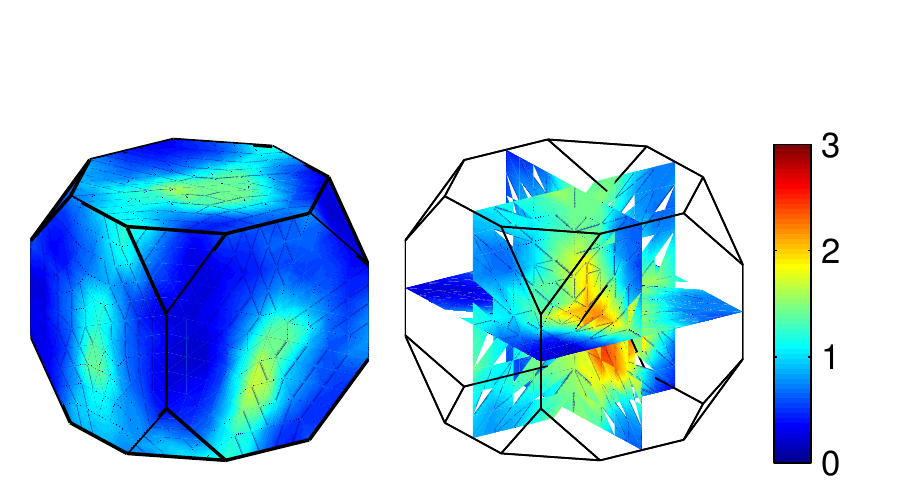}\label{fig:ODF-BCC}}
\caption{Orientation distribution functions for LDX-2101 in multiples of uniform distribution, plotted over the cubic fundamental region in Rodrigues space.}\label{fig:ODF}
\end{figure}

False-color EBSD images of the material microstructure are presented in Figure~\ref{fig:microstruct}. Austenite is red and ferrite is blue. The grains are shaded by crystallographic orientation. The volume fractions of austenite and ferrite are 57\% and 43\%, respectively. The material exhibits a columnar microstructure along the roll direction. The smaller, equiaxed austenite grains are contained in a matrix of larger, irregularly-shaped parent ferrite grains. Annealing twins are present in some of the austenite grains. 
Orientation distribution functions for the two phases are presented in Figure~\ref{fig:ODF}. Ferrite exhibits a weak texture as a result of the rolling process, whereas austenite is more uniformly textured.

LDX-2101 was chosen for this study because it is a dual-phase engineering alloy with desirable properties for both experiments and simulations. Unlike sintered powder specimens, LDX-2101 is an engineering alloy with a host of applications including chemical processing, heat exchangers, and subsea piping~\cite{Alvarez-Armas08a}. The roughly equal volume fractions of the two phases provides good signal for the diffraction experiments. It also helps ensure statistically significant fiber averages for both phases. While the microstructure presents some challenges for three-dimensional spatial modeling, it is not as complex as some other duplex microstructures, such as the colony structures that arise in dual-phase titanium~\cite{Lutjering07a}. 

The selection of LDX-2101 was also based on its mechanical properties. The material systems of primary interest in this work are those whose phases have relatively similar strengths and stiffnesses. In these systems, anisotropy plays a more important role in distinguishing between the two phases. For example the yield behavior of LDX-2101 is qualitatively different from the bimodal yield behavior of the iron-copper systems studied by Han and Dawson~\cite{Han05a} and the duplex stainless steel alloys studied by Baczmanski and Braham~\cite{Baczmanski04a} and Dakhlaoui, et al.~\cite{Dakhlaoui06a, Dakhlaoui07a}. Both phases of LDX-2101 have similar Young's modulus of 200~GPa, but different levels of elastic anisotropy, as quantified by the elastic anisotropy ratio, $r_E$. For cubic crystals, $r_E$ is defined as the ratio of the directional moduli in the crystal $\langle111\rangle$ and $\langle100\rangle$ directions. Austenite ($r_E = 3.2$) has greater elastic anisotropy than ferrite ($r_E = 2.2$). In the model, slip occurs on the \{111\} planes in the $\langle110\rangle$ directions for austenite and on the \{110\} planes in the $\langle111\rangle$ directions for ferrite. Thus, the single-crystal yield surface topology is the same for both phases. In general, slip can also occur on both the \{112\} and \{123\} planes in the $\langle111\rangle$ directions for ferrite, producing a more isotropic yield surface~\cite{Hosford93a}. As determined from experiment and simulation, both the initial slip system strengths and hardening rates are roughly equal between the two phases. Both phases are rate sensitive, with austenite more rate sensitive than ferrite. Details of the model and the procedure used to determine material parameters are given in Sections~\ref{sec:model_formulation} and~\ref{sec:parameters}. It is important to note that the major mechanical difference between the two phases is the level of elastic anisotropy.

\clearpage

\section{Experimental Methodology}
\label{sec:expt_methods}
Monochromatic neutron diffraction was used to measure fiber-averaged lattice strain in LDX-2101 stainless steel specimens subjected to \emph{in-situ} biaxial loading.  Biaxial loading is of interest because biaxial stress states are encountered in a number of engineering applications, including thin plates used in aircraft skins, ship hulls, and pressure vessels. Furthermore, biaxial loading provides an increased level of complexity from uniaxial loading and is a step towards investigating material behavior under generalized multiaxial stress states. 

To date, only a limited number of diffraction experiments have been conducted with \emph{in-situ} biaxial loading. Geandier et al.~\cite{Geandier10a, Geandier14a} and Djaziri et al.~\cite{Djaziri13a, Djaziri14a} performed lattice strain measurements on thin film cruciform specimens. Foecke et al.~\cite{Foecke07a} use lab source x-rays to perform lattice strain measurements on washer-shaped sheet metal specimens undergoing a modified Marciniak biaxial stretching test. Marin et al.~\cite{Marin08a} combined the deep penetration of neutrons with biaxial loading of hollow, cylindrical specimens through simultaneous extension and internal pressurization. Additionally, \emph{in-situ} tension-torsion loading capability has been developed for the VULCAN diffractometer at the Spallation Neutron Source~\cite{Benafan14a}.

%
%

\subsection{Biaxial loading methodology}\label{sec:biaxial_load}

Biaxial loading of thin-walled tubular specimens was achieved using a combination of axial loading and internal pressurization. The advantage of this method over combined tension-torsion loading is that the principal directions of the macroscopic stress remain constant over the entire load history. Furthermore, the principal directions are independent of the ratio of principal macroscopic stress components. The extension-pressurization method also avoids the complicated stress fields associated with cruciform specimens. 

\begin{figure}[h]
\centering
\includegraphics[scale = 0.5]{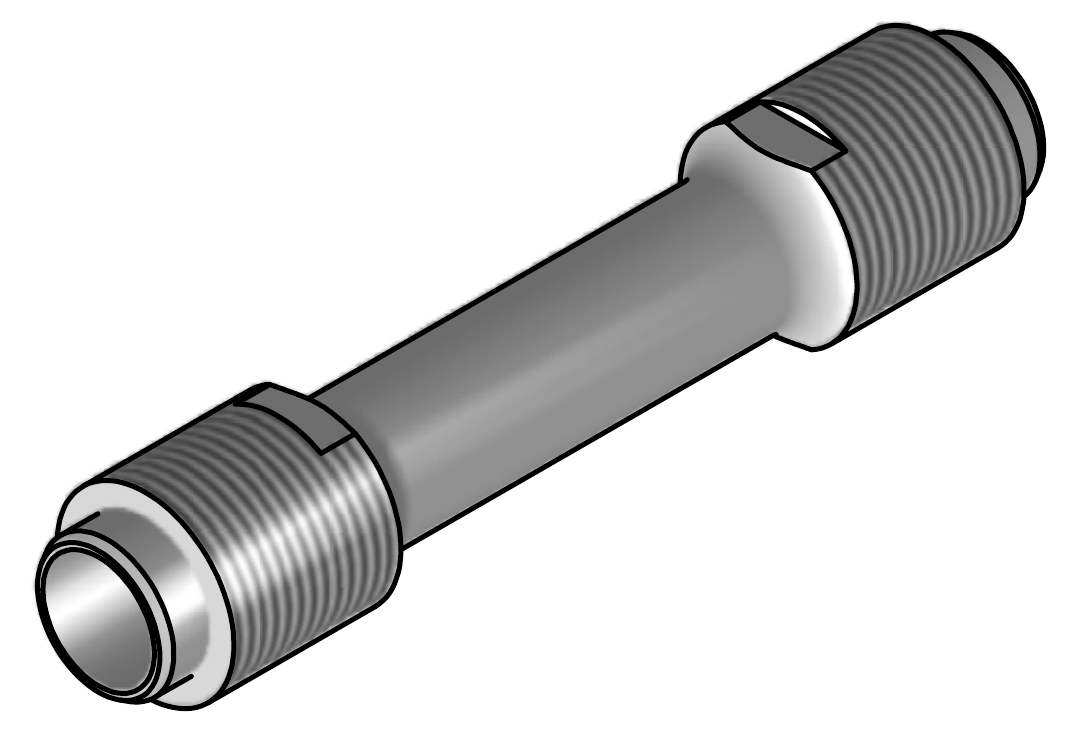}
\caption{Biaxial specimen. \emph{Courtesy of the Canadian Neutron Beam Centre}.}\label{fig:Specimen}
\end{figure}

Specimen geometry is presented in Figure~\ref{fig:Specimen}. The overall specimen length is 127~mm and the macroscopic gauge length is 54~mm. The inner and outer radii  of the gauge section are 8~mm and 9~mm, respectively. The threaded ends of the specimen screw into the load frame grips. Each specimen was instrumented with six strain gauges, four in the axial direction and two in the hoop direction. These strain gauges were used to verify grip alignment and macroscopic loading in the elastic regime. A filler rod was inserted inside the specimen before screwing it into the grips to reduce the amount of hydraulic fluid required to pressurize the specimen. There is sufficient space between the rod and the walls of the specimen so that there is no contact during deformation. As the hydraulic fluid is not transparent to neutrons, the use of the filler rod improves the diffraction signal. It also decreases the pressurization response time, enabling faster loading rates.

An MTS hydraulic load frame was used to apply axial force and a custom-built pressurization system was used to provide internal pressurization. Hoop stress is generated solely from pressurization, whereas both the pressurization and axial force from the load frame contribute to the axial stress. The principal macroscopic stress components are given by:
\begin{equation}\label{eq:radial_stress}
\sigma_{rr} = -\frac{pr_i^2}{r_o^2-r_i^2} \left( \frac{r_o^2}{r^2} - 1 \right)
\end{equation}
\begin{equation}\label{eq:hoop_stress}
\sigma_{\theta\theta} = \frac{pr_i^2}{r_o^2-r_i^2} \left( \frac{r_o^2}{r^2} + 1 \right)
\end{equation}
\begin{equation}\label{eq:axial_stress}
\sigma_{zz} = \frac{pr_i^2}{r_o^2-r_i^2} + \frac{F_z}{A}
\end{equation}
where $p$ is the internal pressure, $F_z$ is the axial force applied by the load frame, and $A$ is the cross-sectional area of the specimen. The inner and outer radii are denoted $r_i$ and $r_o$, respectively. Both the radial and hoop stresses are functions of radial position, $r$. In this work, the macroscopic stress is evaluated at the mean radial position. For the specimens used in this experiment, the radial stress is 5.7\% of the hoop stress and can therefore be approximated as small. The biaxial ratio, $BR$, is defined as the ratio of hoop stress to axial stress.

\begin{figure}[h]
\centering
\subfigure[Full]{\includegraphics[width = 0.49\textwidth]{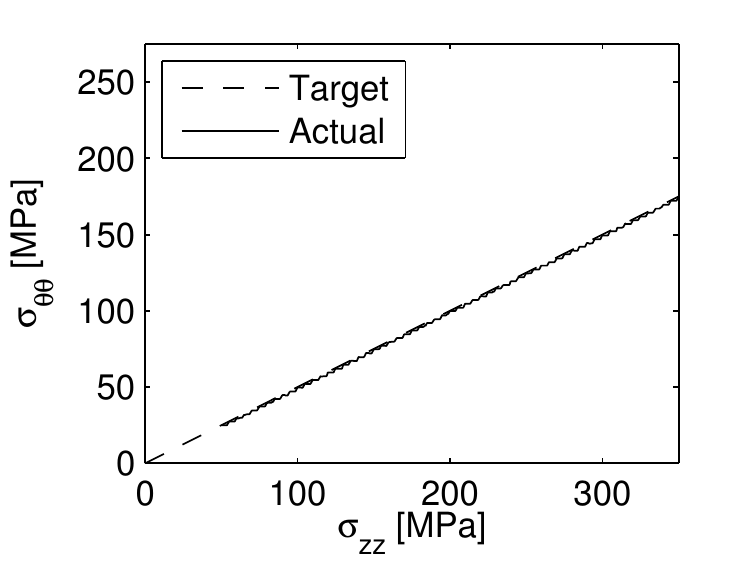}\label{fig:ExpLoadCtrlFull}}
\subfigure[Zoom]{\includegraphics[width = 0.49\textwidth]{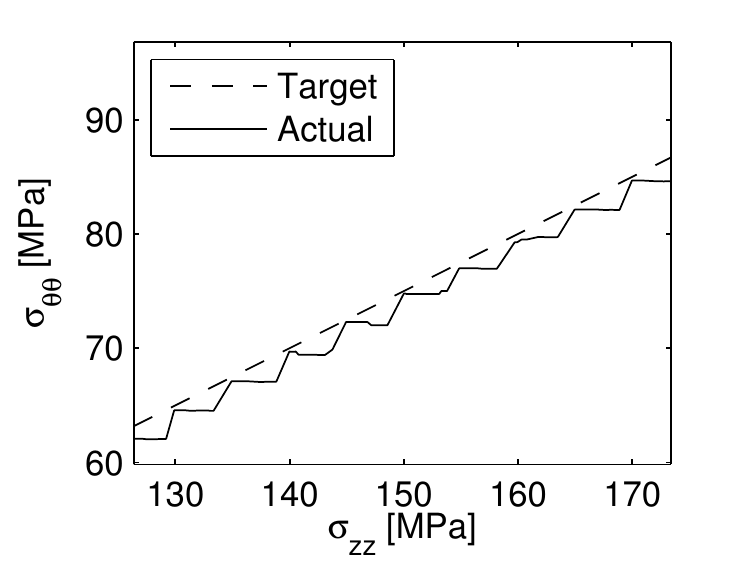}\label{fig:ExpLoadCtrlZoom}}
\caption{Sample load path for $BR = 0.5$. Proportional biaxial load control is attained by dividing each load step into a series of time increments. The load frame waits for a signal from the pressurization system before advancing to the next increment. The resulting stair-step nature is evident in the zoom view of the load history.}\label{fig:ExpLoadCtrl}
\end{figure}

Proportional biaxial loading, in which the biaxial ratio remains constant, was achieved by breaking each load step up into a series of smaller load increments. The pressurization system has a slower response time than the load frame. The load frame waits for a signal from the pressurization controller before proceeding to the next load increment. In this manner, the biaxial ratio remains close to the prescribed value. A sample load history is shown in Figure~\ref{fig:ExpLoadCtrl}. In this example, the biaxial ratio is approximately constant. A closer examination of the load history reveals a stair-step pattern that is the result of alternating ramp and dwell episodes of the load frame.

Specimens were loaded at a constant axial force rate of 100~N/s corresponding to an engineering axial stress rate of 1.9~MPa/s. Axial strain rate varied from $10^{-5}~\mathrm{s^{-1}}$ in the elastic regime to $4 \times 10^{-3}~\mathrm{s^{-1}}$ in fully developed plasticity. It would have been ideal to load at a constant strain rate. However, at the time of the experiment, the control software did not possess the capability to load at constant strain rate and then dwell in load control at a target load. This feature has since been implemented by calculating a new load rate for each increment.

\begin{figure}[h]
\centering
\subfigure[Load path]{\includegraphics{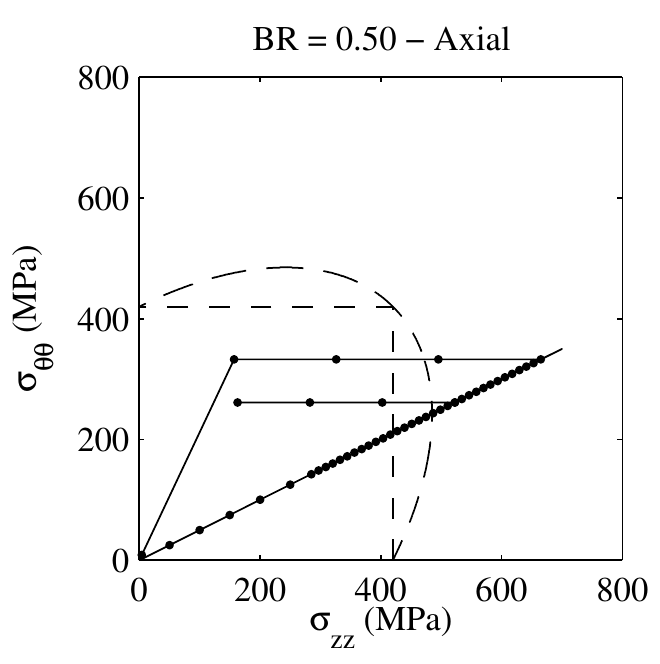}\label{fig:SampLoadPath}}
\subfigure[Axial stress-strain response]{\includegraphics{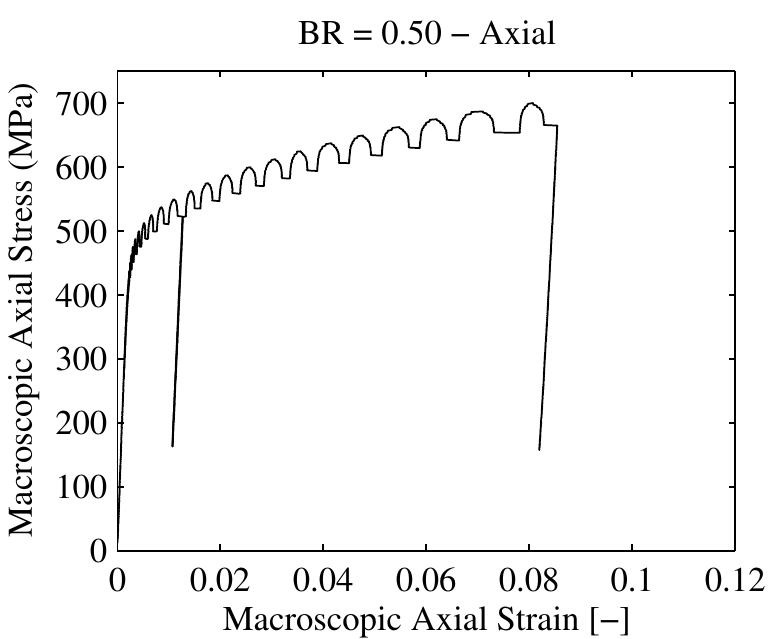}\label{fig:SampStressStrain}}
\caption{Representative (a) load path and (b) axial stress-strain curve for $BR~=~0.5$. Tresca and von Mises yield surfaces are represented with dashed lines, and circles denote data collection points.}\label{fig:SampLoadHist}
\end{figure}

Specimens were loaded under proportional biaxial loading. Five levels of stress biaxiality, ranging from uniaxial ($BR = 0)$ to balanced biaxial ($BR = 1$) were investigated. A sample load history and axial stress-strain curve are shown in Figure~\ref{fig:SampLoadHist} for a biaxial ratio of 0.5. In stress space, proportional loading corresponds to loading radially outward along a straight path from the origin. The loading is paused at target load steps for the collection of diffraction data. Before data collection begins, the load is reduced by 5\% to back off the yield surface in an effort to reduce plastic deformation during collection. Despite the reduction in load, a non-negligible amount of plastic deformation occurs during data collection. Deformation during data collection corresponds to the gently-sloped lines between the unloading and loading episodes in the stress-strain curve. Data collection times range from about thirty minutes to an hour per load step. Additionally, two elastic unloading episodes were performed to obtain supplemental information that was used to validate the measurements. One unloading episode was performed after the elasto-plastic transition, and the other at the end of the load sequence. During each of these unloading episodes, only the axial component of stress was reduced, while the hoop component of stress remained constant.

%
%

\subsection{{\it In situ} loading, neutron diffraction experimental configuration}\label{sec:exp_setup}

\begin{figure}[h]
\centering
\includegraphics[width = 0.6\textwidth]{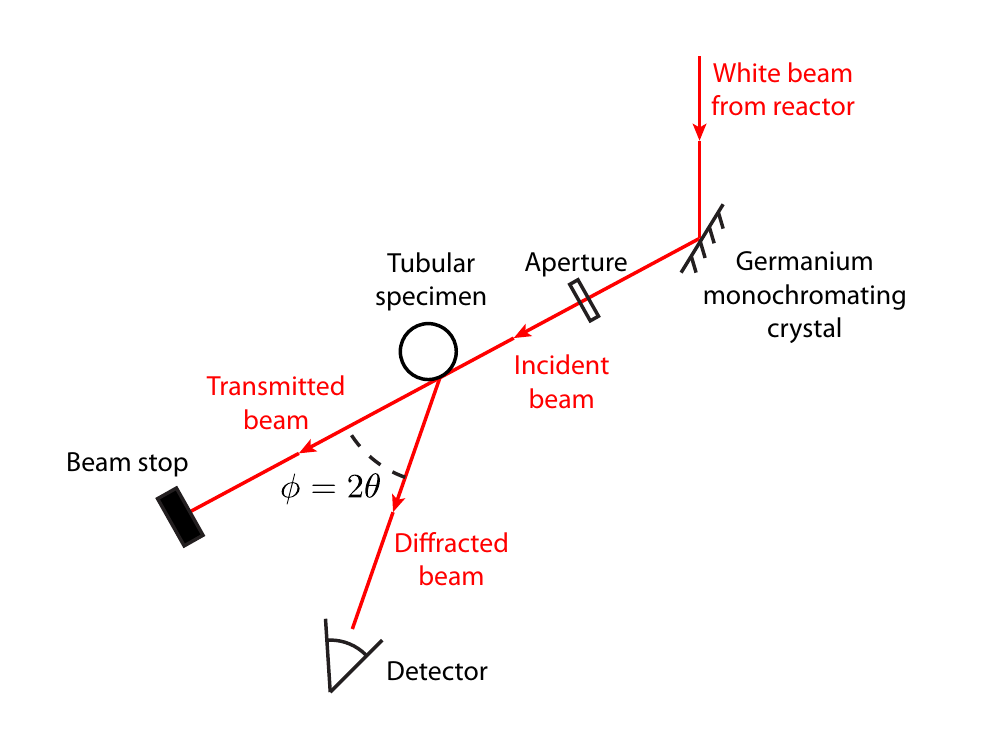}
\caption{Top-view schematic of the diffraction geometry. The monochromatic incident beam is created from the reactor beam using a germanium crystal monochromator. The incident beam passes through the detector shielding and impinges on the wall of the tubular specimen. The diffracted beam from the specimen is sampled by the detector. The detector angle, $\phi$, is twice the Bragg angle, $\theta$.}
\label{fig:DiffractionSchematic}
\end{figure}

Experiments were conducted at the L3 beamline at the Canadian Neutron Beam Centre. A schematic of the experimental geometry is presented in Figure~\ref{fig:DiffractionSchematic}. The [331] reflection of a germanium crystal monochromator was used to produce a 0.172~nm wavelength beam. The beam impinged on the specimen, mounted in the load frame. The diffracted beam was sampled by a 32-wire neutron detector. The angular distance between wires was 0.0824\textdegree~for a total span of 2.55\textdegree~in the plane spanned by the incident and diffracted beams. Each wire registered a count when a neutron passed through it. The incident beam, load frame, and detector could be positioned so as to sample different crystallographic fibers. Diffracted intensity distributions were measured one diffraction peak at a time. 

\begin{figure}[h]
	\centering
	\subfigure[Full view.]{\includegraphics[width = 3.0in]{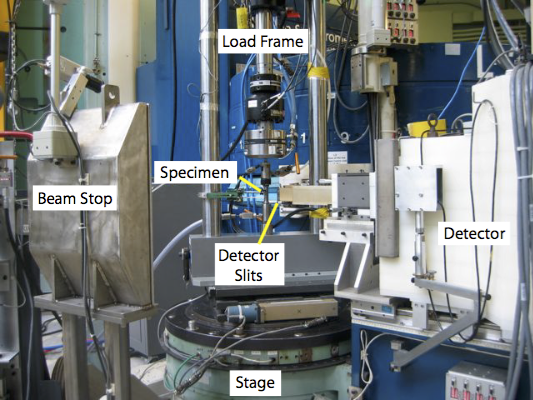}}
	\subfigure[Zoom view.]{\includegraphics[width = 3.0in]{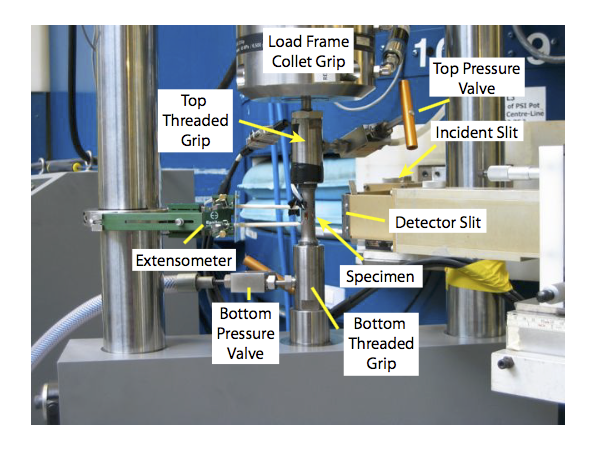}}
	\caption{Experimental setup for hoop and radial lattice strain measurements.}
	\label{fig:exp_setup_hr}
\end{figure}

\begin{figure}[h]
	\centering
	\subfigure[Full view.]{\includegraphics[width = 3.0in]{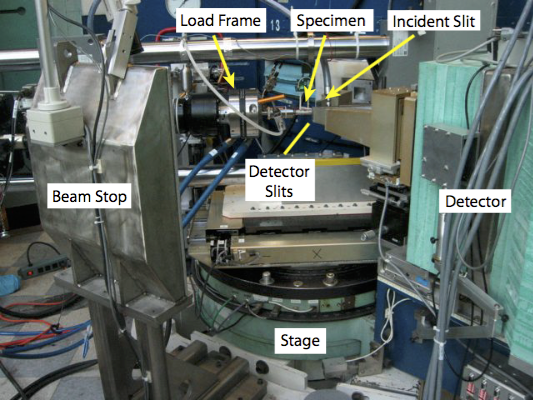}}
	\subfigure[Zoom view.]{\includegraphics[width = 3.0in]{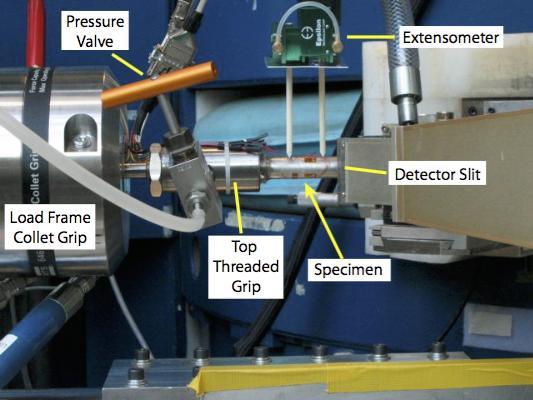}}
	\caption{Experimental setup for axial lattice strain measurements.}
	\label{fig:exp_setup_axial}
\end{figure}

Diffraction peaks associated with scattering vectors aligned with the three principal loading directions were measured. The load frame was mounted vertically for the measurement of hoop and radial lattice strains and horizontally for the axial lattice strain measurement. Because of the need to manually rotate the load frame, two different specimens were used for each biaxial ratio, one for hoop/radial lattice strain measurement and the other for axial lattice strain measurement. The experimental setups for the hoop/radial and axial lattice strain measurements are shown in Figures~\ref{fig:exp_setup_hr} and~\ref{fig:exp_setup_axial}, respectively. Aluminum, which is mostly transparent to neutrons, was used for the filler rod in the hoop/radial configuration, so as not to obstruct the incident beam as it passed through the specimen during hoop lattice strain measurements. A cadmium coated aluminum rod was used to provide shielding in the axial configuration. 

\begin{figure}[h]
\centering
\subfigure[Hoop]{\includegraphics[scale = 0.6]{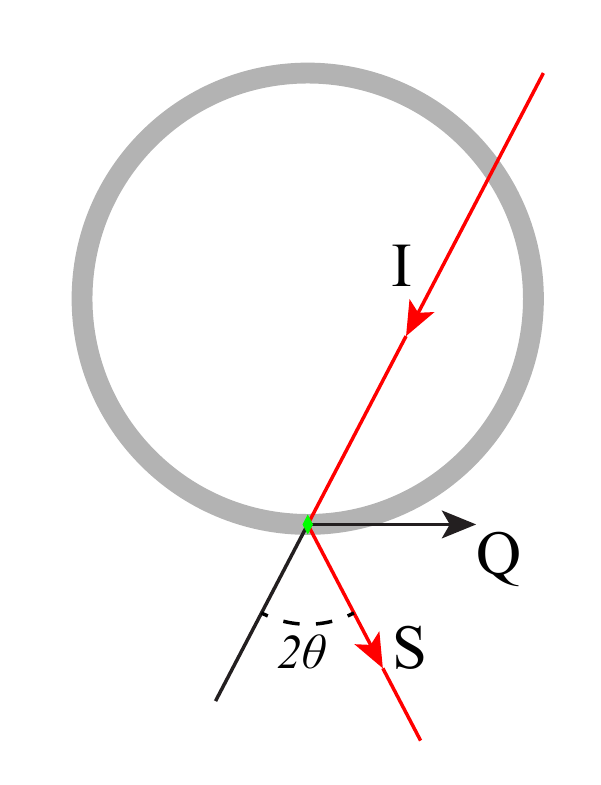}\label{fig:DiffractGeoHoop}}
\subfigure[Radial]{\includegraphics[scale = 0.6]{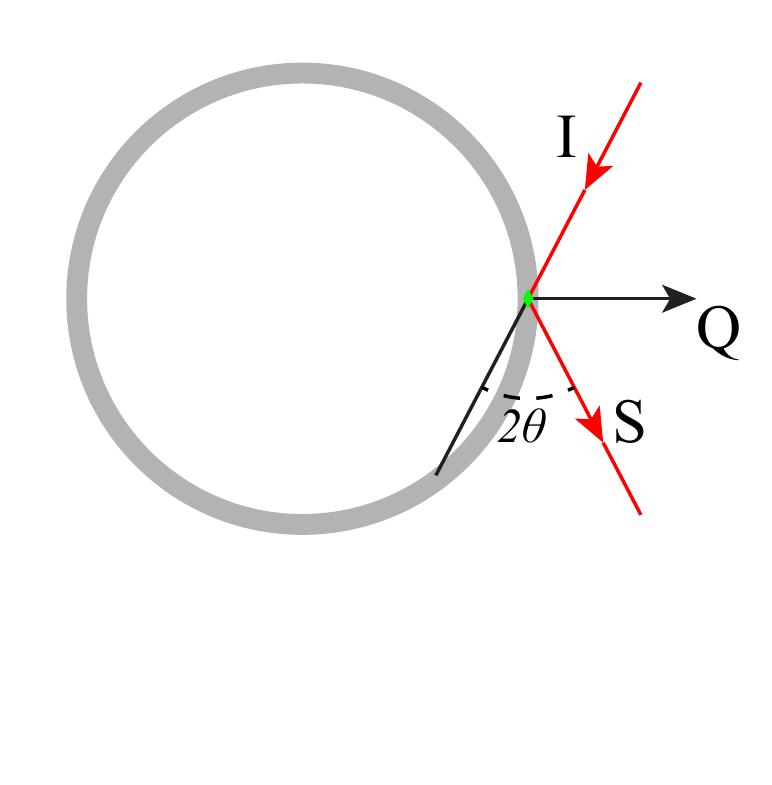}\label{fig:DiffractGeoRadial}}
\subfigure[Axial]{\includegraphics[scale = 0.6]{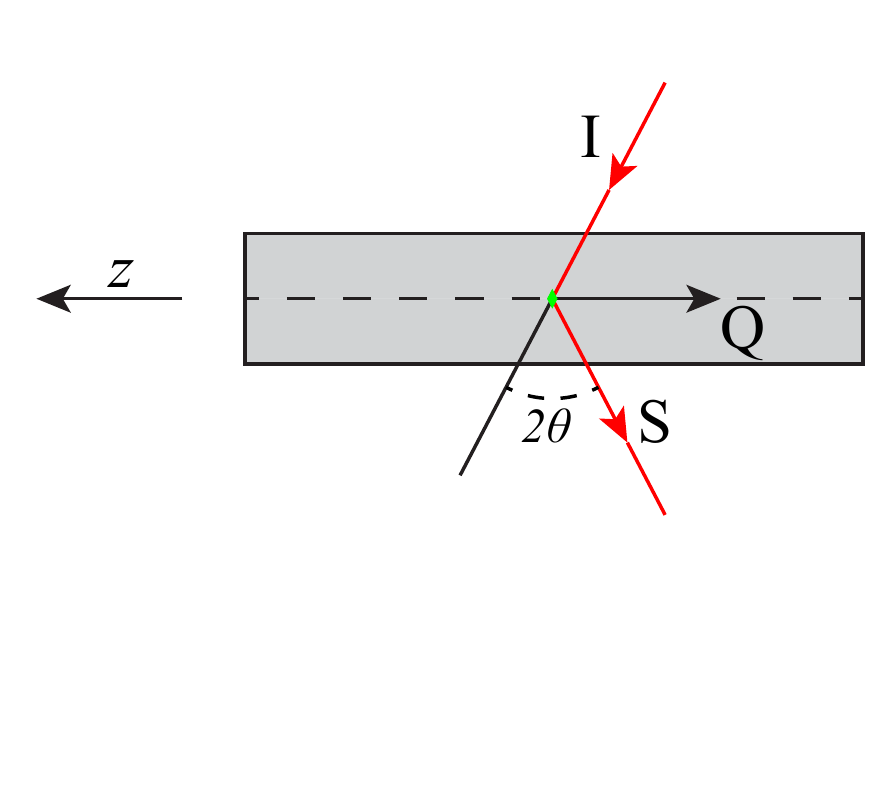}\label{fig:DiffractGeoAxial}}
\caption{Top view of diffraction geometries, depicting the relationship between the specimen, incident beam I, diffracted beam S, and scattering vector Q. Diffraction volumes are represented by the green diamond at the intersection of the incident and diffracted beams.}\label{fig:diffract_geo}
\end{figure}

The diffraction volume was defined by a pair of incident and detector slits. For the hoop/radial lattice strain measurements, 0.75~mm x 30~mm slits were used, resulting in a pencil shaped diffraction volume. For the axial lattice strain measurements, 4~mm slits were used. The diffraction volume extended through the top and bottom walls of the specimen. Diffraction geometries are illustrated in Figure~\ref{fig:diffract_geo}.

%
%

\subsection{Lattice strain measurements}\label{sec:exp_method}

The specimen was mounted in the load frame and connected to the hydraulic pressurization system. An extensometer was attached to the specimen to measure macroscopic axial strain. 
The specimen was loaded according to the load history described in Section~\ref{sec:biaxial_load}. Neutron diffraction data were acquired at target load steps. For hoop and radial lattice strains, wall scans were performed at each load step, before data acquisition, to locate the center of the specimen wall. The specimen was then positioned such that the center of the gauge volume coincided with the center of the wall thickness. Wall scans were not needed for the axial lattice strain measurements.

Diffracted beams were measured for the FCC \{200\}, \{111\}, and \{220\} crystal planes and BCC \{200\}, \{220\}, and \{211\} crystal planes.  Measurements were taken for a total of 18 crystallographic fibers (6 crystal directions x 3 sample directions). Five levels of stress biaxiality, ranging from uniaxial ($BR = 0$) to balanced biaxial ($BR = 1$) were investigated. Monitor counts for the incident beam were chosen such that the diffraction peak height was at least 300 counts, ensuring sufficient signal for determining the center of each peak. The monitor was manually adjusted during the experiment to account for changes in diffracted intensity due to texture evolution. Data collection times per load step, including wall scans, were on the order of one hour for hoop/radial lattice strain measurements and thirty minutes for axial lattice strain measurements. Total run time for each load history was about three days for hoop/radial measurements and one and a half days for axial measurements.

%
%
The procedure for extracting the lattice strains from neutron count (peak) profiles is detailed in Appendix~\ref{sec:exp_data_reduction}.  From the peak profiles, lattice strain histories are determined for the 18 crystallographic fibers  and 5 levels of stress biaxiality specified above.  
Sources of uncertainty in the experiment include diffraction peak fitting, positioning of the detector, and alignment of the detector slits. For these measurements, lattice strain uncertainty is dominated by the uncertainty in diffraction peak fitting.  Consequently, only the lattice strain uncertainty due to peak fitting is considered in the uncertainty analysis. This analysis therefore provides a lower, but reasonable, estimate of the uncertainty.  Typical lattice strain uncertainty is $\pm200$~$\mu\epsilon$ at the 95\% confidence interval.
\clearpage

\section{Modeling Methodology}
\label{sec:sim_methods}
Deformation of LDX-2101 polycrystalline aggregates was simulated using \textbf{\textit{FEpX}}.
 \textbf{\textit{FEpX}} is a finite element package for simulating elasto-plastic deformation of virtual single crystals and polycrystalline aggregates~\cite{Dawson14a}. 
 In this section, we provide a brief overview of the \textbf{\textit{FEpX}} capabilities, 
 summarize the single crystal constitutive equations, and 
 describe the special procedures used to impose the boundary conditions
 consistent with the  biaxial loading experiments.

\subsection{Finite element framework}\label{sec:fe_framework}

\fepx\, was developed for modeling the response of virtual microstructures over large strain motions and incorporates the following capabilites:
\begin{itemize}
\item{nonlinear kinematics capable of handling motions with both large strains and large rotations;}
\item{anisotropic elasticity based on cubic or hexagonal crystal symmetry;}
\item{anisotropic plasticity based on rate-dependent slip on a restricted number of systems for cubic or hexagonal symmetry;} 
\item{evolution of state variables for crystal lattice orientation and slip system strengths.} 
\end{itemize}
To accommodate these behaviors the finite element formulation has incorporated a number a numerical features, such as:
\begin{itemize}
\item{higher-order, isoparametric elements with quadrature for integrating over the volume;} 
\item{implicit update of the stress in integrations over time;}
\item{monotonic and cyclic loading under quasi-static conditions.}
\end{itemize}

The virtual microstructures are instantiated to define grains and to discretize the grains into multiple 10-node tetrahedral finite elements. 
The local behaviors associated with the material within an element correspond to those of a single crystal. 
A phase and initial crystallographic orientation are assigned to each element. 
Each element's state, characterized by its elastic strain, crystallographic orientation, and slip system hardness, evolves independently from that of the other elements. 
Velocity, traction, or mixed boundary conditions are applied to each surface of the virtual specimen. 
Compatibility of the deformation is enforced by the smoothness of motion guaranteed by the continuity of the interpolation functions. 
The deformation satisfies quasi-static equilibrium, which is enforced via the weak form of conservation of linear momentum. 

An implicit time integration scheme is used to solve for the velocity field and material state.  An estimated material state at the end of each time increment is used to evaluate the constitutive equations. Implicit integration ensures stability. The solution algorithm for each time increment begins with initializing an initial guess of the velocity field. The deformed geometry at the end of the time increment is then estimated, based on the velocity field. The velocity gradient is computed and used to solve for the crystal state at each quadrature point.  Iteration continues on the geometry, crystal state, and velocity field until the velocity solution is converged. The solution then advances to the next time increment.
%
%

\subsection{Constitutive model for elastoplastic behavior}\label{sec:model_formulation}

The constitutive model employs a kinematic decomposition of the deformation into a sequence of deformations due to crystallographic slip, rotation, and elastic stretch. Using this decomposition, the deformation gradient, $\boldsymbol{F}$, can be represented as
\begin{equation}\label{eqn:kinematic_decomp}
\boldsymbol{F} = \boldsymbol{V}^e  \boldsymbol{R}^*  \boldsymbol{F}^p 
\end{equation}
where $\boldsymbol{F}^p$, $\boldsymbol{R}^*$, and  $\boldsymbol{V}^e$ correspond to crystallographic slip, rotation, and elastic stretch, respectively.  This decomposition defines a reference configuration $\mathscr{B}_0$, a deformed configuration $\mathscr{B}$, and two intermediate configurations $\bar{\mathscr{B}}$ and $\hat{\mathscr{B}}$. The state equations are written in the intermediate $\hat{\mathscr{B}}$ configuration defined by the relaxation of the elastic deformation from the current $\mathscr{B}$ configuration. Elastic strains, $\boldsymbol{\epsilon}^e$, are required to be small, which allows the elastic stretch tensor to be written as
\begin{equation}
\boldsymbol{V}^e = \boldsymbol{I} + \boldsymbol{\epsilon}^e
\end{equation}
The velocity gradient, $\boldsymbol{L}$, is calculated from the deformation gradient using the relationship
\begin{equation}
\boldsymbol{L} = \dot{\boldsymbol{F}} \boldsymbol{F}^{-1}
\end{equation}
The velocity gradient can be decomposed into a symmetric deformation rate, $\boldsymbol{D}$, and skew-symmetric spin rate, $\boldsymbol{W}$
\begin{equation}
\boldsymbol{L} = \boldsymbol{D} + \boldsymbol{W}
\end{equation}
A generic symmetric tensor, $\boldsymbol{A}$, can be additively decomposed into mean and deviatoric components
\begin{equation}\label{eqn:vol_dev_decomp}
\boldsymbol{A} = \sfrac{1}{3} \mathrm{tr} (\boldsymbol{A}) \boldsymbol{I} + \boldsymbol{A}^\prime
\end{equation}
where $\sfrac{1}{3} \mathrm{tr} (\boldsymbol{A})$ is the scalar mean component and $\boldsymbol{A}^\prime$ is the tensorial deviatoric component.
Utilizing Equations~\ref{eqn:kinematic_decomp}-\ref{eqn:vol_dev_decomp} the volumetric deformation rate, deviatoric deformation rate, and spin rate can be expressed as
\begin{equation}
\mathrm{tr} \left( \boldsymbol{D} \right) = \mathrm{tr} \left( \dot{\boldsymbol{\epsilon}}^e \right)
\end{equation}
\begin{equation}
\boldsymbol{D}^\prime = \dot{\boldsymbol{\epsilon}}^{e\prime} + \hat{\boldsymbol{D}}^p + \boldsymbol{\epsilon}^{e\prime} \cdot \hat{\boldsymbol{W}}^p - \hat{\boldsymbol{W}}^p \cdot \boldsymbol{\epsilon}^{e\prime}
\end{equation}
\begin{equation}\label{eqn:spin_rate}
\boldsymbol{W} = \hat{\boldsymbol{W}}^p + \boldsymbol{\epsilon}^{e\prime} \cdot \hat{\boldsymbol{D}}^p - \hat{\boldsymbol{D}}^p \cdot \boldsymbol{\epsilon}^{e\prime}
\end{equation}
where $\hat{\boldsymbol{D}}^p$ and $\hat{\boldsymbol{W}}^p$ are the plastic deformation and spin rates.

Constitutive equations relate the stress to the deformation. The Kirchhoff stress, $\boldsymbol{\tau}$, in the $\hat{\mathscr{B}}$ configuration is related to the Cauchy stress, $\boldsymbol{\sigma}$, in the current configuration $\mathscr{B}$ by the determinant of the elastic stretch tensor
\begin{equation}
\boldsymbol{\tau} = \mathrm{det} (\boldsymbol{V^e}) \boldsymbol{\sigma}
\end{equation}
The Kirchhoff stress is related to the elastic strain through anisotropic Hooke's law
\begin{equation}
\mathrm{tr} \left( \boldsymbol{\tau} \right) = 3K \mathrm{tr} \left( \boldsymbol{\epsilon}^e \right)
\end{equation}
\begin{equation}
\boldsymbol{\tau}^\prime = \mathscr{C}^* \boldsymbol{\epsilon}^{e\prime}
\end{equation}
where $K$ is the bulk modulus and $\mathscr{C}^*$ is the fourth-order stiffness tensor.

Plastic deformation due to crystallographic slip occurs on a restricted set of slip systems. For FCC crystals, slip occurs on the \{111\} planes in the [110] directions. For BCC crystals, slip on the \{110\} planes in the [111] directions is considered. The plastic deformation rate and plastic spin rate are given by 
\begin{equation}
\hat{\boldsymbol{D}}^p = \sum_\alpha \dot{\gamma}^\alpha \hat{\boldsymbol{P}}^\alpha
\end{equation}
\begin{equation}
\hat{\boldsymbol{W}}^p = \dot{\boldsymbol{R}^*} \boldsymbol{R}^{*T} + \sum_\alpha \dot{\gamma}^\alpha \hat{\boldsymbol{Q}}^\alpha
\end{equation}
where $\hat{\boldsymbol{P}}^\alpha$ and $\hat{\boldsymbol{Q}}^\alpha$ are the symmetric and skew symmetric components of the Schmid tensor $\hat{\boldsymbol{T}}^\alpha$, and $\dot{\gamma}^\alpha$ is the shear rate on the $\alpha$-slip system. The Schmid tensor is defined as the dyad of the slip direction, $\hat{\mathbf{s}}^\alpha$, and slip plane normal, $\hat{\mathbf{m}}^\alpha$
\begin{equation}
\hat{\boldsymbol{T}}^\alpha = \hat{\mathbf{s}}^\alpha \otimes \hat{\mathbf{m}}^\alpha
\end{equation}
The slip system shear rate for a given slip system is related to the critical resolved shear stress on that slip system, $\tau^\alpha$, by a power law relationship
\begin{equation}
\dot{\gamma}^\alpha = \dot{\gamma}_0 \left( \frac{\vert \tau^\alpha \vert}{g^\alpha} \right)^\frac{1}{m} \mathrm{sgn} \left( \tau^\alpha \right)
\end{equation}
where $\dot{\gamma}_0$ is a reference slip system shear rate and $g^\alpha$ is the slip system strength. The resolved shear stress is the projection of the deviatoric stress onto the slip system
\begin{equation}
\tau^\alpha = \boldsymbol{\tau}^\prime : \hat{\boldsymbol{P}}^\alpha
\end{equation}

A modified Voce hardening law is used to describe slip system strength evolution 
\begin{equation} \label{eqn:Voce}
\dot{g}^\alpha = h_0 \left( \frac{g_s - g^\alpha}{g_s - g_0} \right)^{n^\prime} \sum_\alpha \dot{\gamma}^\alpha
\end{equation}
where $h_0$ is the reference hardening rate, $g_s$ is the saturation strength, and $n^\prime$ is the hardening exponent. The hardening law is isotropic; at a given material point, all slip systems harden at the same rate. Equation~\ref{eqn:spin_rate} describes the crystal reorientation.

\subsection{Triaxial load control}\label{sec:triaxial}

Two modes of triaxial load control were implemented in \textbf{\textit{FEpX}}, a constant load rate mode to model the experiments, and a constant engineering strain rate mode for additional investigations into material behavior. 
Details for the two modes are provided in Appendix~\ref{sec:sim-bc_implemetation}.
 The constant load rate mode is capable of modeling the dwell episodes that occurred during neutron data acquisition. 

Both control modes are designed for cube domains where the principal loading axes are aligned with the cube axes. Symmetry boundary conditions are applied to three orthogonal mesh surfaces. On these surfaces, normal velocities and tangential tractions are prescribed to be zero. The other three orthogonal surfaces are the control surfaces. On each control surface, a uniform normal velocity is prescribed. Tangential tractions are prescribed to be zero. There is an iteration on the applied surface velocities so that either the desired load or the desired ratio between loads is obtained. 

While iterating on applied boundary conditions increases computation time by a factor of about two to six, the present surface velocity method possesses distinct advantages over prescribing traction boundary conditions on the control surfaces. Velocity convergence issues are frequently encountered with the traction method because the macroscopic velocity is sensitive to the macroscopic load. Convergence is particularly problematic when loading at a constant load rate because the velocity changes by several orders of magnitude. The surface velocity method significantly reduces convergence issues by enabling reasonable initial guesses of the velocity field for each iteration.

Constant strain rate control for biaxial loading was implemented by Marin et al.~\cite{Marin12a}. Their implementation utilized applied tractions on the control surfaces. Iterations were performed on a single scaling parameter for the applied tractions, so as to maintain a constant velocity in the primary control direction. Both Marin et al.'s method and the surface velocity method require boundary condition iterations. However, the surface velocity method has the advantage of improved convergence.
\clearpage

\section{Simulation of the Biaxial Loading Experiments}
\label{sec:simulations}
To simulate the  response of the  LDX-2101 material to biaxial loading, it is first necessary to build a virtual sample and assign attributes to it.  The attributes include the mechanical properties of each phase plus values for the initial state, namely the strengths of the phases and the orientations of the crystal lattices of the grains.   This section documents the choices made in constructing a suitable virtual sample.  In reaching this sample, a comprehensive  study was carried out to examine the sensitivities of the lattice strains to the sample instantiation and the single crystal elastic and plastic constitutive parameters.  This study is reported in a separate paper~\cite{pos_daw_parmstudy}.   The study showed that the macroscopic stress-strain and fiber-averaged lattice strain responses for LDX-2101 were relatively insensitive to microstructure and justified the use of the simplified mesh to extract these particular quantities of interest.  Further, it demonstrated that the single crystal constitutive parameters were capable of predicting the mechanical response well for the uniaxial loading case.  

\subsection{Virtual microstructure instantiation}\label{sec:instantiation}

LDX-2101 microstructure presents several challenges for mesh generation. The first challenge is the irregular shapes of the ferrite grains. As discussed in Section~\ref{sec:material}, austenite grains transform from the ferrite matrix along grain boundaries. This leaves behind irregularly-shaped ferrite grains. The surface concavity of the ferrite grains poses challenges for most meshing algorithms, including those used by Neper~\cite{Quey11a}, a software package for virtual polycrystal generation and meshing. Furthermore, it is challenging to obtain meaningful grain size distribution statistics from two-dimensional EBSD data without serial sectioning, as conventional stereographic analysis relies on the assumption of ellipsoidal grains~\cite{Rollett07a}. Average grain sizes and grain size distributions are therefore only estimates, determined by eye, to reproduce the observed microstructure qualitatively, rather than quantitatively. Another software package DREAM.3D~\cite{Groeber14a} includes methods for generating virtual, voxelated microstructures from grain statistics. It has the ability to model dual phase materials formed by the precipitation of a secondary phase along grain boundaries. However, the DREAM.3D algorithm cannot, at present, reproduce the columnar phase structure observed in LDX-2101. The second challenge to microstructure modeling is the disparate grain size between the two phases. Austenite grains are smaller, on the order of 20-50~$\mu$m diameter, whereas the larger ferrite grains can extend over 100~$\mu$m. The disparate grain sizes poses challenges for mesh refinement. Furthermore, about half of the austenite grains contain annealing twins, which adds subgrain detail.

To address these challenges, idealized microstructures were created using hexagonal prismatic building blocks. Meshes were constructed by tessellating a hexagonal prismatic base mesh, comprised of multiple tetrahedral elements. The hexagonal prismatic base regions formed building blocks that were then stitched together to form grains. Annealing twins are just on the order of mesh resolution and were not represented in the virtual microstructure.

\begin{figure}[h]
	\centering
	\subfigure[Three-dimensional base structure.]{\includegraphics[width=0.49\linewidth]{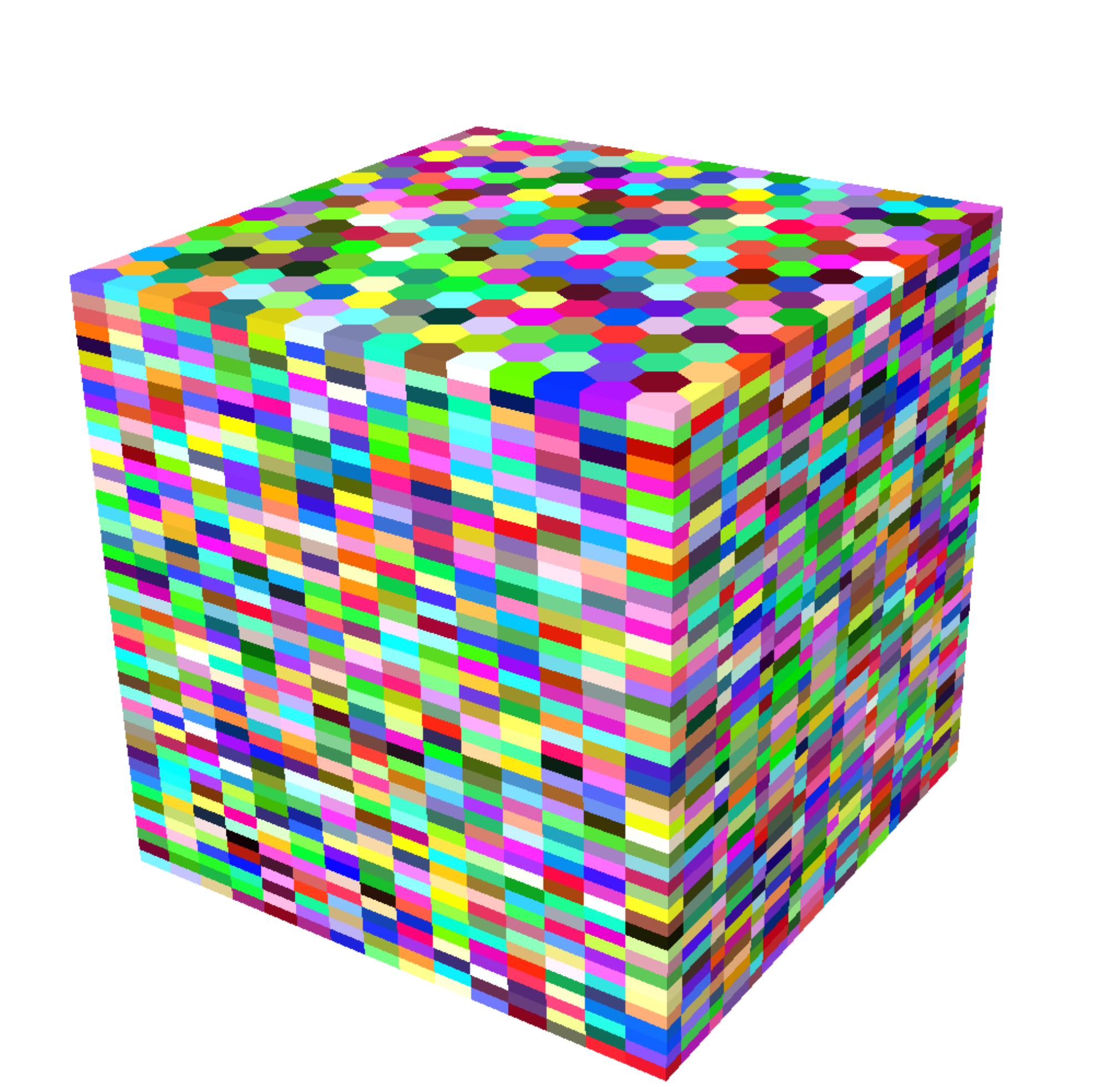}\label{fig:base_mesh}}
	\subfigure[Two-dimensional parent microstructure.]{\includegraphics[width=0.49\linewidth]{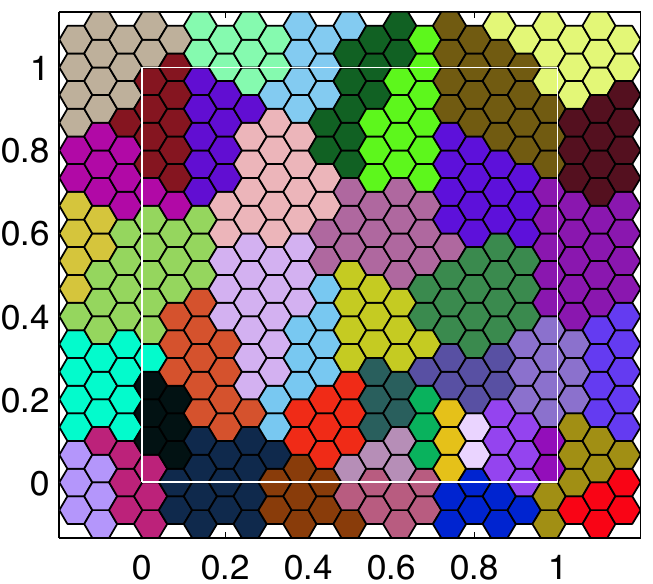}\label{fig:ferrite2D}}
	\subfigure[Two-dimensional microstructure. Austenite regions are colored white.]{\includegraphics[width=0.49\linewidth]{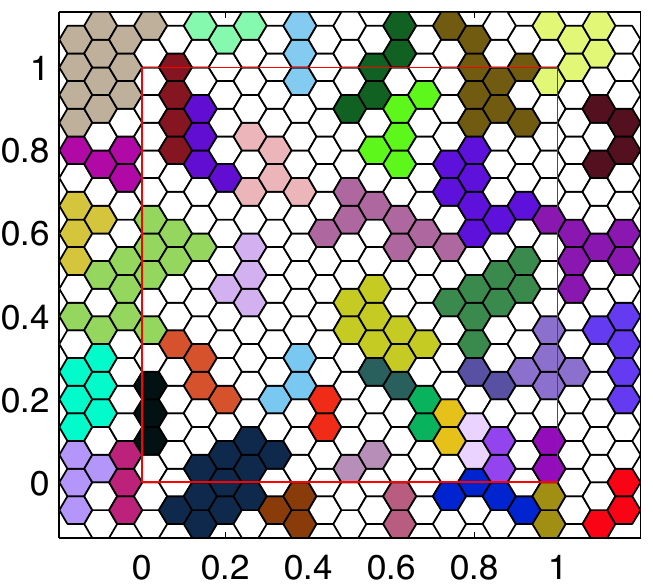}\label{fig:microstruct2D}}
	\subfigure[Three-dimensional microstructure.]{\includegraphics[width=0.49\linewidth]{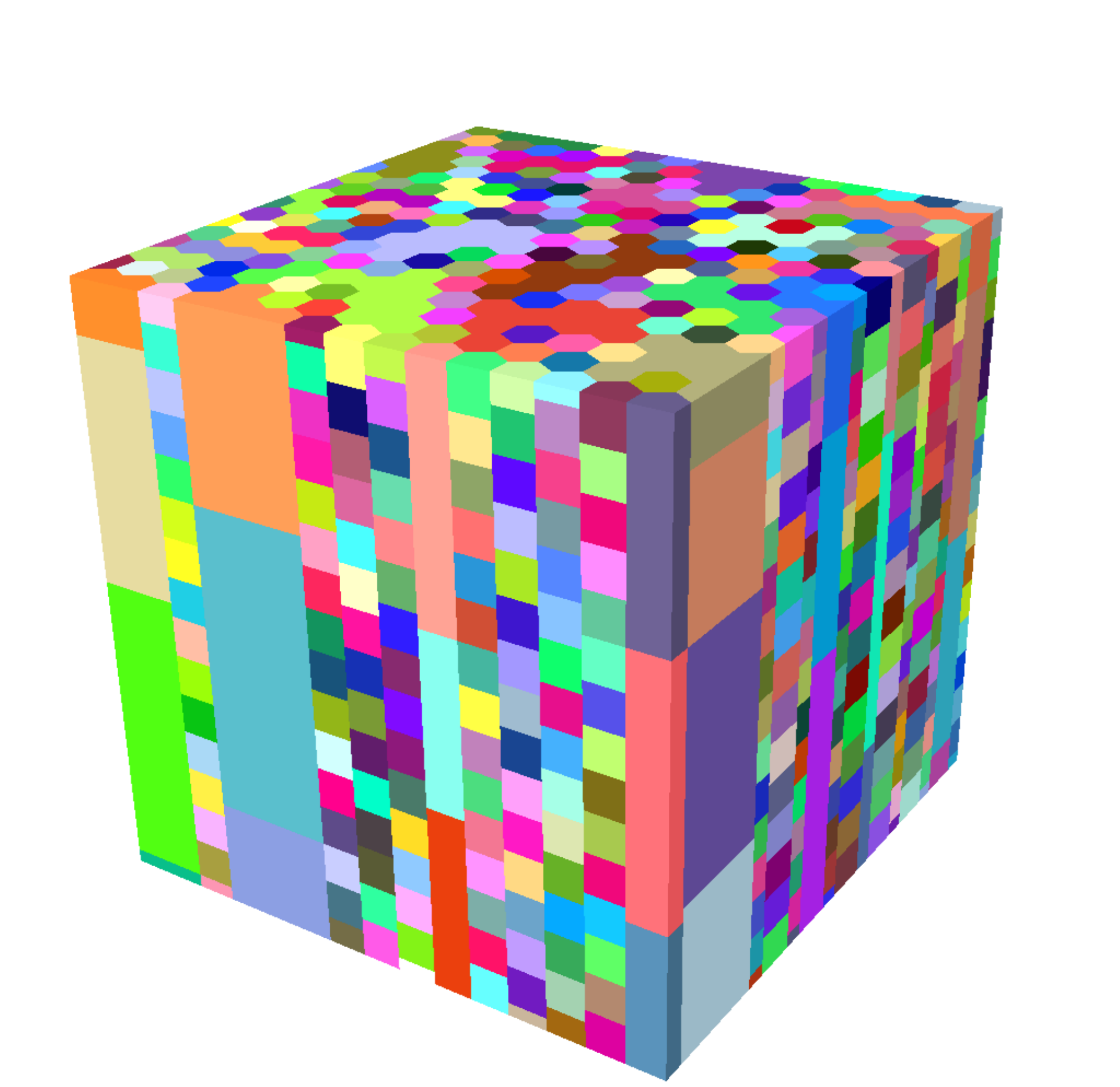}\label{fig:microstruct3D}}
	\caption{Microstructure generation mimics the material processing. Following the generation of the underlying mesh \subref{fig:base_mesh}, a two-dimensional microstructure is created. The two-dimensional parent ferrite microstructure is generated using mapped Voronoi tessellation \subref{fig:ferrite2D}. Austenite regions are then randomly transformed out along grain boundaries, until the desired volume ratio between the two phases is obtained \subref{fig:microstruct2D}. The two-dimensional microstructure is then extruded to form a three-dimensional microstructure \subref{fig:microstruct3D}. }
	\label{fig:MicrostructGen}
\end{figure}

Microstructure generation mimics the material processing, as illustrated in Figure~\ref{fig:MicrostructGen}. Following the generation of the underlying mesh, a two-dimensional microstructure is created. The two-dimensional parent ferrite microstructure is generated using mapped Voronoi tessellation. Austenite regions are then randomly transformed out along grain boundaries, until the desired volume ratio between the two phases is obtained. The two-dimensional microstructure is then extruded to form a three-dimensional microstructure. Each resulting columnar region is divided into grains based on a grain size distribution. All the grains in a column are assigned to the same phase as that of the parent two-dimensional region. Grain orientations are assigned by randomly sampling the experimentally-determined orientation distribution function for each phase.

\begin{figure}[h]
	\centering
	\subfigure[Grain]{\includegraphics[width=0.49\linewidth]{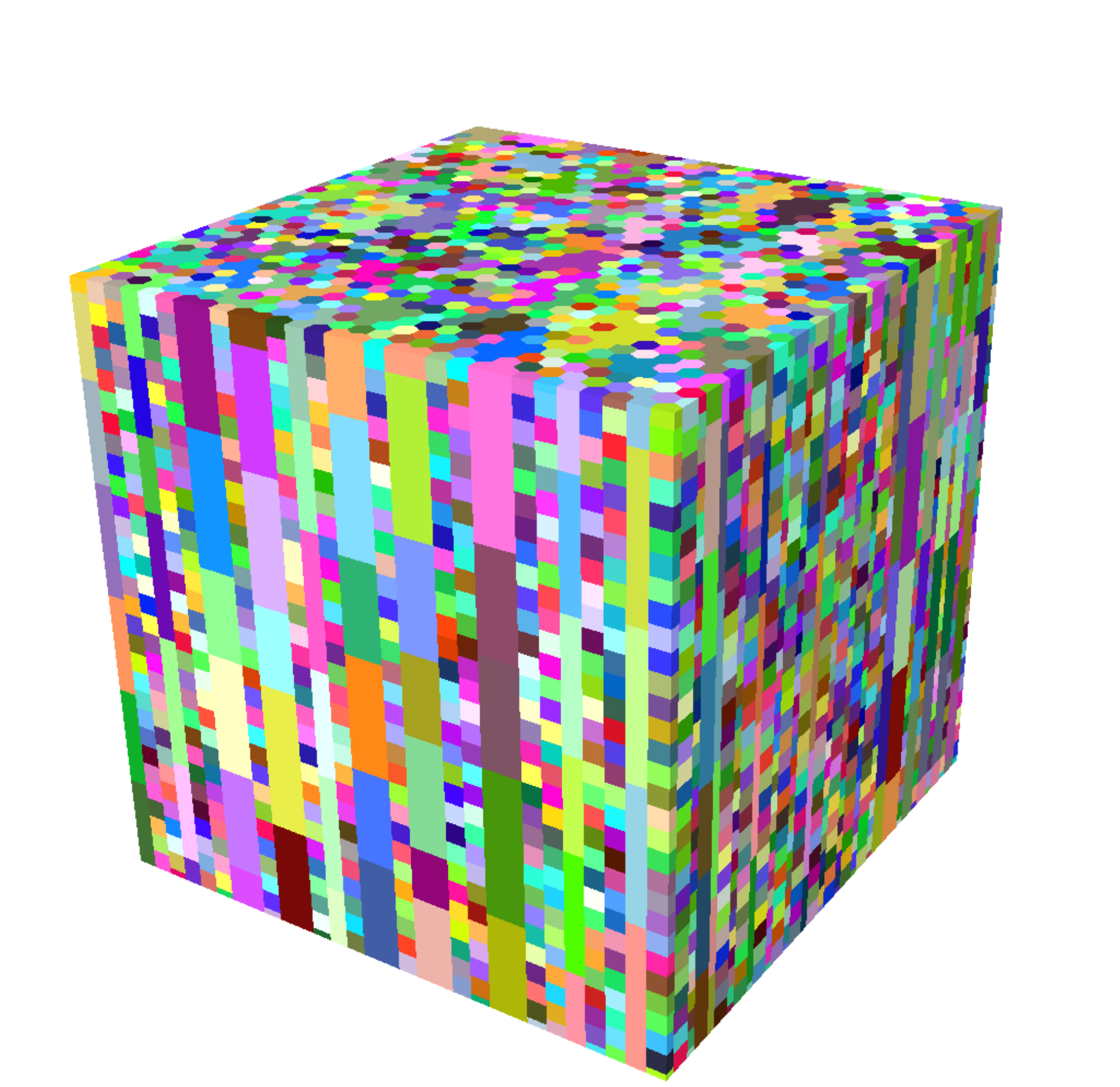}\label{fig:hex30-90-1-2-grain.pdf}}
	\subfigure[Phase]{\includegraphics[width=0.49\linewidth]{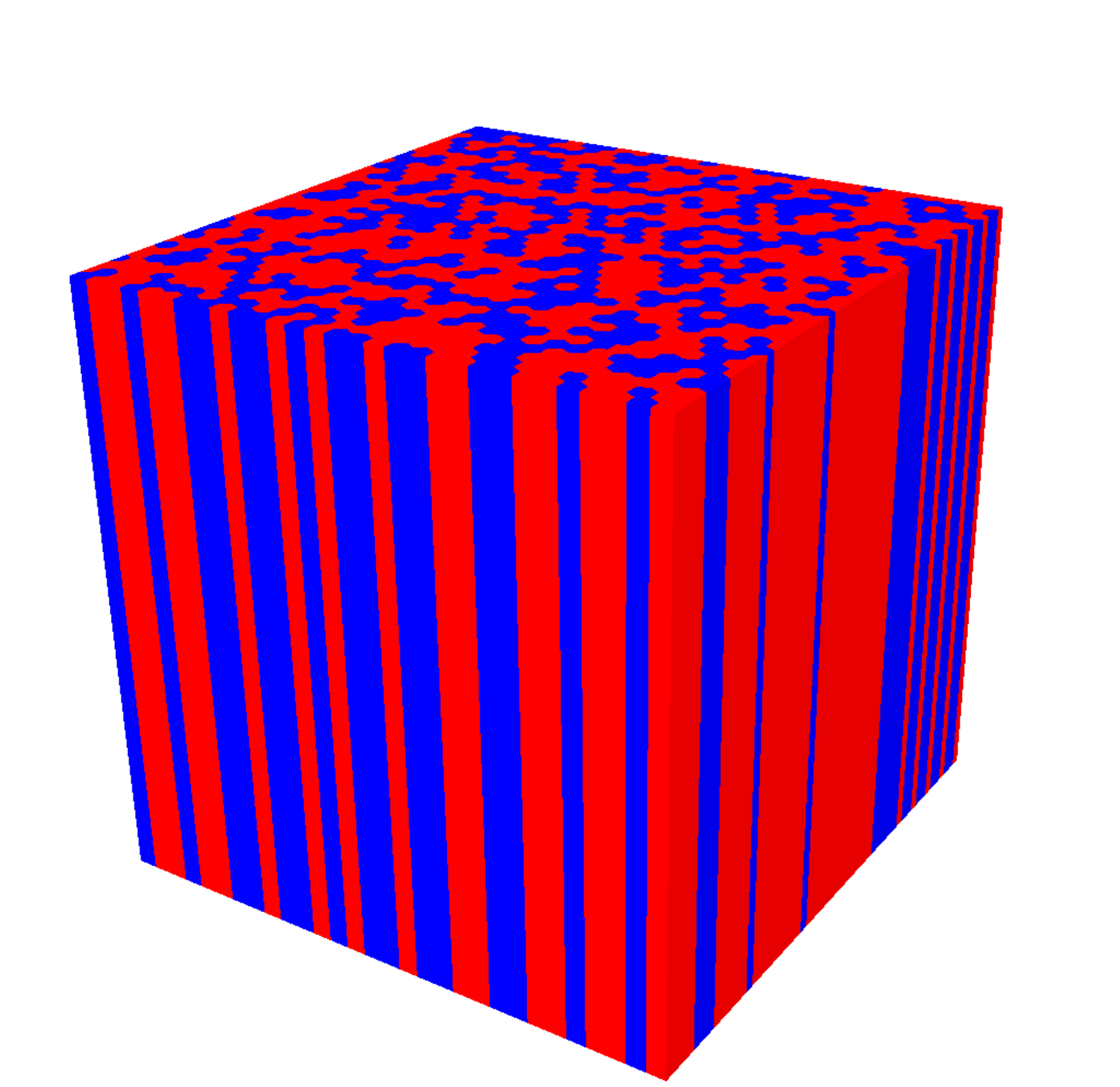}\label{fig:hex30-90-1-2-phase}}
	\caption{Representative virtual microstructure for LDX-2101, exhibiting a columnar phase structure with elongated, irregularly-shaped ferrite grains.}
	\label{fig:StandardMesh}
\end{figure}

This method of mesh generation effectively addressed the issue of meshing concave regions. It also ensures good mesh quality. However, it lacks adaptive mesh refinement around small features. The entire domain must be meshed with the same level of refinement as the smallest feature that is to be resolved. Large ferrite grains are meshed with the same refinement as small austenite grains, which makes the simulation more computationally intensive. Increased computation time is the tradeoff for having a representative microstructure with good element quality. A representative mesh consisting of 1,701,000 elements, 2,321,325 nodes, 18,673 austenite grains, and 769 ferrite grains is shown in Figure~\ref{fig:StandardMesh}. The level of mesh refinement for the smaller austenitic grains is sufficient to resolve intragranular stress and orientation gradients and is comparable to the level of mesh refinement in prior studies~\cite{Han05a, Wong10a, Marin12a}.

\begin{figure}[h]
	\centering
	\subfigure[Grain]{\includegraphics[width=0.49\linewidth]{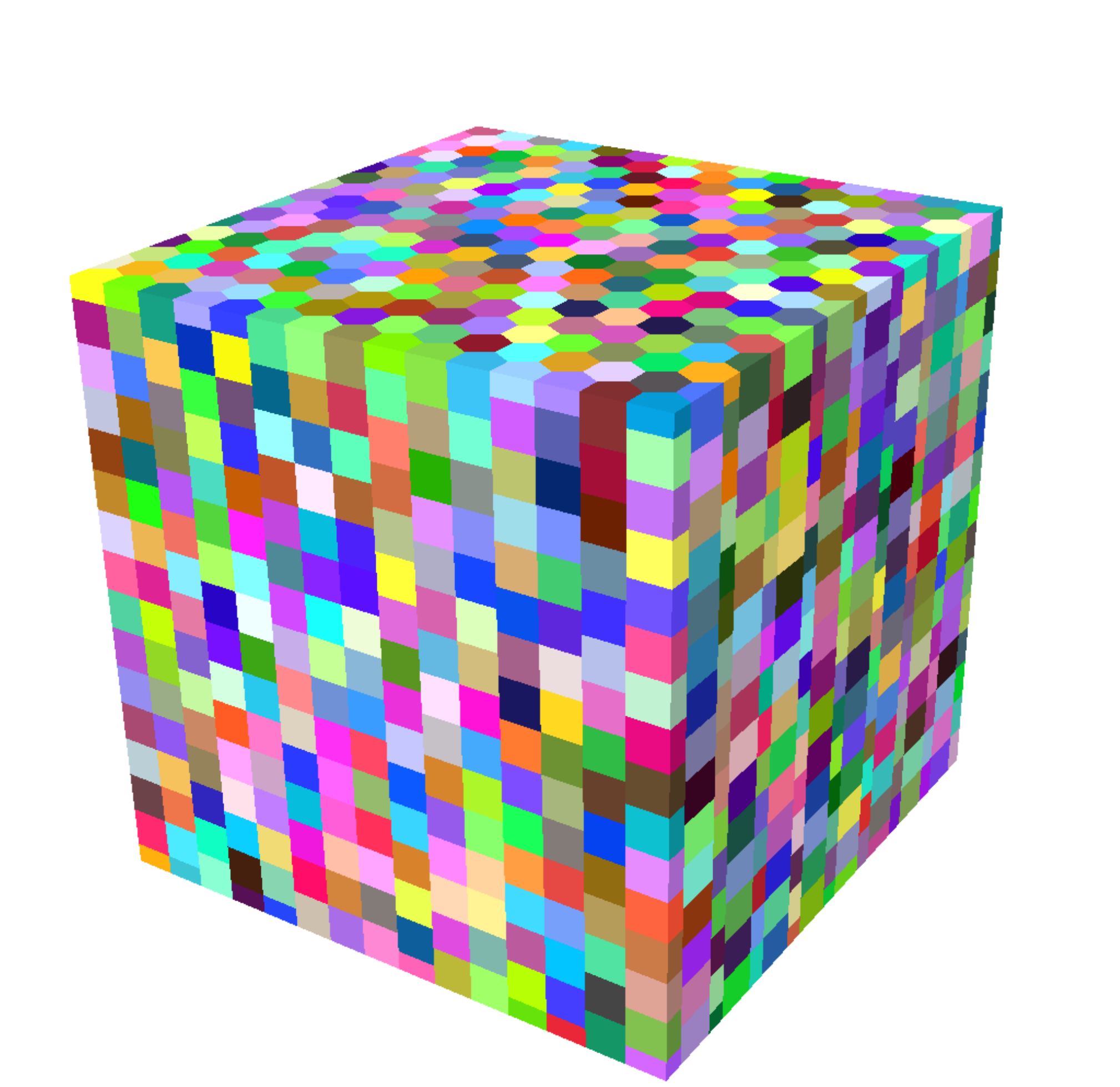}\label{fig:hex15-30-1-1-grain.pdf}}
	\subfigure[Phase]{\includegraphics[width=0.49\linewidth]{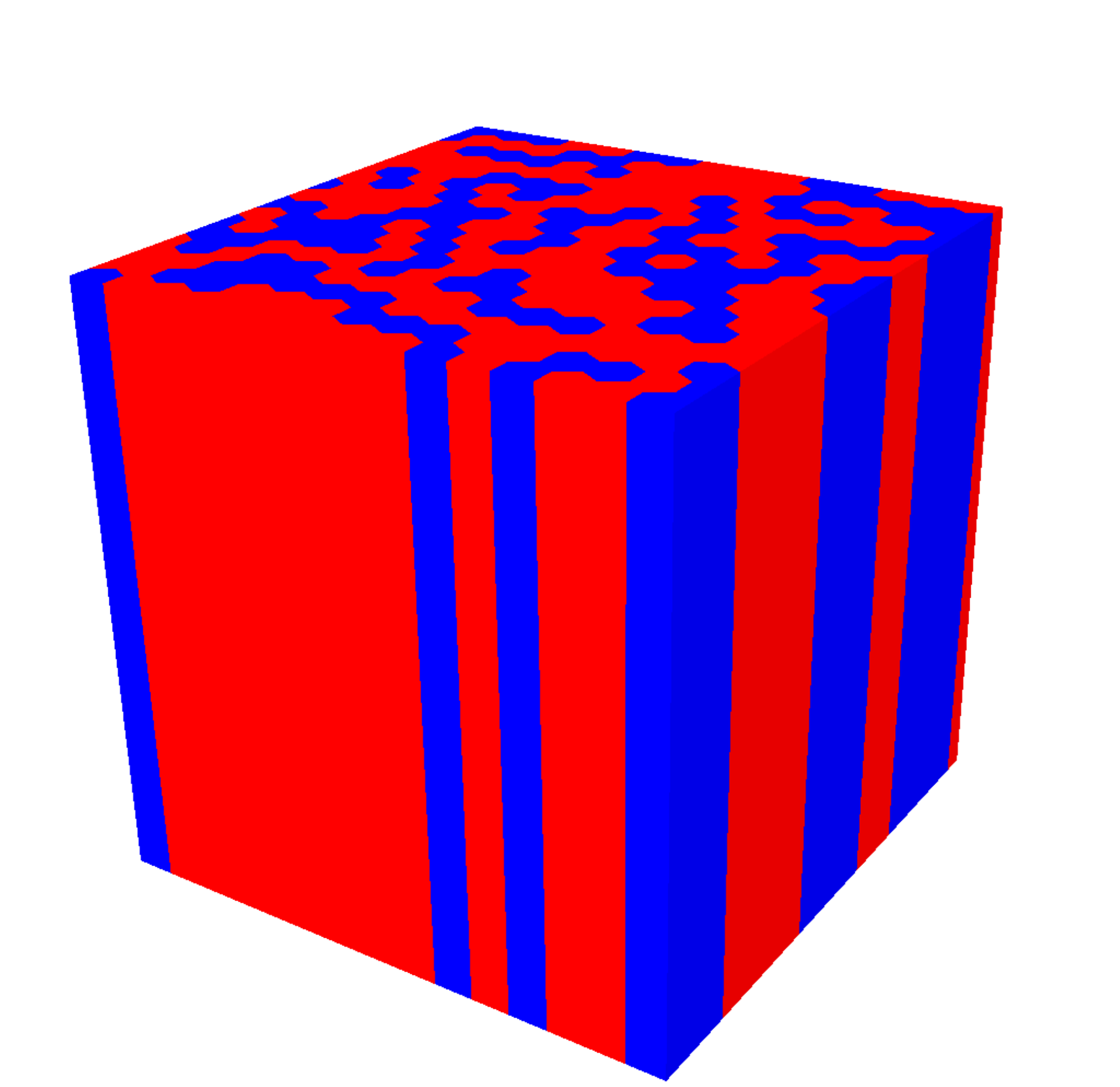}\label{fig:hex15-30-1-1-phase}}
	\caption{Equiaxed virtual microstructure used for the simulation-experiment comparison.}
	\label{fig:SimExpMesh}
\end{figure}

 Modeling the neutron experiments is computationally intensive, due to the complex load history and the large number of grains required to obtain meaningful fiber-averaged statistics.
This necessitated simplification of the microstructure.  A simplified microstructure was built consisting of equiaxed austenite and ferrite grains, but retaining major features of the phase structure as described above.  This microstructure was used for the simulation-experiment comparison in Section~\ref{sec:results}.   The mesh discretization of the microstructure consisted of 137,700 elements and 193,431 nodes, and was comprised of 2,320 austenite grains and 1,753 ferrite grains (Figure~\ref{fig:SimExpMesh}).
\clearpage

\subsection{Material parameter identification}\label{sec:parameters}

There are several challenges inherent in characterizing the material parameters of a dual phase system like LDX-2101, in which the constituent phases cannot readily be physically separated and tested individually in a load frame. First, there are twice as many parameters to optimize. Secondly, the parameters cannot be determined from macroscopic data alone. Data pertaining to the mechanical responses of the individual phases are also required. These data are provided by a subset of the lattice strain measurements, acquired using neutron diffraction. It should be emphasized that only the uniaxial lattice strain data were used for fitting the material parameters. The other four sets of lattice strain data, corresponding to biaxial loading, had no role in the parameter selection. 

Material parameters were derived from multiple references and experiments. Tabulated values were used for the single crystal elastic constants. Uniaxial strain rate jump tests were employed to characterize the rate sensitivity. Macroscopic stress-strain data for monotonic uniaxial loading at constant strain rate were used to characterize the remaining plasticity parameters, assuming equal values for both phases. Lattice strain data for uniaxial loading were then used to determine unique values for each phase.

This approach to fitting material parameters is similar to the methodology developed by Baczmanski and Braham~\cite{Baczmanski04a} for duplex stainless steel. They only compared lattice strains for one crystallographic fiber per phase. However, the lattice strain response of one fiber is not representative of the entire phase. Wong and Dawson~\cite{Wong10a} demonstrated that for even single phase materials, the location and nature of inflection points differ from fiber to fiber and depend on directional strength-to-stiffness. It is therefore critical to compare experimental and simulated lattice strain responses across several fibers per phase, as was done by Baczmanski, Dakhlaoui, and colleagues in subsequent works~\cite{Dakhlaoui06a, Dakhlaoui07a}.

Single crystal elastic constants, procured from references, are listed in Table~\ref{tab:ElasticConstants}. Ledbetter's elastic constants for austenitic stainless steel~\cite{Ledbetter01a} were used for austenite and Simmons and Wang's constants for pure BCC iron were used for ferrite~\cite{Simmons71a}. While alloying affects elastic moduli, these values provided reasonable estimates of the elastic constants for LDX-2101. These constants produced a good match between simulation and experiment for both fiber-averaged lattice strains in the elastic region and the macroscopic Young's modulus. 
\begin{table}[h]
	\centering
	\caption{Single crystal elastic constants using the strength of materials convention ($\tau_{44} = c_{44} \gamma_{44}$).}
	\begin{tabular} {c c c c}		
		phase & $c_{11}$ & $c_{12}$ & $c_{44}$ \\
		& (GPa) & (GPa) & (GPa)  \\ \hline
		FCC & 205 & 138 & 126 \\ 
		BCC & 237 & 141 & 116
	\end{tabular}	
	\label{tab:ElasticConstants}
\end{table}

The plasticity parameters for LDX-2101 used in subsequent simulations are presented in Table~\ref{tab:PlasticityParams}. Agreement between simulated and experimental lattice strains across a range of biaxial ratios, presented in Section~\ref{sec:results}, provides additional confidence in the simulation and choice of material parameters.
\begin{table}[h]
	\centering
	\caption{Plasticity parameters.}	
	\begin{tabular} {c c c c c c c}
		phase & $m$ & $n^\prime$ & $\dot{\gamma}_0$ & $h_0$ & $g_0$ & $g_s$ \\
		& & & ($\mathrm{s^{-1}})$ & (MPa) & (MPa) & (MPa) \\ \hline
		FCC & 0.020 & 1 & $10^{-4}$ & 336 & 192 & 458 \\ 
		BCC & 0.013 & 1 & $10^{-4}$ & 336 & 192 & 458
	\end{tabular}
	\label{tab:PlasticityParams}
\end{table}
\clearpage

\section{Comparisons Of Measured and Computed Lattice Strains }
\label{sec:results}
We now compare the results from the simulation and experiment. The comparison provides confidence in the model, which then serves as a complementary tool to the neutron diffraction experiments to further investigate the deformation of LDX-2101. First, background on lattice strain analysis is summarized. The macroscopic stress-strain response and axial fiber-averaged lattice strain data for uniaxial loading are then presented and discussed. Lastly, the data across all biaxial ratios are presented.


\subsection{Lattice strain analysis background}

The coupling of experimental lattice strain measurements, conducted \emph{in-situ} using either X-ray or neutron diffraction, and micromechanical deformation simulations provides a comprehensive toolset for investigating material behavior. Experiments ground simulations in reality and can uncover new phenomena. Simulation data can, in turn, be used to guide experiments and interpret experimental data. For example, Schuren et al.~\cite{Schuren14a} developed a methodology for using finite element simulation data to quantify uncertainty in lattice strain pole figures. Another example is forward-projection modeling of diffraction experiments, using finite element simulations to generate synthetic diffraction images, which enables the comparison between experiment and simulation to be made at the detector~\cite{Wong13a, Obstalecki14a}. Simulations are also capable of probing materials in ways that are complementary to the experiment, providing access to a wealth of micromechanical data, such as plastic deformation rates and slip system hardnesses, which are inaccessible from the experiment alone. The combined experiment-simulation approach has been used to study both single and dual phase materials under uniaxial loading, including iron-copper~\cite{Han05a}, low and high carbon steel~\cite{Oliver04a}, and duplex stainless steel~\cite{Baczmanski04a, Dakhlaoui06a, Dakhlaoui07a, Jia08a, Hedstrom10a}. Furthermore, Marin et al.~\cite{Marin12a} investigated austenitic stainless steel under biaxial loading. 

The majority of duplex stainless steel simulations, including those conducted by Baczmanski and Braham~\cite{Baczmanski04a}, Dakhlaoui et al.~\cite{Dakhlaoui06a, Dakhlaoui07a} and Jia et al.~\cite{Jia08a} were conducted using elasto-viscoplastic self consistent modeling. Self consistent models are useful for texture evolution and capturing average mechanical response. However, they do not capture local grain neighborhood effects or variations in state across a grain. Finite element models, like the one used by Hedstrom et al.~\cite{Hedstrom10a} provide a more detailed, accurate representation of material microstructure and mechanical behavior. Baczmanski and Braham, Dakhlaoui et al., and Jia et al. compared their simulated data with fiber-averaged lattice strains measured using neutron diffraction, while Hedstrom et al. compared to grain-averaged lattice strains measured using high-energy X-ray diffraction.

Marin et al.~\cite{Marin12a} provide a framework for analyzing the lattice strain responses produced by macroscopic biaxial loading in terms of stress decomposition. The macroscopic stress is first decomposed into mean and deviatoric components. For cubic and some hexagonal close-packed materials, the mean component is coupled to the volumetric deformation, which is purely elastic. The deviatoric stress drives the anisotropic elasto-plastic deformation. As the biaxial ratio, defined as the ratio of in-plane principal stress components, increases from uniaxial ($BR=0$) to balanced biaxial ($BR=1$), the ratio of mean to deviatoric stress increases, resulting in a more isotropic lattice strain response. 

Furthermore, they showed that the eigenvalues of the deviatoric stress tensor can be used to identify similarities between biaxial stress states and trends in the lattice strain responses. The eigenvalues of the deviatoric stress can be normalized by the largest eigenvalue magnitude, reordered, and represented in the form $\alpha_1 : \alpha_2 : 1$. In this representation, each deviatoric stress state in the range of $BR=0$ to $BR=1$ is related to another deviatoric stress state in the same range through a permutation of the sample axes. For example, the ratio of normalized eigenvalues for uniaxial and balanced biaxial loading are both $-\sfrac{1}{2} : -\sfrac{1}{2} : 1$. Similarly, the ratio of normalized eigenvalues for $BR=0.25$ and $BR=0.75$ are both $-\sfrac{2}{7} : -\sfrac{5}{7} : 1$. For $BR=0.5$ the normalized eigenvalue ratio is $0 : -1 : 1$, which is similar to itself. Between each pair of biaxial stress states, there are qualitative similarities in the lattice strain responses, after a sign change and permutation of axes. Differences between the lattice strain responses exist due material anisotropy and different ratios of mean to deviatoric stress, however, the similarities are still quite apparent. For uniaxial and balanced biaxial loading, two of the normalized eigenvalues are equal. The deviatoric stress is isotropic in the plane spanned by these two eigenvectors. In-plane lattice strains exhibit similar features to one another. The biaxial ratio of 0.5 exhibits the largest range of relative eigenvalues and is therefore the most distinct stress state in relation to both uniaxial and balanced biaxial stress states. Its lattice strain responses are also the most distinct from those of uniaxial and balanced biaxial loading.

\subsection{Uniaxial loading results}\label{sec:sim_exp_uniaxial}

\begin{figure}[h]
\centering
\includegraphics{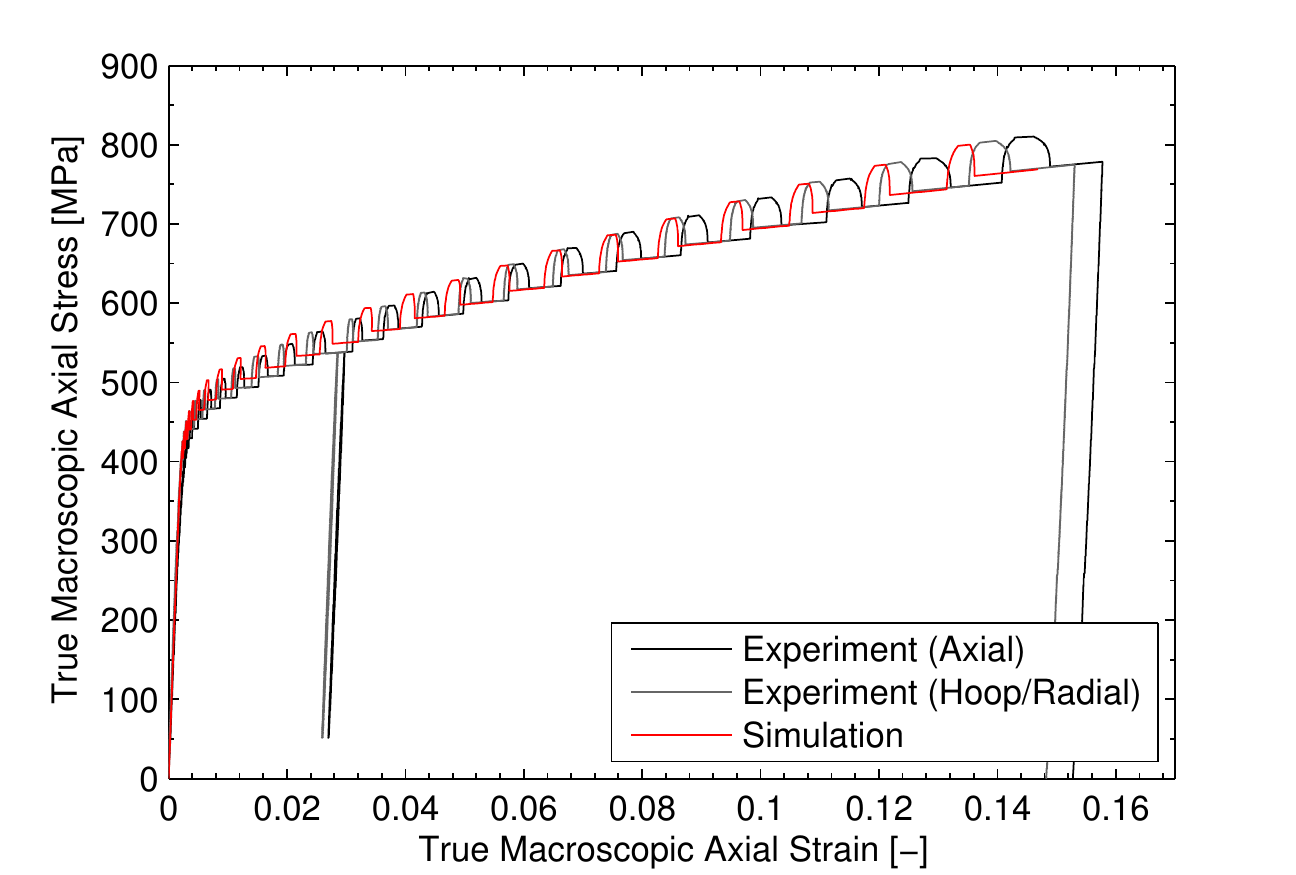}
\caption{Macroscopic stress-strain response for uniaxial loading ($BR~=~0$).}
\label{fig:MacroStressStrainBR000}
\end{figure}

The macroscopic true stress-strain response for uniaxial loading is presented in Figure~\ref{fig:MacroStressStrainBR000}. Two sets of experimental data are presented, corresponding to the two specimens used to acquire hoop/radial lattice strain data and axial lattice strain data. The simulation models the ramp, short dwell, unload, and long dwell episodes of the experimental load history with great fidelity. The elastic unloading and reloading episode is omitted. Both the simulation and experiment dwell in load control during the neutron data acquisition portion of the load history. Plastic deformation continues during the unloading and dwell episodes even though the load is reduced. Dwell episodes correspond to the slightly-sloped linear regions of the stress-strain curve linking unloading and loading. The slight increase in true stress during the dwell episodes is due to the reduction in cross-sectional area. The nominal times for the data collection dwell episodes are thirty minutes for the simulation and axial lattice strain measurement, and sixty minutes for the hoop/radial lattice strain measurement. The amount of plastic deformation during the dwell is comparable between the simulation and both experiments. In fact, the total deformation for the axial specimen is slightly greater than that of the hoop/radial specimen. The amount of plastic deformation occurring during the dwell episodes are similar, even though the dwell times are different, because the deformation rate slows considerably as the material hardens. Most of the plastic deformation occurs early in the dwell.

\begin{figure}[h]
\centering
\subfigure[Lattice strain.]{\includegraphics[width = 3.0in]{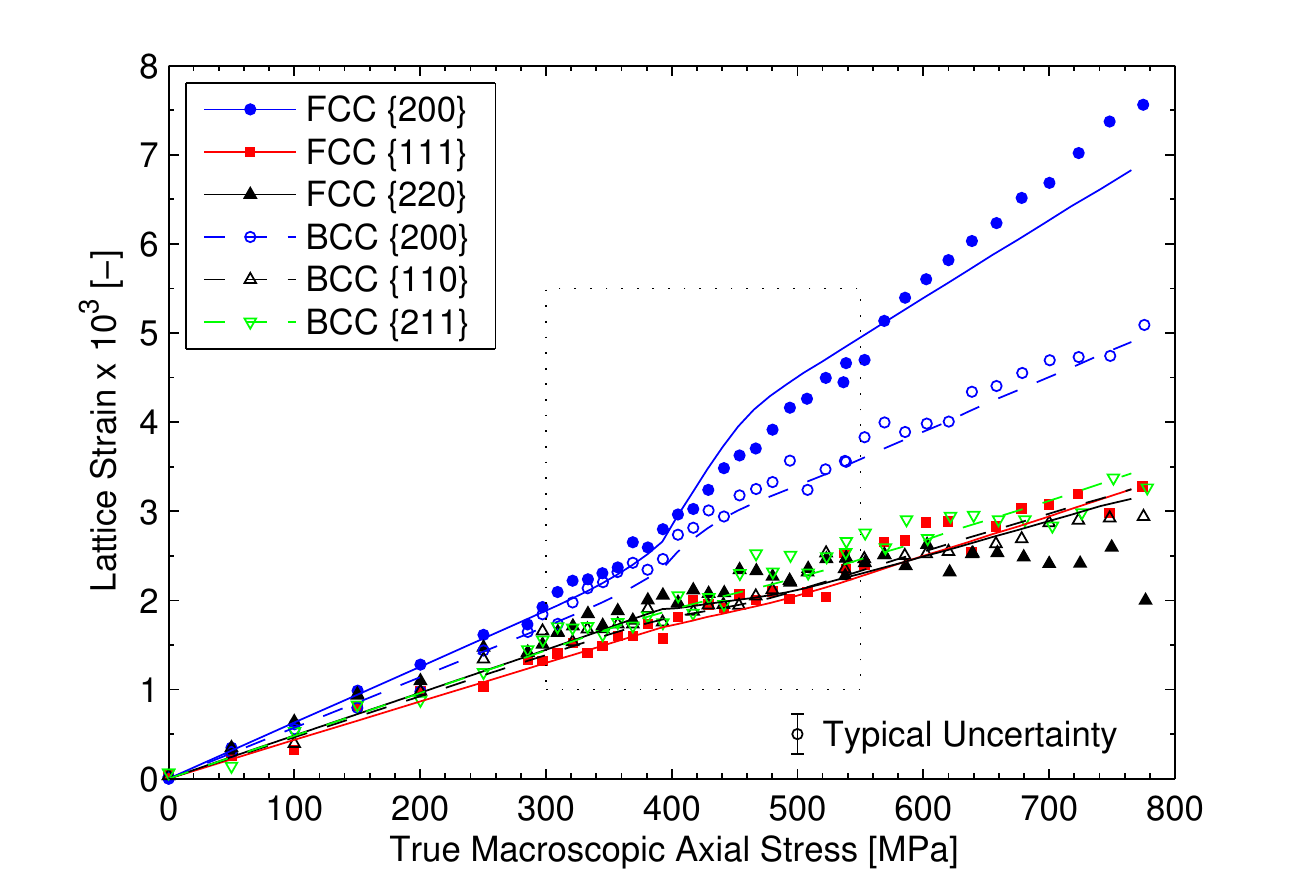}\label{fig:AxialLSBR000}}
\subfigure[Effective plastic deformation rate ($D_{\mathit{eff}}^p$).]{\includegraphics[width = 3.0in]{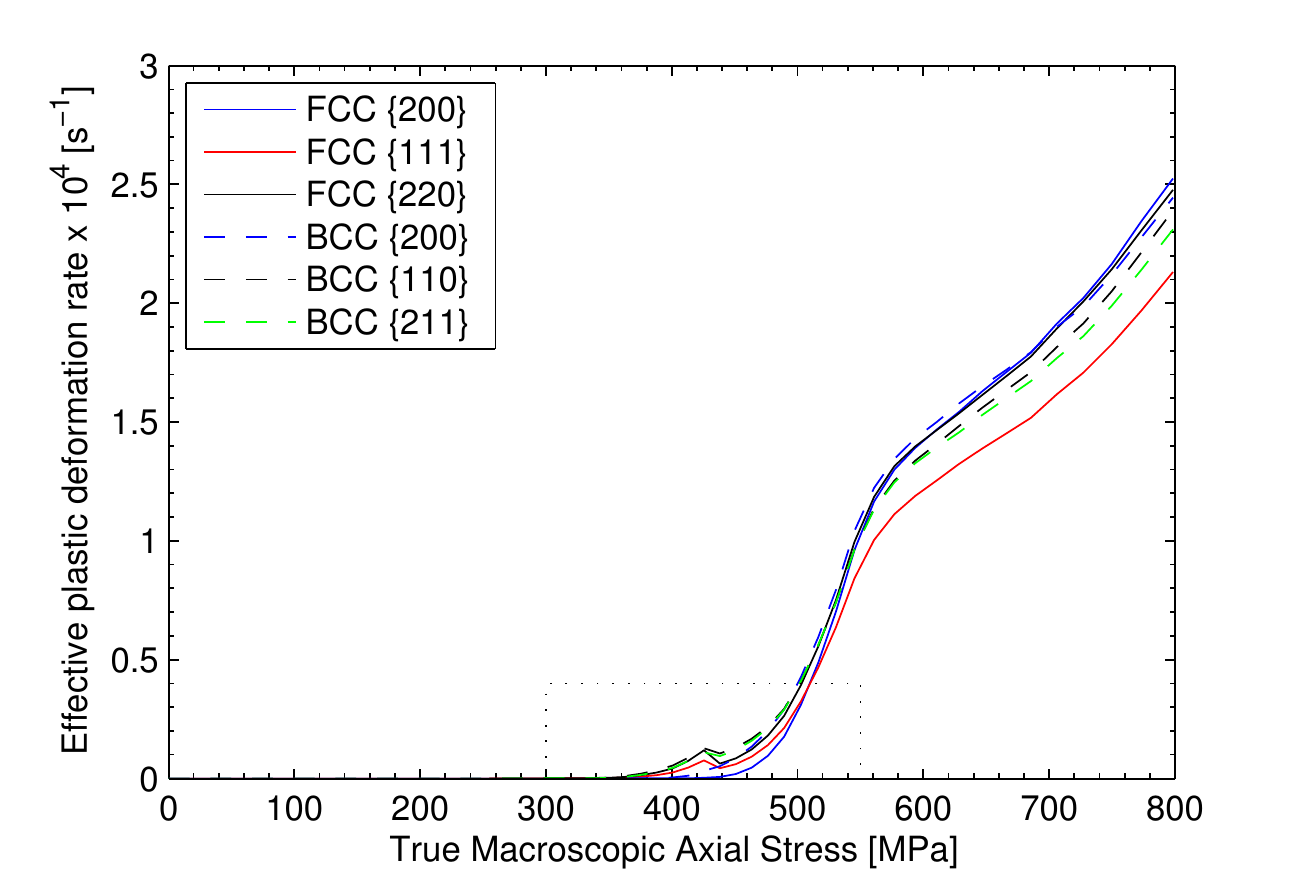}\label{fig:AxialDpeffBR000}}
\caption{Axial fiber-averaged lattice strains and simulated effective plastic deformation rates for uniaxial loading. Dotted regions are enlarged in Figure~\ref{fig:AxialLSAndDpeffBR000Zoom}.}
\label{fig:AxialLSAndDpeffBR000}
\end{figure}

Axial lattice strains are presented in Figure~\ref{fig:AxialLSBR000}. There is good agreement between simulation and experiment in the elastic region, which extends up to 300~MPa. The simulation captures the relative lattice strain magnitudes and most of the key inflection points. However, there are some discrepancies in the fully-developed plastic region. For example, the experimental data for the FCC \{200\} fiber exhibit only upward inflection, whereas the simulated data contain both upwards and downwards inflections. The experimental lattice strain for the FCC \{220\} fiber plateaus in the plastic regime, whereas the simulated lattice strain continues to increase. It should be noted, however, that there is significantly greater than average uncertainty, on the order of $\pm500\mu\epsilon$ for the FCC \{220\} fiber in the fully developed plastic region. The high uncertainty is due to a decrease in signal caused by texture evolution. The difference between simulated and experimental lattice strains for the FCC \{220\} fiber is not far outside the measurement uncertainty.

\begin{figure}[h]
\centering
\subfigure[Lattice strain.]{\includegraphics[width = 3.0in]{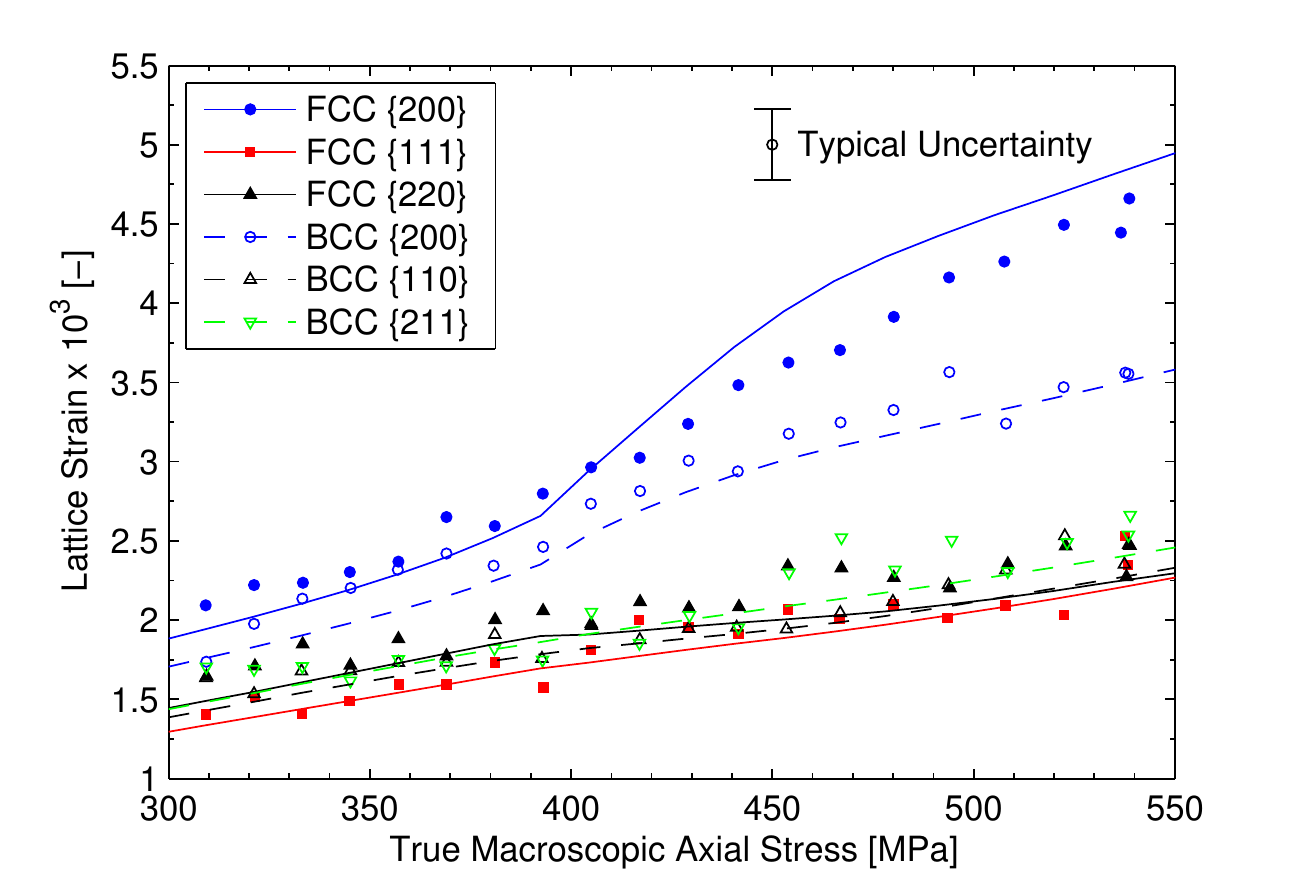}\label{fig:AxialLSBR000Zoom}}
\subfigure[Effective plastic deformation rate ($D_{\mathit{eff}}^p$).]{\includegraphics[width = 3.0in]{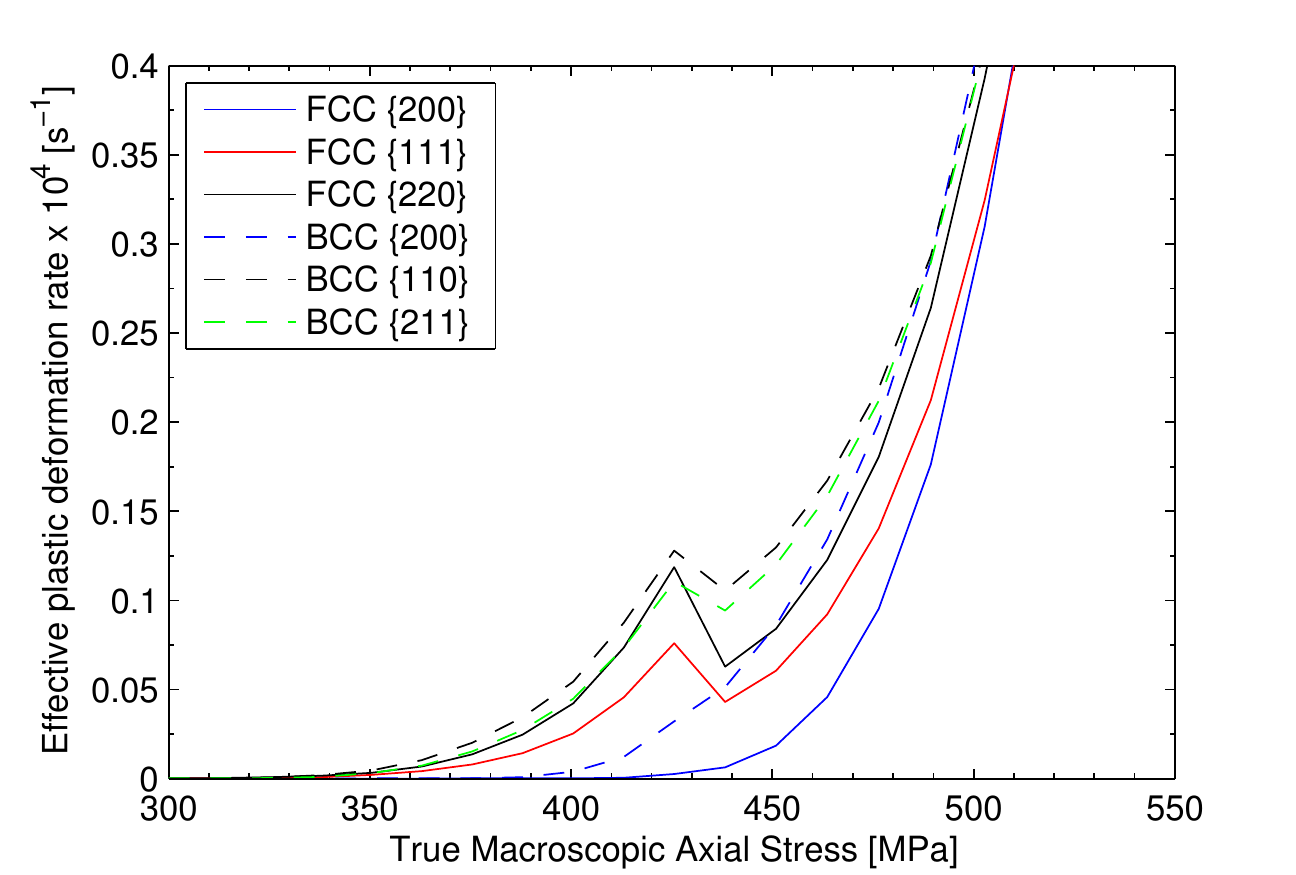}\label{fig:AxialDpeffBR000Zoom}}
\caption{Zoom view of axial fiber-averaged lattice strains and simulated effective plastic deformation rates for uniaxial loading during the elasto-plastic transition.}
\label{fig:AxialLSAndDpeffBR000Zoom}
\end{figure}

Inflections in lattice strain curvature for uniaxial loading can be explained in terms of strength-to-stiffness ratio~\cite{Wong10a}, which governs the initiation of yielding. The initiation of yielding can be detected by examining the effective plastic deformation rate $D^p_{\mathit{eff}}$, recovered from the simulation, as a function of macroscopic stress (Figure~\ref{fig:AxialDpeffBR000}). The plastic deformation rate for all fibers is zero in the elastic regime. Nonzero values of $D^p_{\mathit{eff}}$ indicate that crystals along a fiber have yielded. The plastic deformation rate continues to increase in the plastic regime in order to maintain a constant load rate. Enlarged views of the boxed regions in Figure~\ref{fig:AxialLSAndDpeffBR000}, encompassing the elasto-plastic transition, are presented in Figure~\ref{fig:AxialLSAndDpeffBR000Zoom}. The sudden decrease in $D^p_{\mathit{eff}}$ between 425 and 438~MPa is an artifact of the simulation corresponding the the first load step where dwell episodes were modeled. The important features of the $D^p_{\mathit{eff}}$ plot are the stresses at which $D^p_{\mathit{eff}}$ becomes nonzero.

The FCC \{111\}, FCC \{220\}, BCC \{110\}, and BCC \{211\} fibers all have relatively low strength-to-stiffness ratios, between 11 and 13~TPa$^{-1}$, calculated using Wong and Dawson's Schmid factor formulation of strength-to-stiffness. As a result, crystals along these fibers are among the first to yield at around 375~MPa. The initiation of yield is evident from nonzero values of $D^p_{\mathit{eff}}$. After a crystal yields, it is unable to carry additional incremental load during the elasto-plastic transition, with the exception of a small amount due to strain hardening. Therefore, during the elasto-plastic transition, the rate of change of lattice strain with respect to the macroscopic load decreases after crystals along a fiber yield. This change is manifest as a downward inflection in the lattice strain curve. The lattice strain for all these fibers exhibit subtle downward inflections around 400~MPa. Since yielded crystals are significantly less able to carry incremental load, the remaining elastic crystals must carry a larger percentage. As a result, there is an upward inflection in lattice strain associated with fibers whose crystals remain mostly elastic, namely the BCC \{200\} and FCC \{200\} fibers. Both these fibers have relatively high values of strength-to-stiffness, 19 and 26~TPa$^{-1}$, respectively. The FCC \{200\} fiber, which has the highest strength-to-stiffness of the six fibers shown, is the last to yield.

Lattice strain curvature changes again at the end of the elasto-plastic transition, around 500~MPa. At this point, all crystals have yielded, so hardening must occur for the aggregate to carry additional incremental load. The changes in lattice strain curvature at the end of the elasto-plastic transition are a result of the incremental load being accommodated by strain hardening rather than being borne by the subset of crystals that remain elastic. In the fully-developed plastic regime, the lattice strain response is dominated by hardening and the reorientation of crystal stresses into the vertices of the single-crystal yield surface~\cite{Ritz10a}.

\begin{figure}[h]
	\centering
	\subfigure[FCC \{200\}.]{\includegraphics[width=0.49\linewidth]{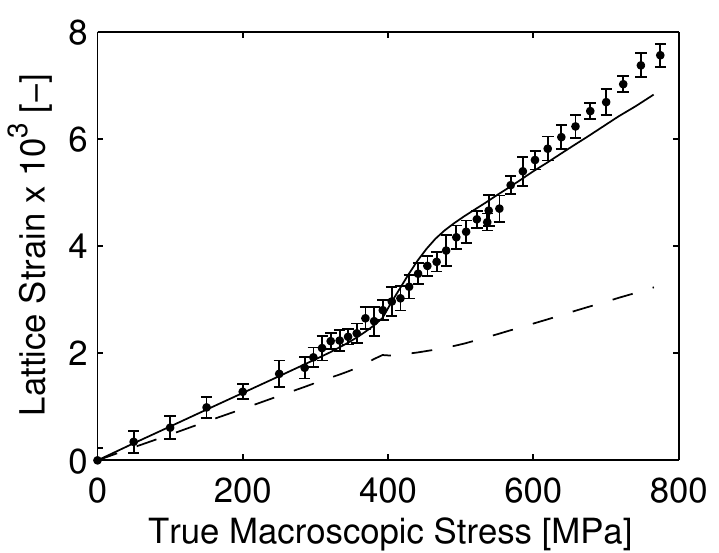}\label{fig:AnisotropyFCC200}}
	\subfigure[BCC \{200\}.]{\includegraphics[width=0.49\linewidth]{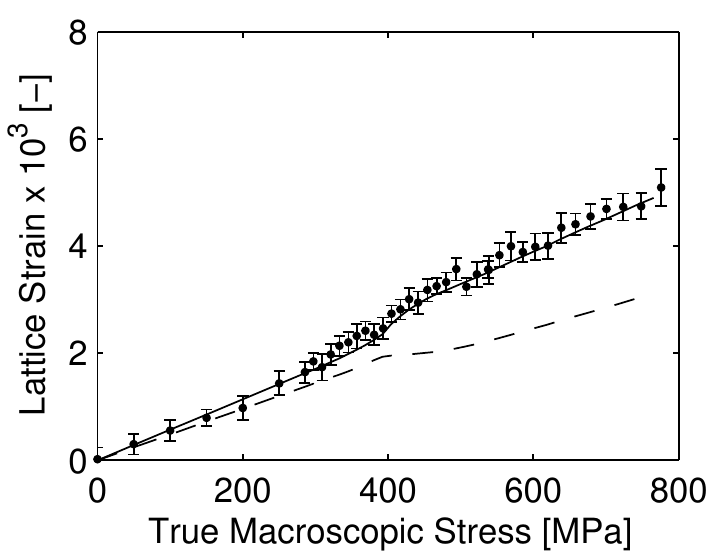}\label{fig:AnisotropyBCC200}}
	\subfigure[FCC \{111\}.]{\includegraphics[width=0.49\linewidth]{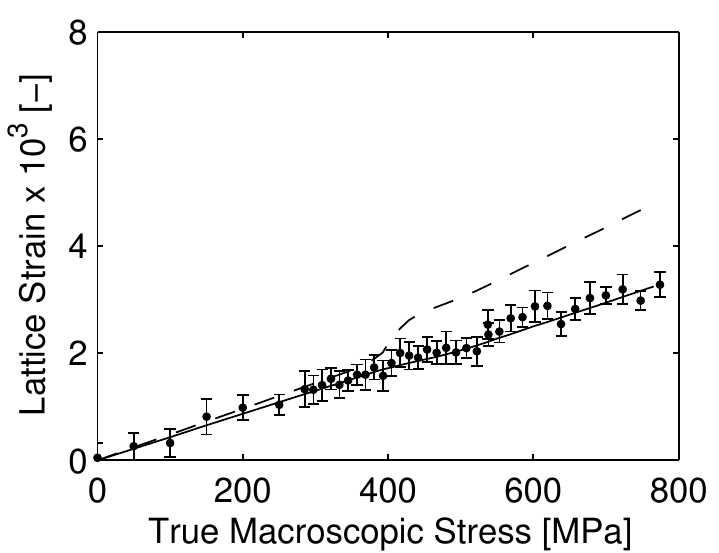}\label{fig:AnisotropyFCC222}}
	\subfigure[BCC \{110\}.]{\includegraphics[width=0.49\linewidth]{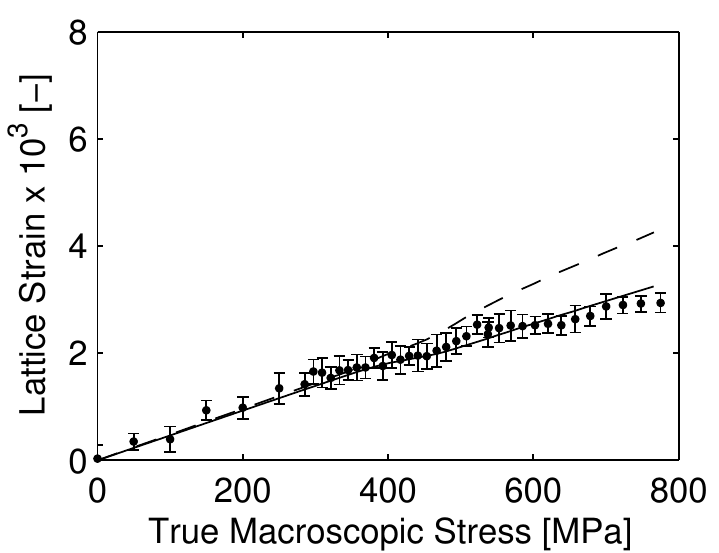}\label{fig:AnisotropyBCC220}}
	\subfigure[FCC \{220\}.]{\includegraphics[width=0.49\linewidth]{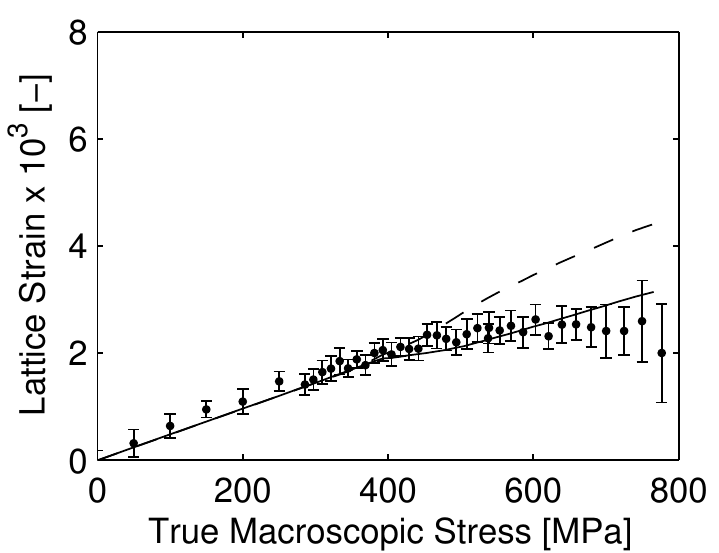}\label{fig:AnisotropyFCC220}}
	\subfigure[BCC \{211\}.]{\includegraphics[width=0.49\linewidth]{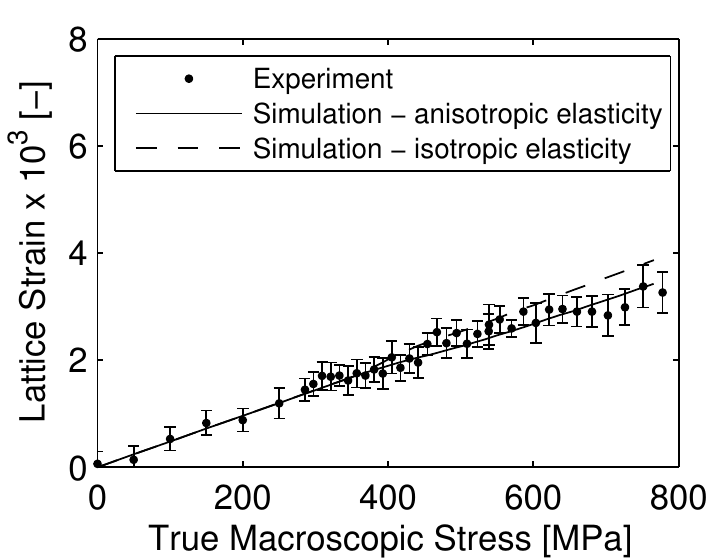}\label{fig:AnisotropyBCC211}}
	\caption{Effect of elastic anisotropy on axial lattice strain response for uniaxial macroscopic loading.}
	\label{fig:LSElasAniso}
\end{figure}

Elastic anisotropy plays a key role in governing the lattice strain response. The major difference between the two phases is the single crystal elastic constants. To demonstrate the importance of elastic anisotropy in governing both the elastic and plastic response, a simulation in which both phases were elastically isotropic was conducted. The single crystal elastic constants, presented in Table~\ref{tab:elas_const_iso}, were chosen such that the Voigt-averaged Young's and bulk moduli were the same as for the anisotropic case. Elastic anisotropy has a significant effect on the mechanical response, as evidenced by lattice strains (Figure~\ref{fig:LSElasAniso}). Changing the level of elastic anisotropy changes the relative fiber-averaged strength-to-stiffness ratios. The strength-to-stiffness ratio is 18~TPa$^{-1}$ for the FCC \{111\} fiber and 12~TPa$^{-1}$ for the other five fibers shown. As a result, the nature of the inflection points are different for the elastically anisotropic and isotropic cases. For example, the FCC \{200\} fiber exhibits initial upward concavity for the anisotropic case, and initial downward concavity for the isotropic case. Similar reversals can be seen in the other lattice strain plots. These changes in concavity imply that the order in which crystals yield is affected by elastic anisotropy, as was demonstrated by~\cite{Wong10a}. Not only does elastic anisotropy play a role in the elasto-plastic transition, it also affects the mechanical response in the fully-developed plastic regime. 

\begin{table}[h]
	\centering
	\caption{Single crystal elastic constants for the isotropic elasticity study. The strength of materials convention ($\tau_{44} = c_{44} \gamma_{44}$) is used.}
	\begin{tabular} {c c c c}		
		phase & $c_{11}$ & $c_{12}$ & $c_{44}$ \\
		& (GPa) & (GPa) & (GPa)  \\ \hline
		FCC & 266 & 107 & 80 \\ 
		BCC & 282 & 118 & 82
	\end{tabular}	
	\label{tab:elas_const_iso}
\end{table}

Furthermore, differences in elastic anisotropy between the two phase produce the differences in mechanical responses observed in the experiment. For example, the measured lattice strain responses are different for the FCC \{200\} and BCC \{200\} fibers and for the FCC \{220\} and BCC \{110\} fibers. These differences are captured by the anisotropic simulation, but not by the isotropic one. The simulated lattice strains for the isotropic case are nearly identical for each pair of fibers. Thus, elastic anisotropy is primarily responsible for the differences in lattice strain behavior observed between the two phases.


\subsection{Biaxial loading results}\label{sec:sim_exp_biaxial}

\begin{figure}[h]
\centering	
\subfigure[BR=0.25]{\includegraphics[width=0.49\linewidth]
{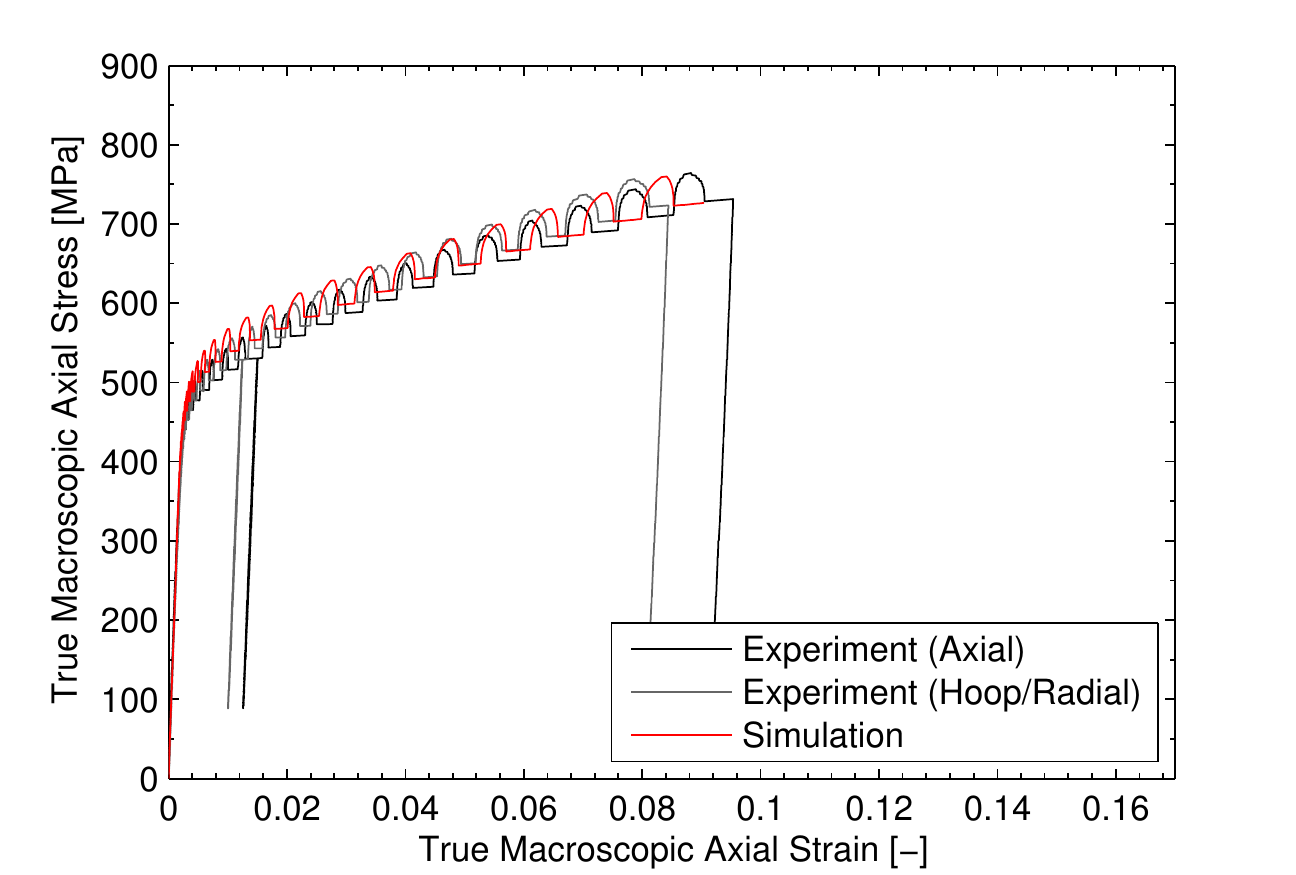}\label{fig:MacroStressStrainBR02}}
\subfigure[BR=0.50]{\includegraphics[width=0.49\linewidth]
{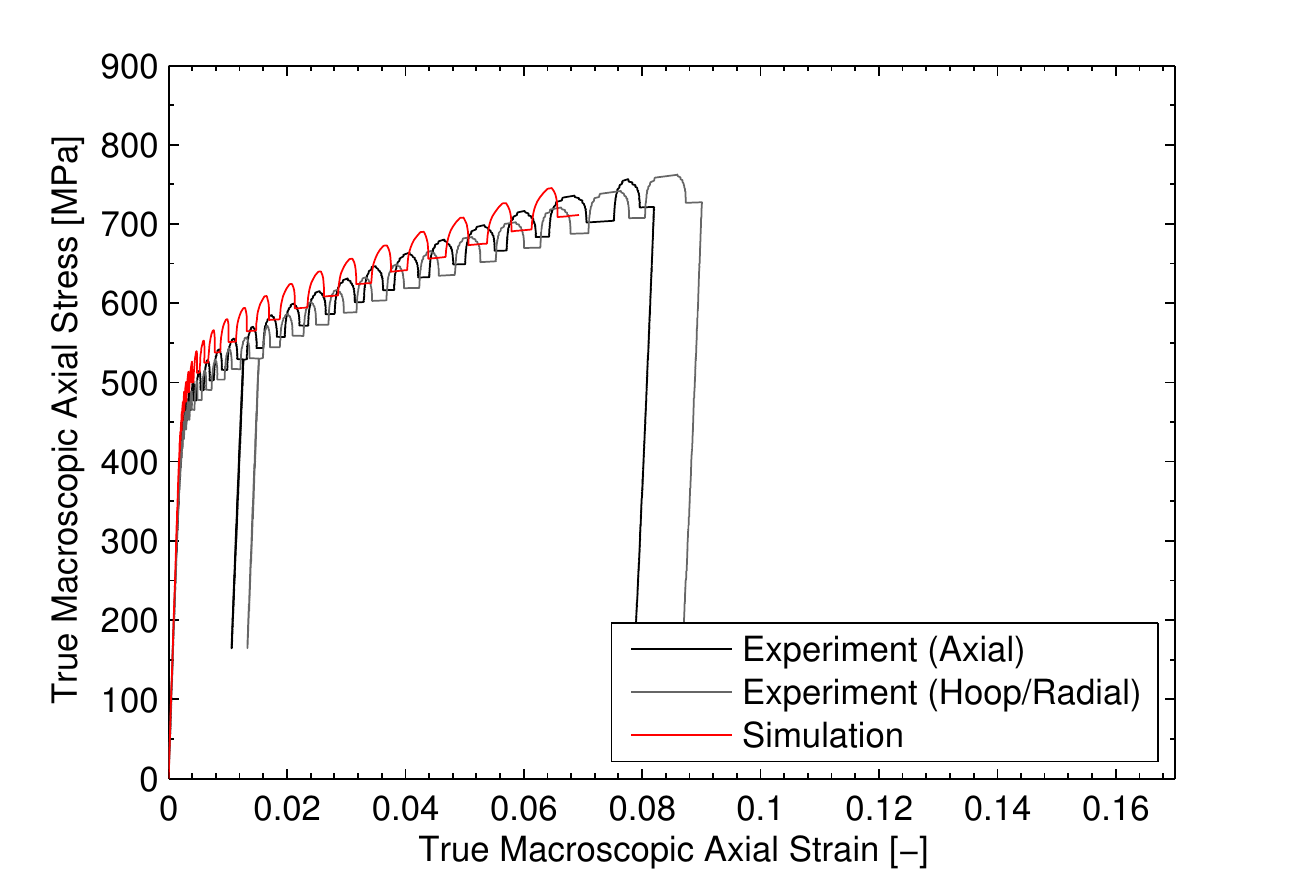}\label{fig:MacroStressStrainBR050}}
\subfigure[BR=0.75]{\includegraphics[width=0.49\linewidth]
{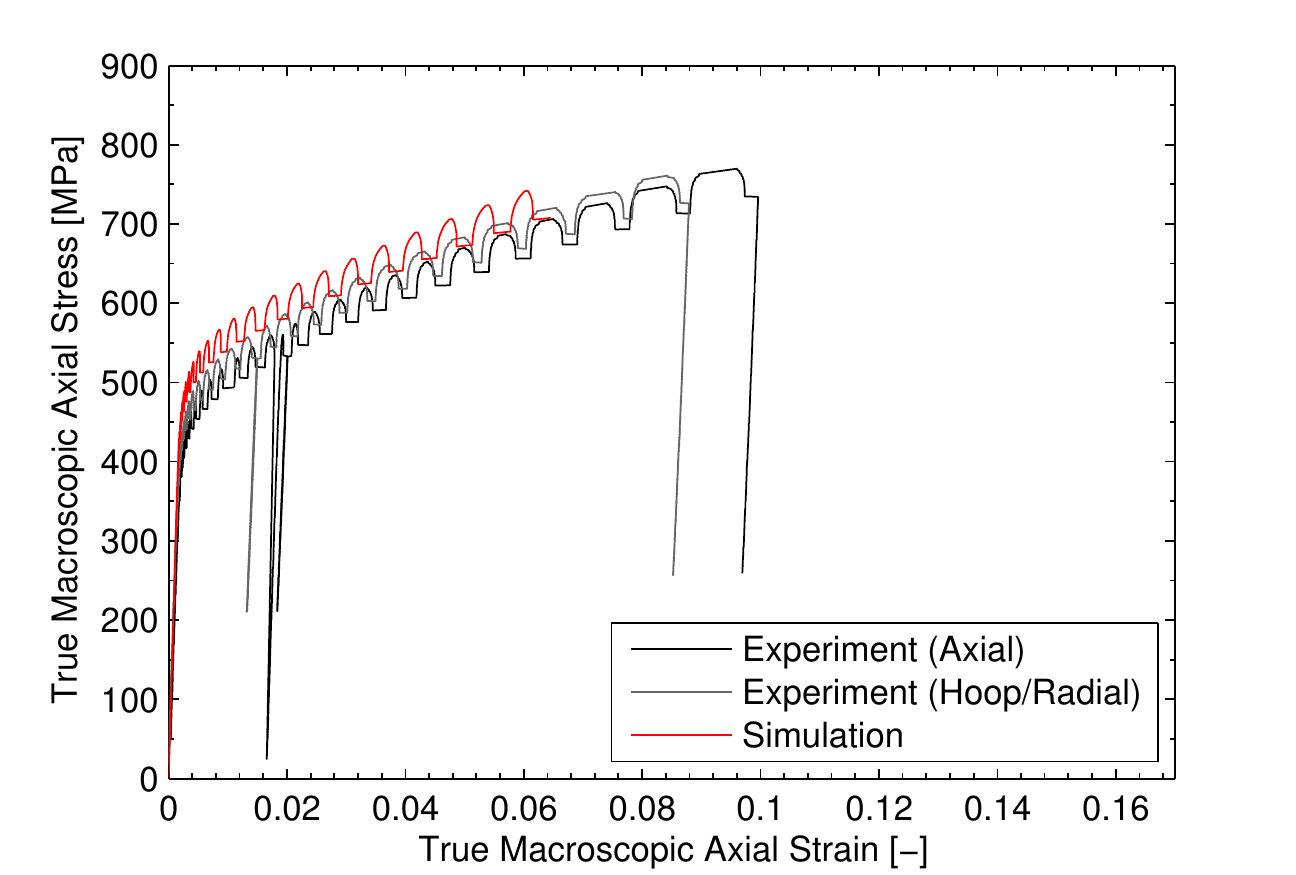}\label{fig:MacroStressStrainBR075}}
\subfigure[BR=1.00]{\includegraphics[width=0.49\linewidth]
{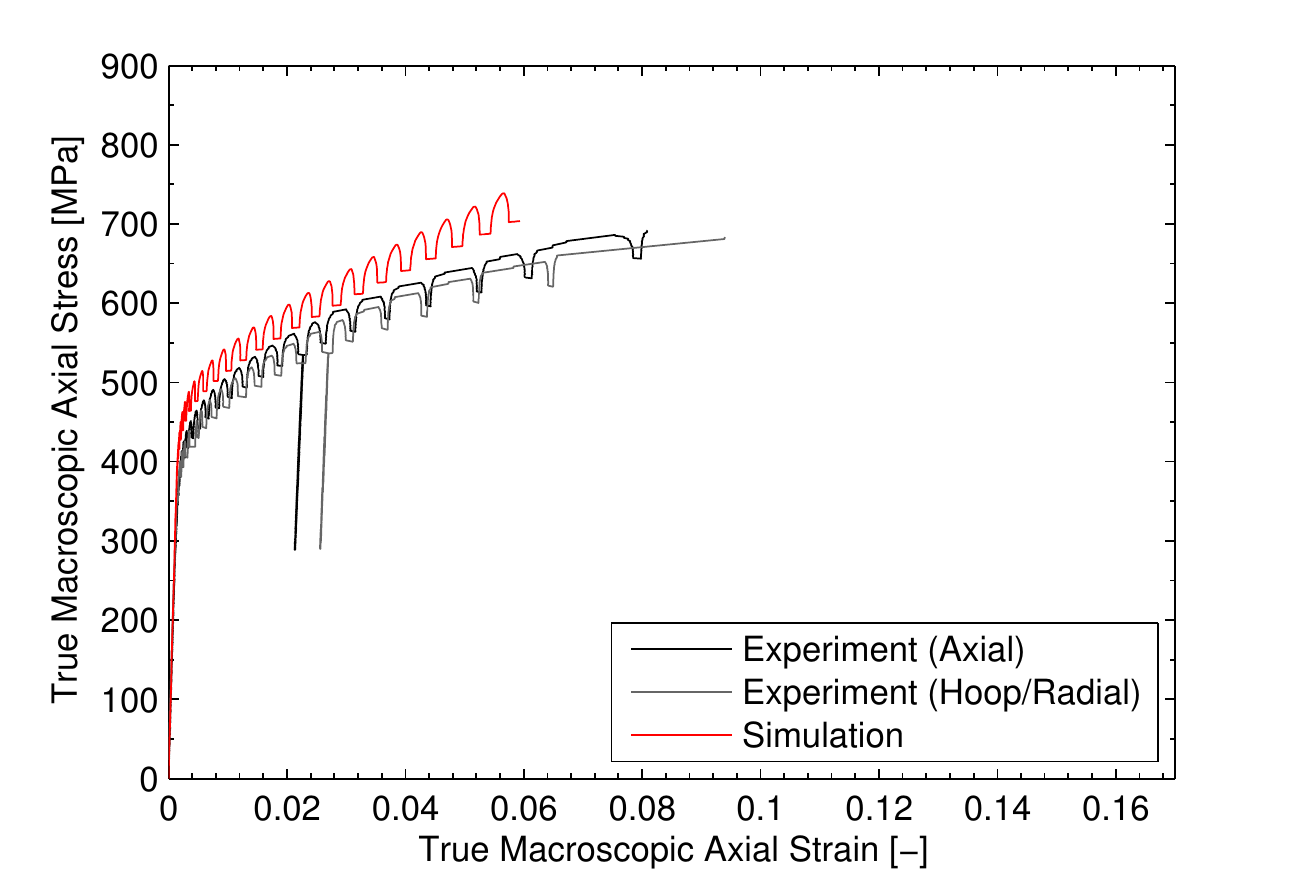}\label{fig:MacroStressStrainBR100}}
\caption{Macroscopic stress-strain responses for various biaxial ratios}
\label{fig:MacroStressStrainCurves}
\end{figure}

Macroscopic true axial stress-strain responses for the biaxial experiments are presented in Figures~\ref{fig:MacroStressStrainBR025}-\ref{fig:MacroStressStrainBR100}. The specimens for balanced biaxial loading rupture. In general, the simulated and experimental macroscopic stress-strain curves exhibit good agreement. As biaxial ratio increases from uniaxial to balanced biaxial, the ratio of mean to deviatoric stress also increases. Since deviatoric stress drives plastic deformation, the amount of plastic deformation occurring during the dwell episodes decreases with increasing biaxial ratio. The uniaxial tests achieve strains of 15\%, whereas for $BR=0.75$, the maximum strain is 8\%. The amount of plastic deformation that occurs during the ramp increases with increasing biaxial ratio. As biaxial ratio increases, there is increased demand on the pressurization system. The strain rate during the ramp decreases, and so it takes longer to ramp. Since the stress is on the yield surface during most of the ramp, the longer ramp times result in greater amount of plastic deformation during this stage of the load history.


\begin{figure}[h]
\centering	
\subfigure[FCC \{200\}]{\includegraphics[width=0.49\linewidth]{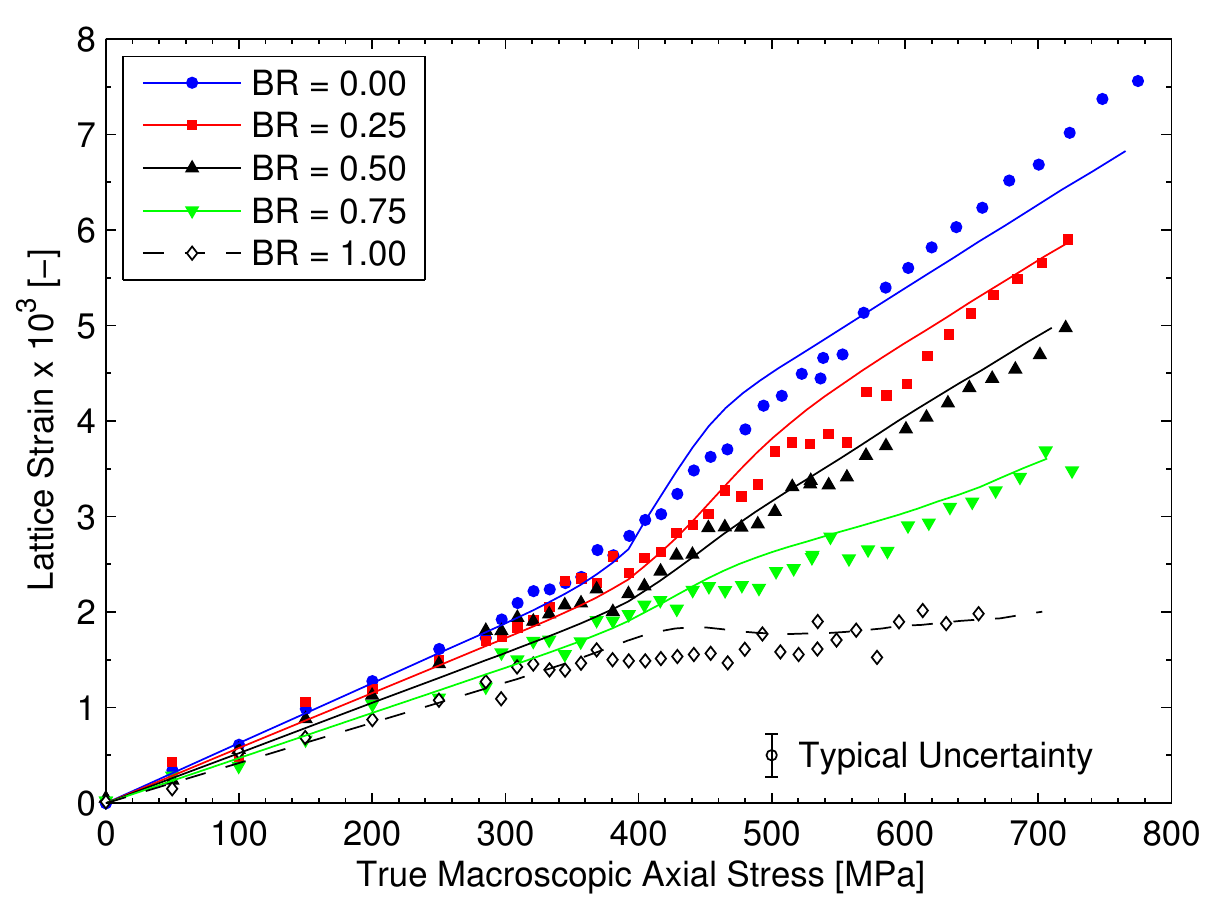}\label{fig:AxialFCC200}}
\subfigure[FCC \{111\}]{\includegraphics[width=0.49\linewidth]{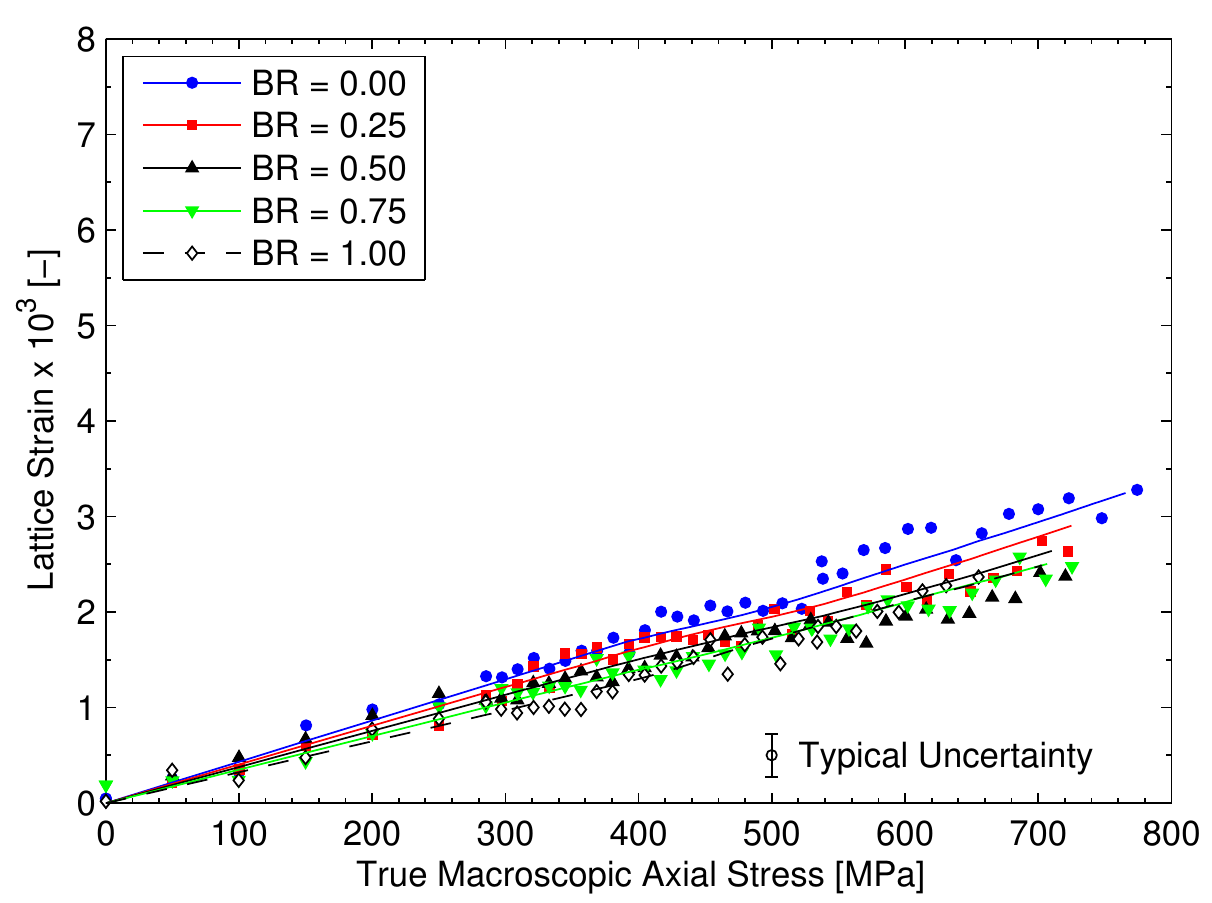}\label{fig:AxialFCC111}}
\subfigure[FCC \{220\}]{\includegraphics[width=0.49\linewidth]{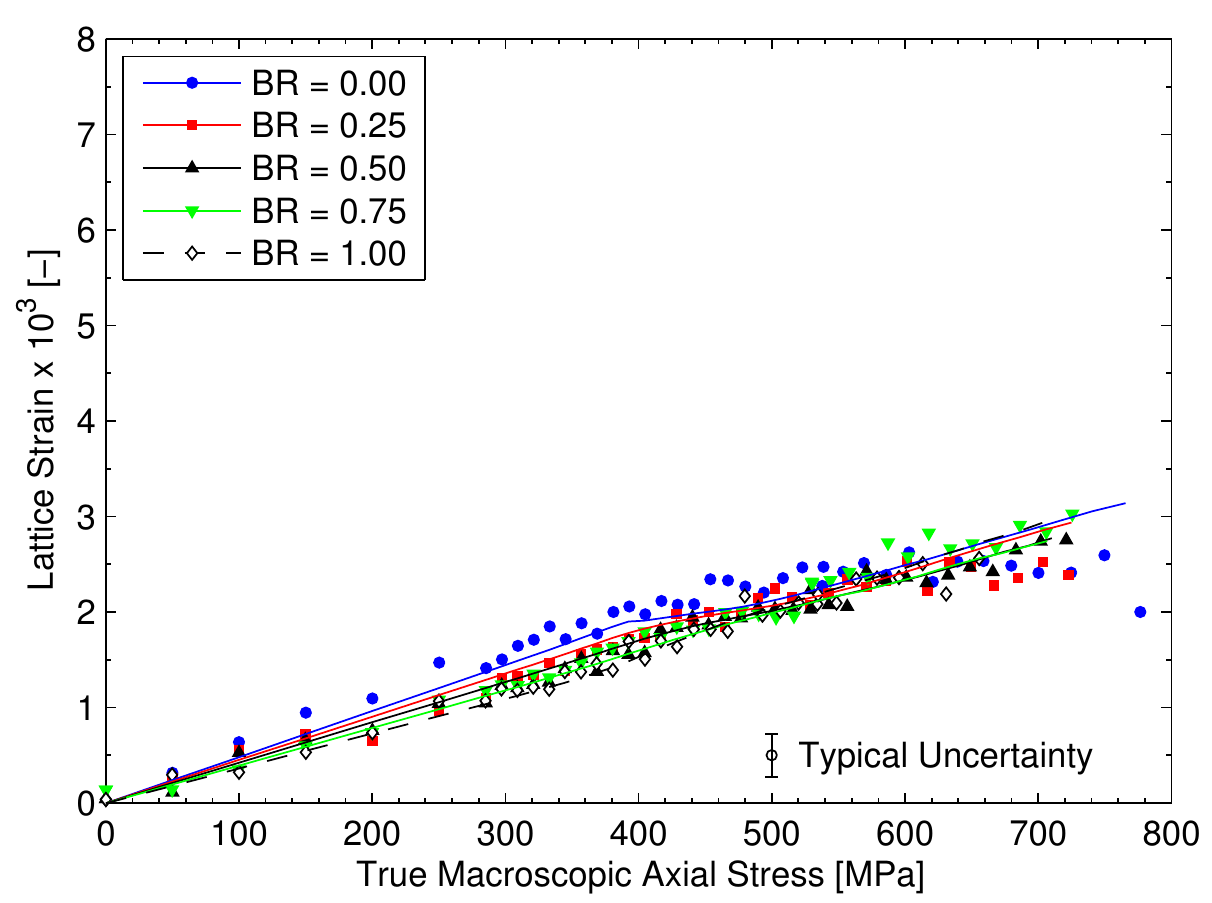}\label{fig:AxialFCC220}}
\subfigure[BCC \{200\}]{\includegraphics[width=0.49\linewidth]{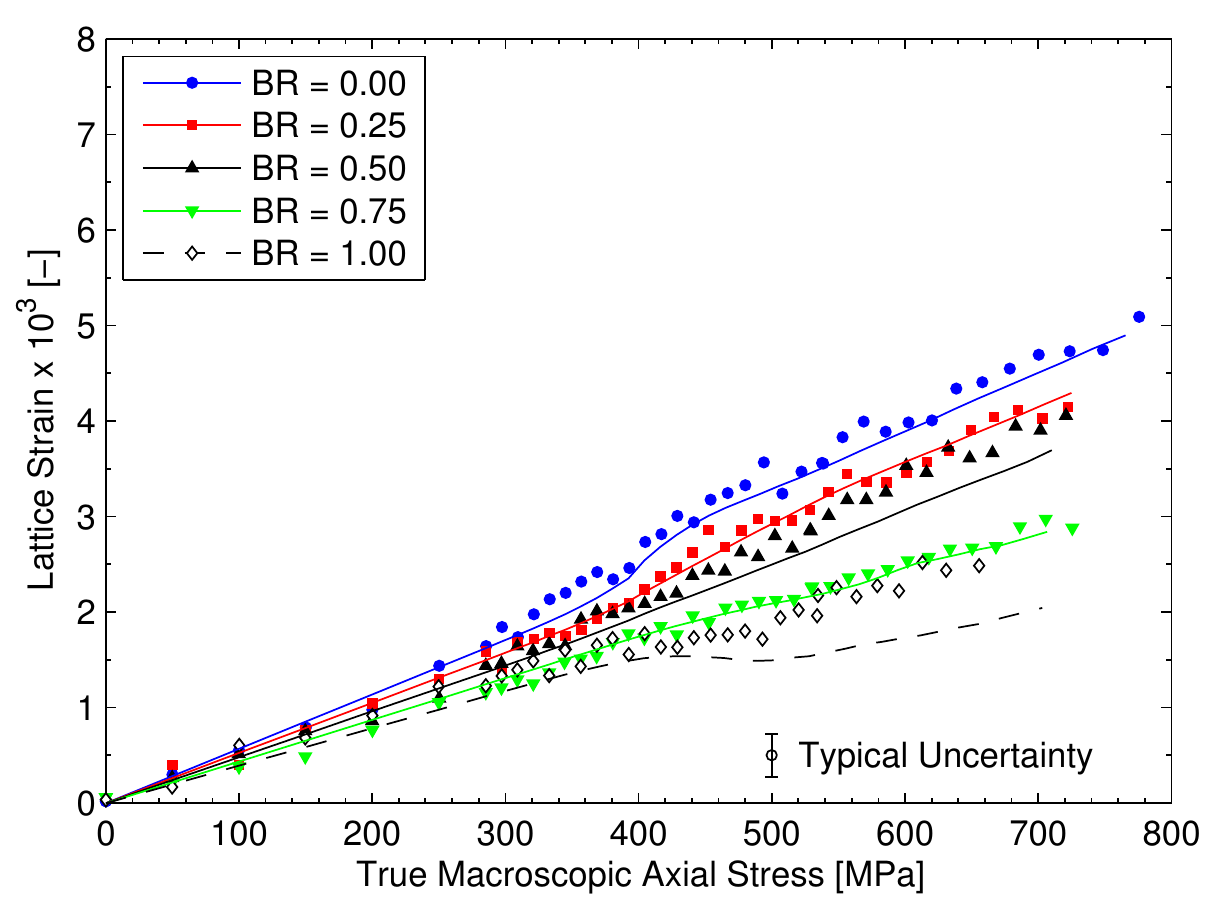}\label{fig:AxialBCC200}}
\subfigure[BCC \{110\}]{\includegraphics[width=0.49\linewidth]{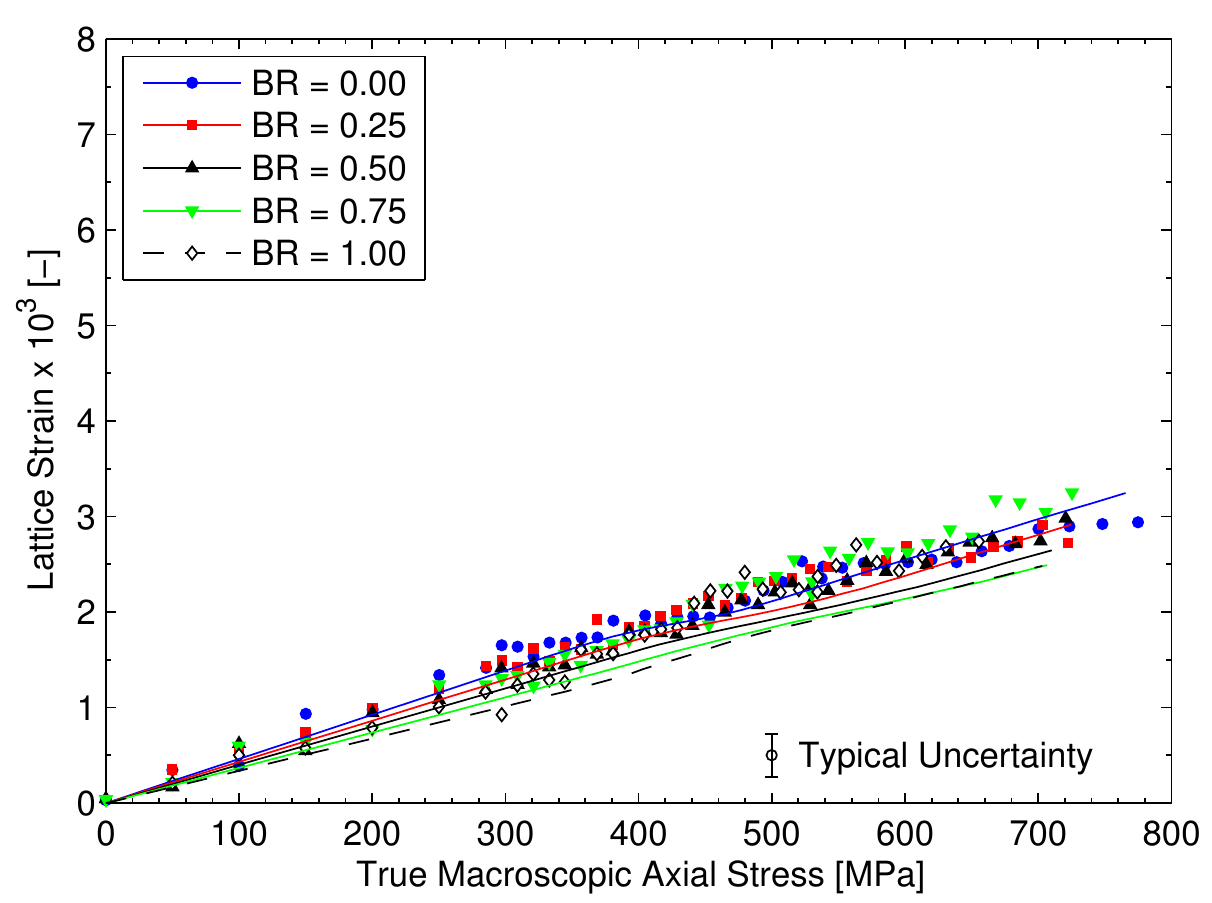}\label{fig:AxialBCC110}}
\subfigure[BCC \{211\}]{\includegraphics[width=0.49\linewidth]{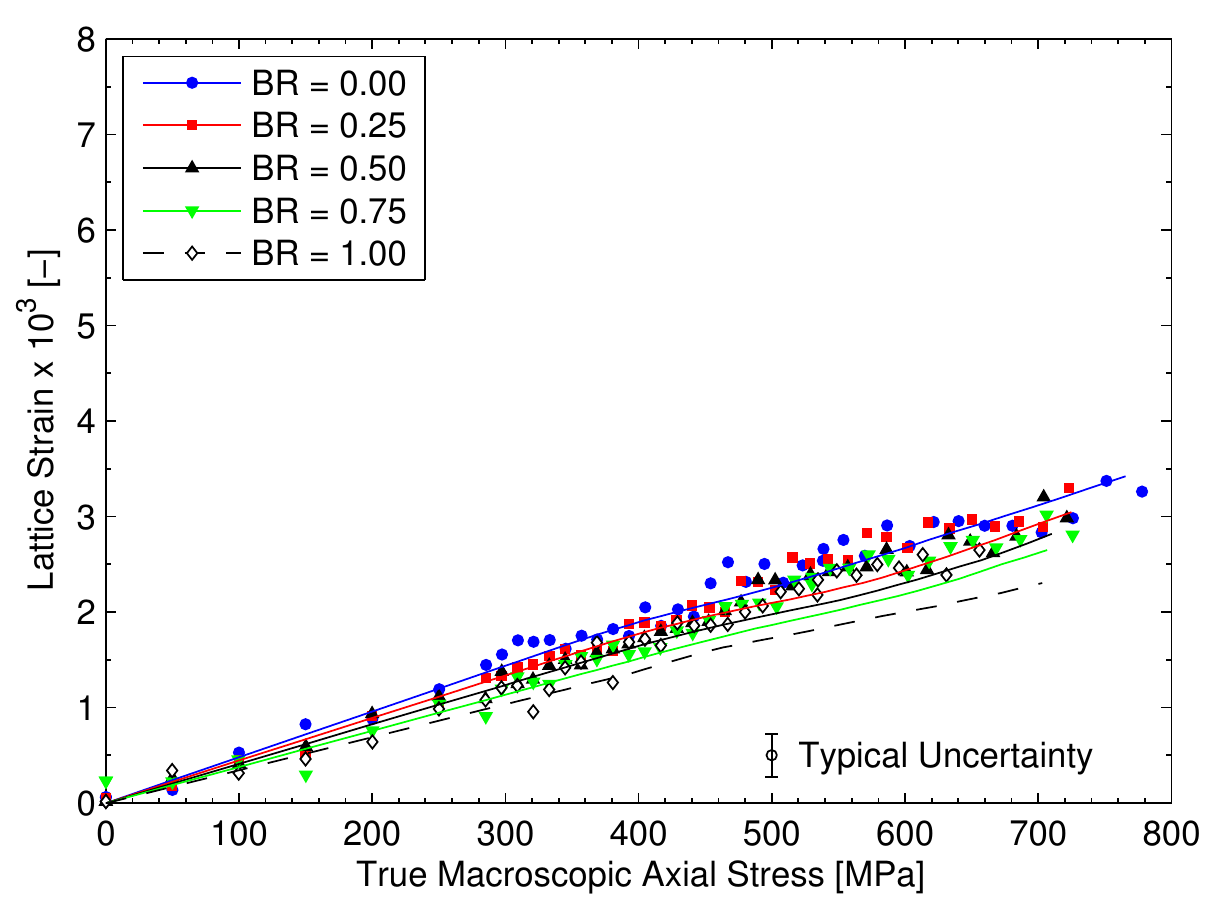}\label{fig:AxialBCC211}}
\caption{Axial lattice strains for both phases.}
\label{fig:axial_lattice_strains}
\end{figure}

\begin{figure}[h]
\centering	
\subfigure[FCC \{200\}]{\includegraphics[width=0.49\linewidth]{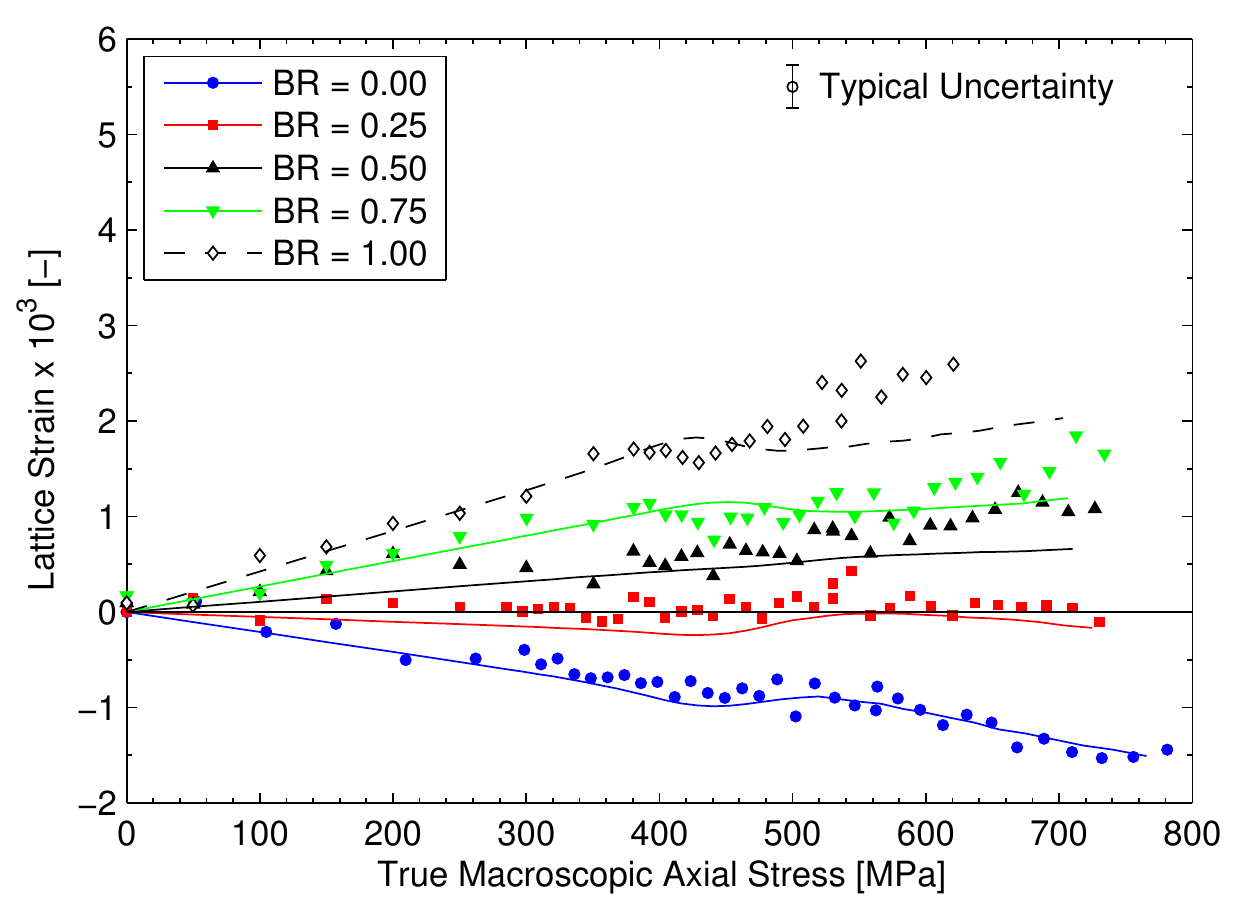}\label{fig:HoopFCC200}}
\subfigure[FCC \{111\}]{\includegraphics[width=0.49\linewidth]{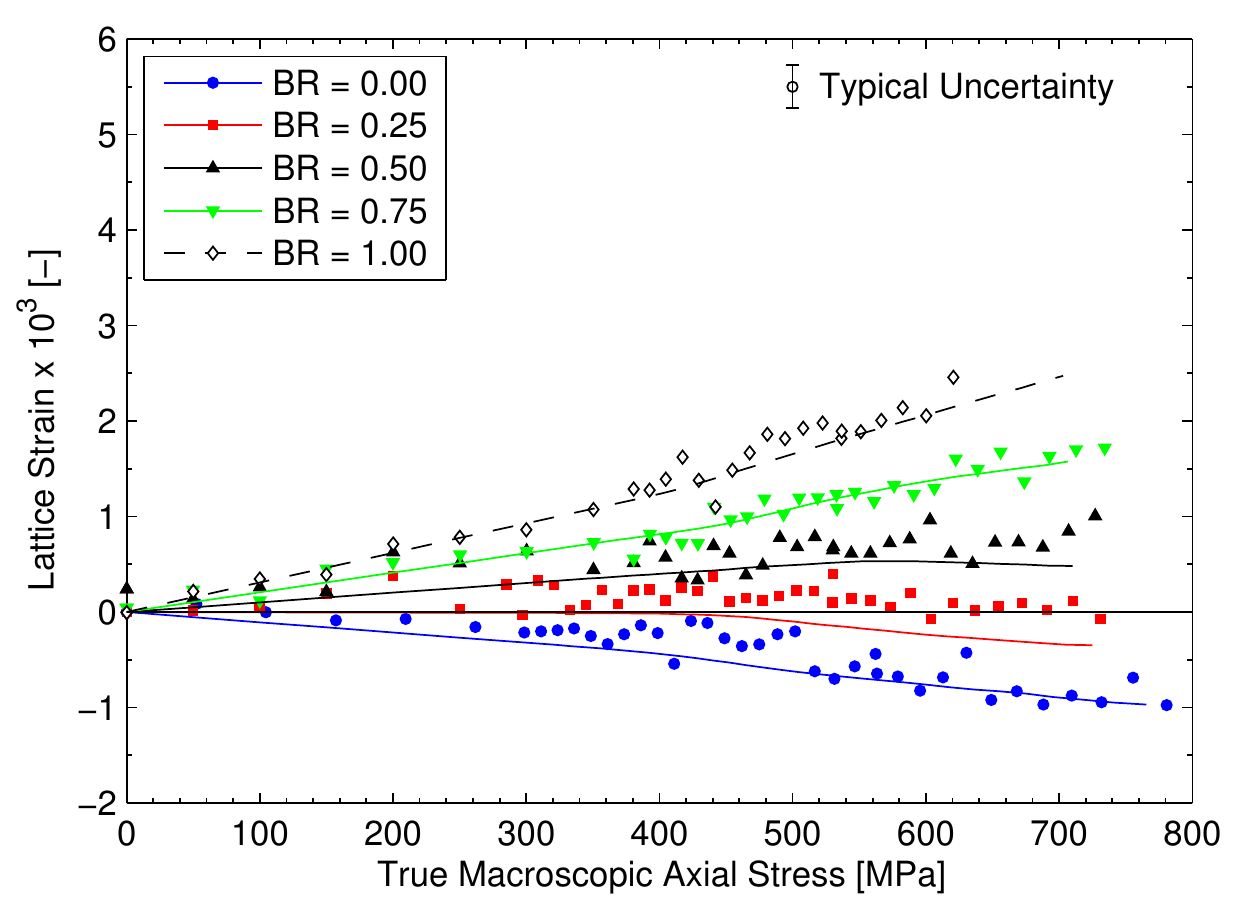}\label{fig:HoopFCC111}}
\subfigure[FCC \{220\}]{\includegraphics[width=0.49\linewidth]{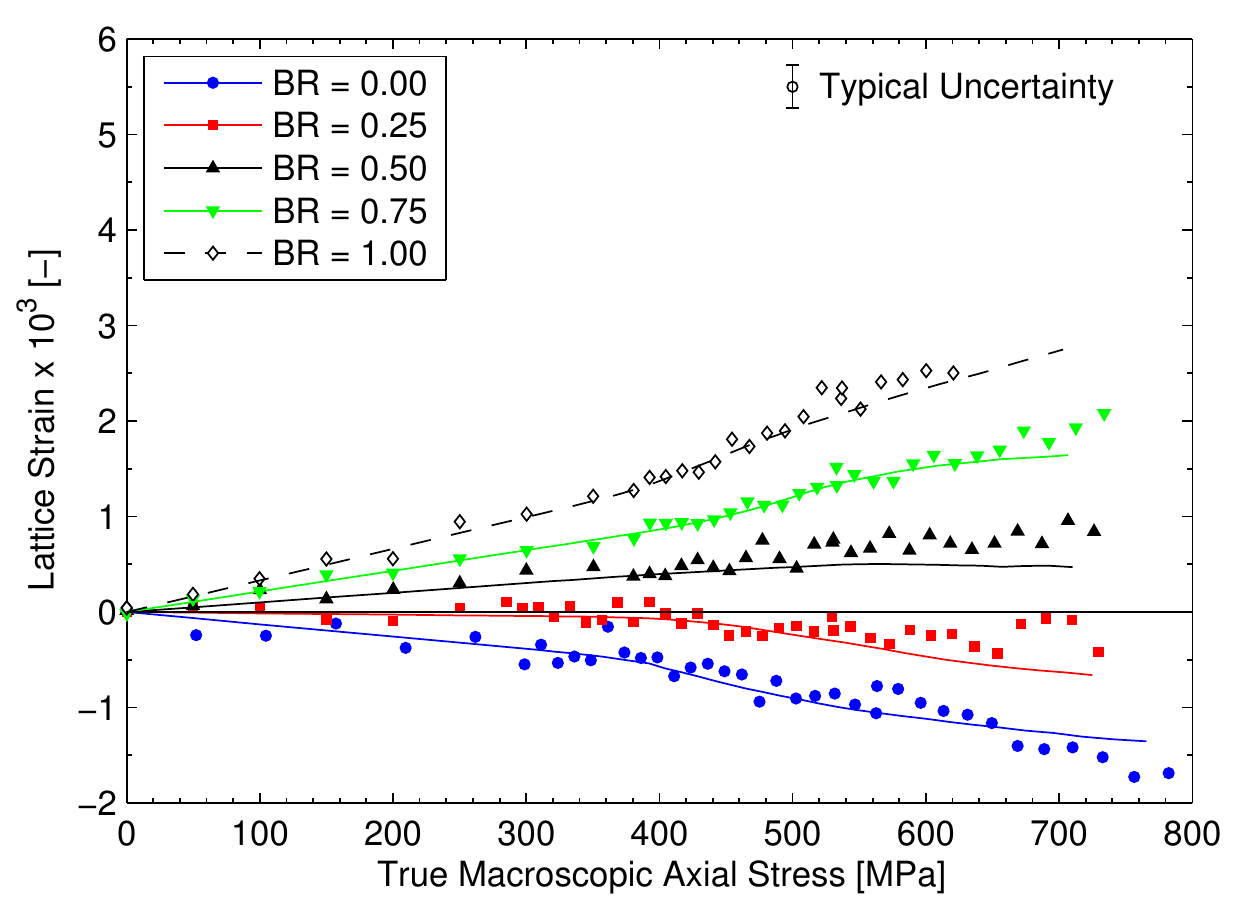}\label{fig:HoopFCC220}}
\subfigure[BCC \{200\}]{\includegraphics[width=0.49\linewidth]{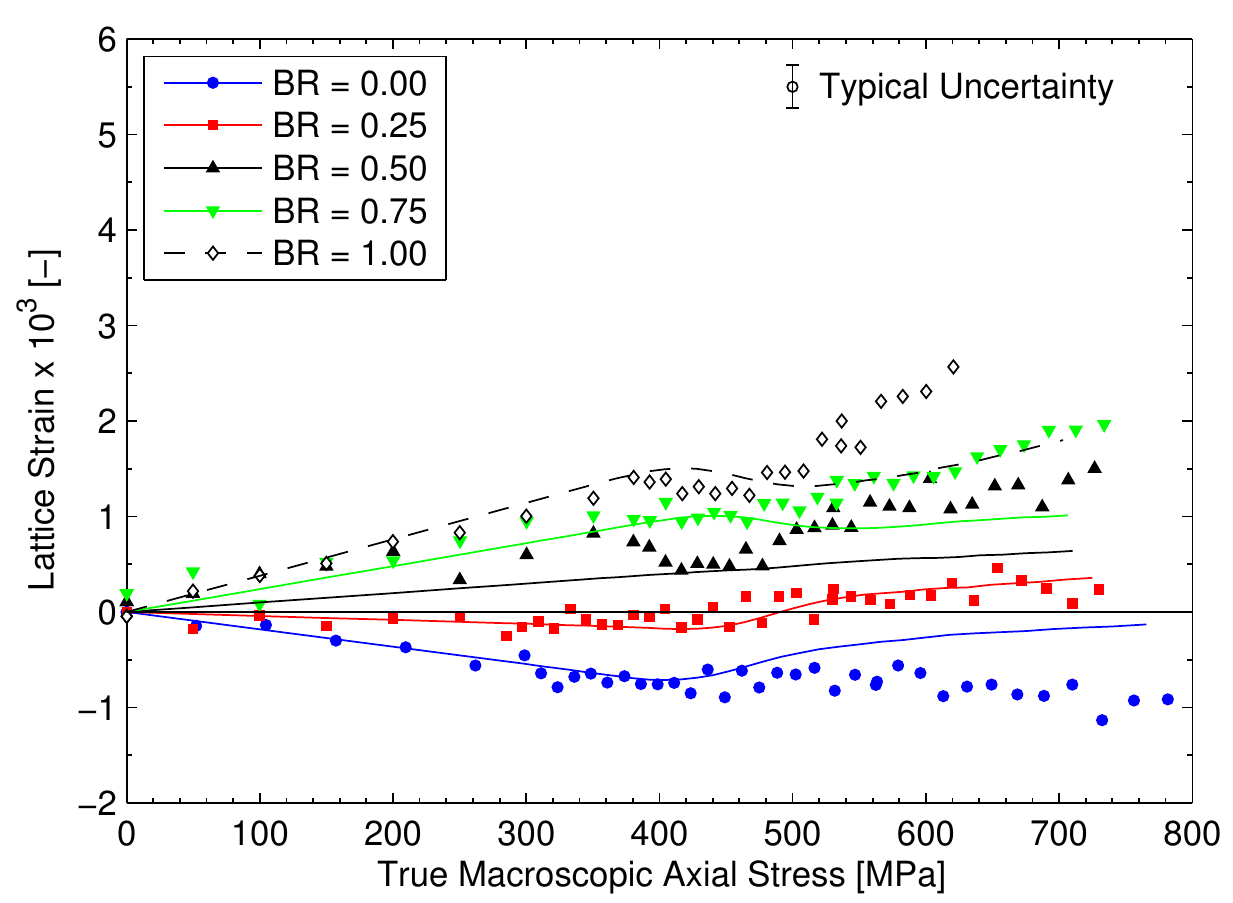}\label{fig:HoopBCC200}}
\subfigure[BCC \{110\}]{\includegraphics[width=0.49\linewidth]{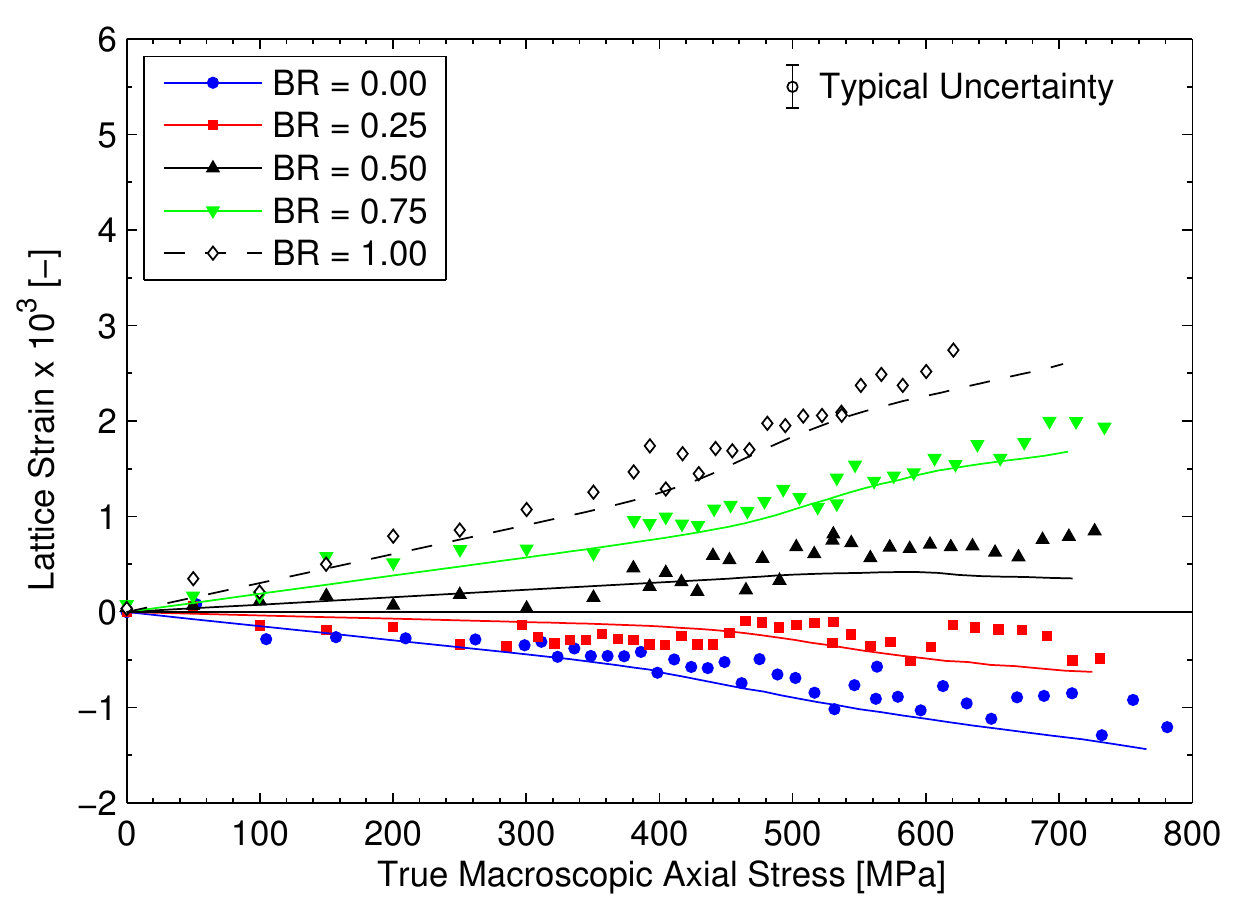}\label{fig:HoopBCC110}}
\subfigure[BCC \{211\}]{\includegraphics[width=0.49\linewidth]{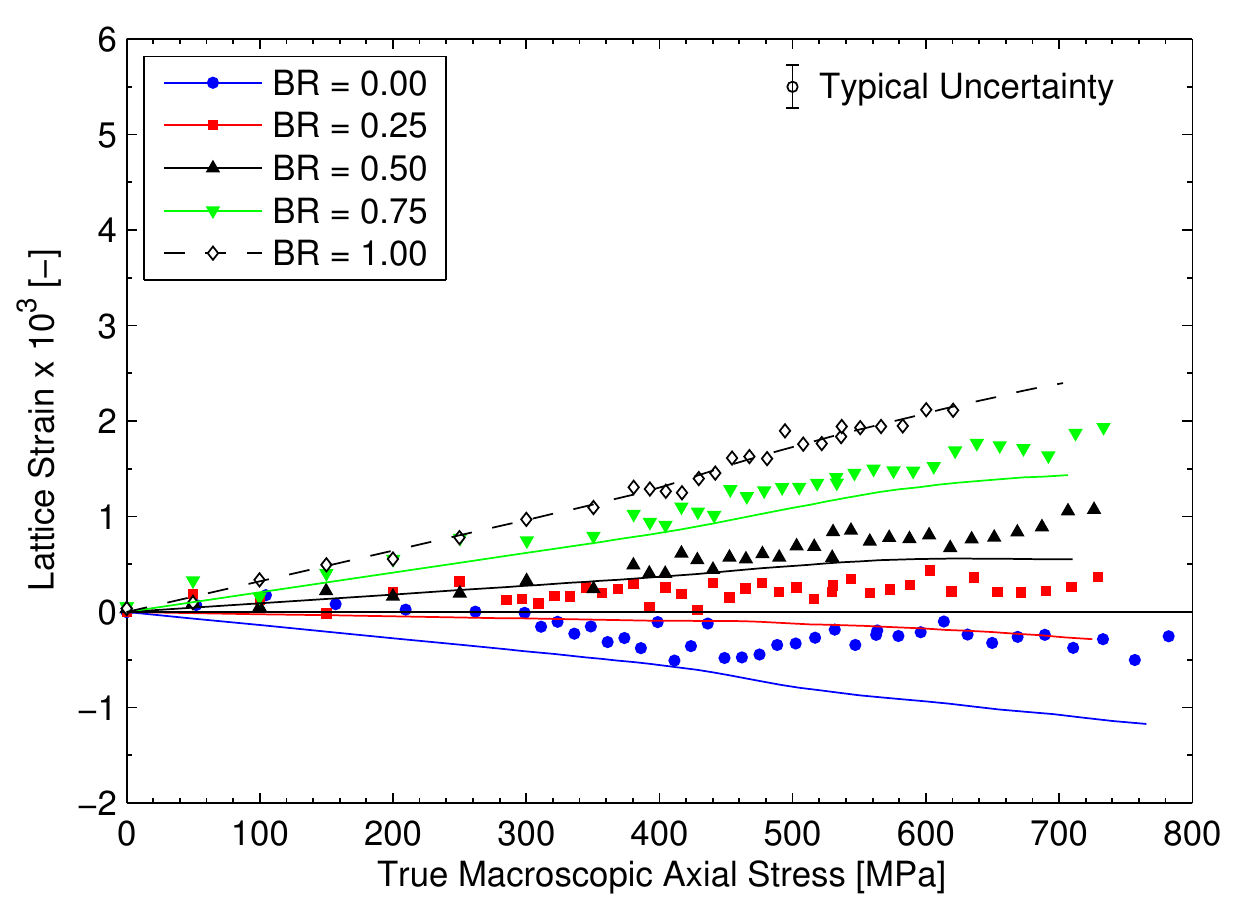}\label{fig:HoopBCC211}}
\caption{Hoop lattice strains for both phases.}
\label{fig:hoop_lattice_strains}
\end{figure}

\begin{figure}[h]
\centering	
\subfigure[FCC \{200\}]{\includegraphics[width=0.49\linewidth]{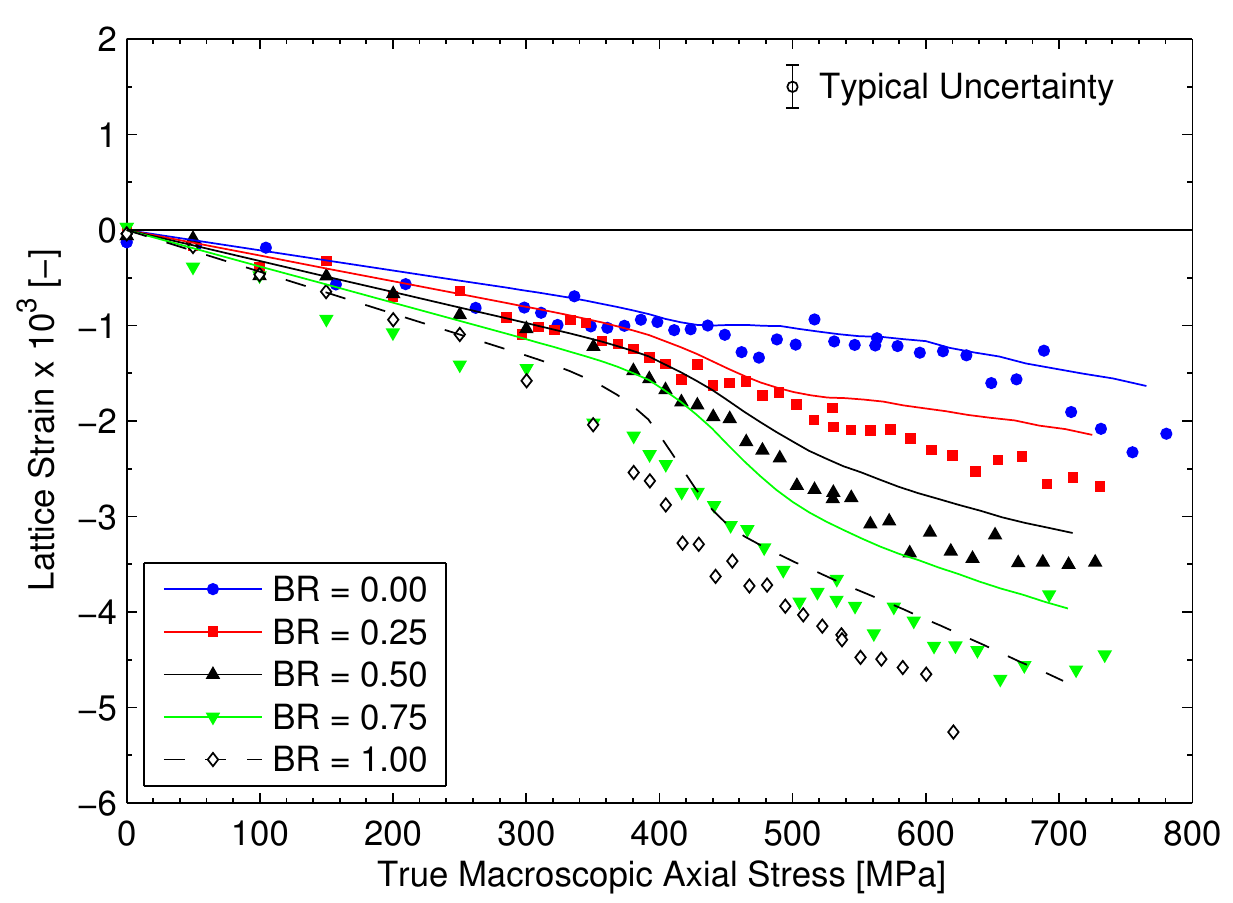}\label{fig:RadialFCC200}}
\subfigure[FCC \{111\}]{\includegraphics[width=0.49\linewidth]{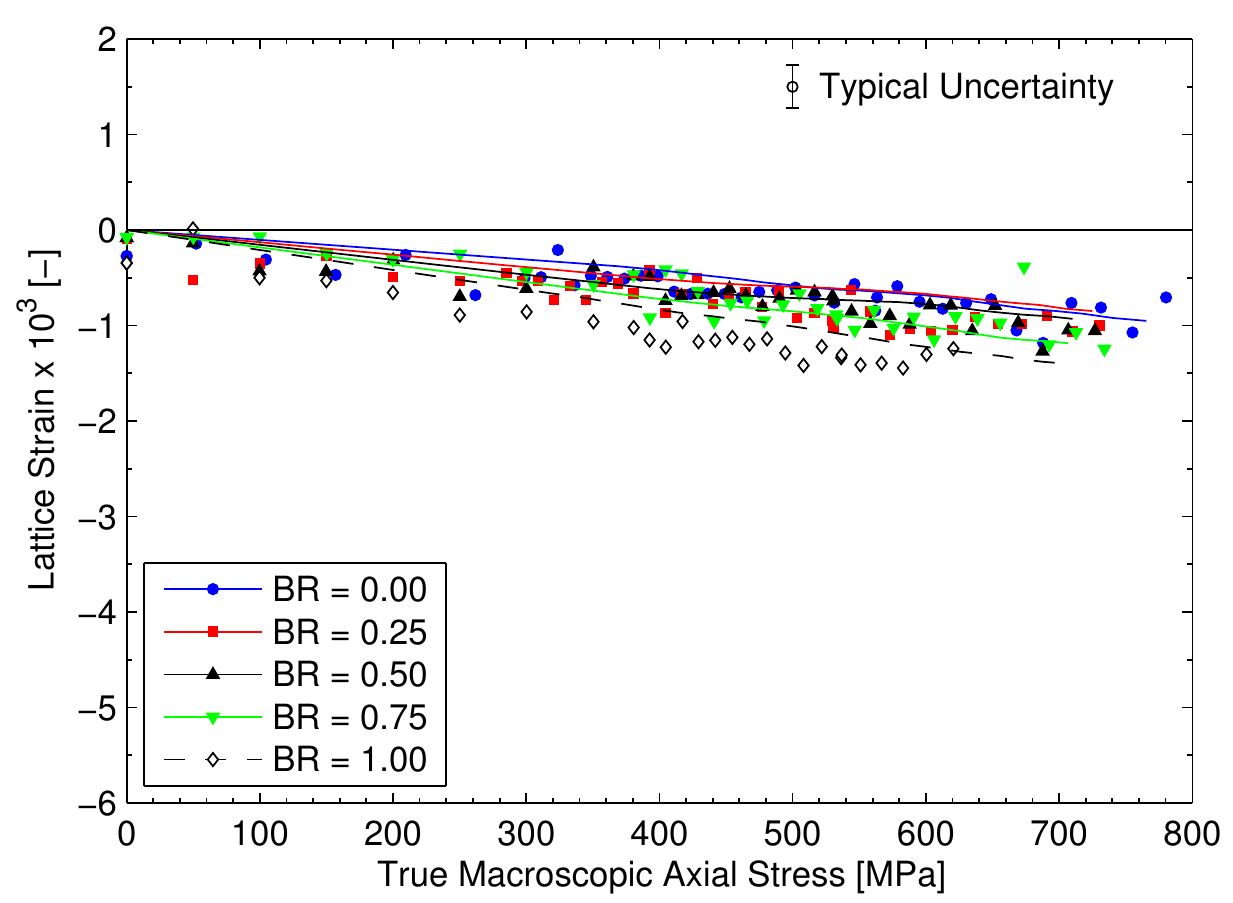}\label{fig:RadialFCC111}}
\subfigure[FCC \{220\}]{\includegraphics[width=0.49\linewidth]{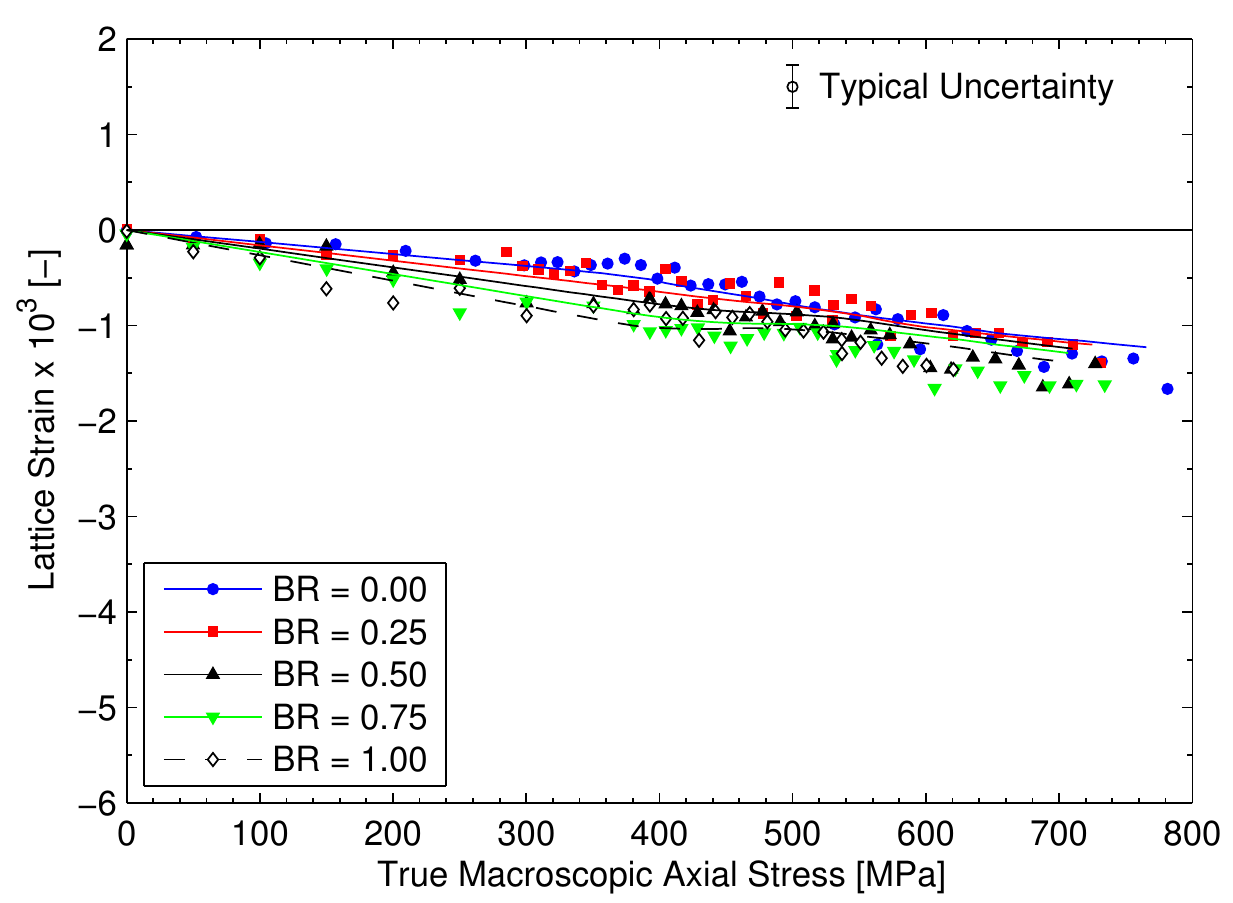}\label{fig:RadialFCC220}}
\subfigure[BCC \{200\}]{\includegraphics[width=0.49\linewidth]{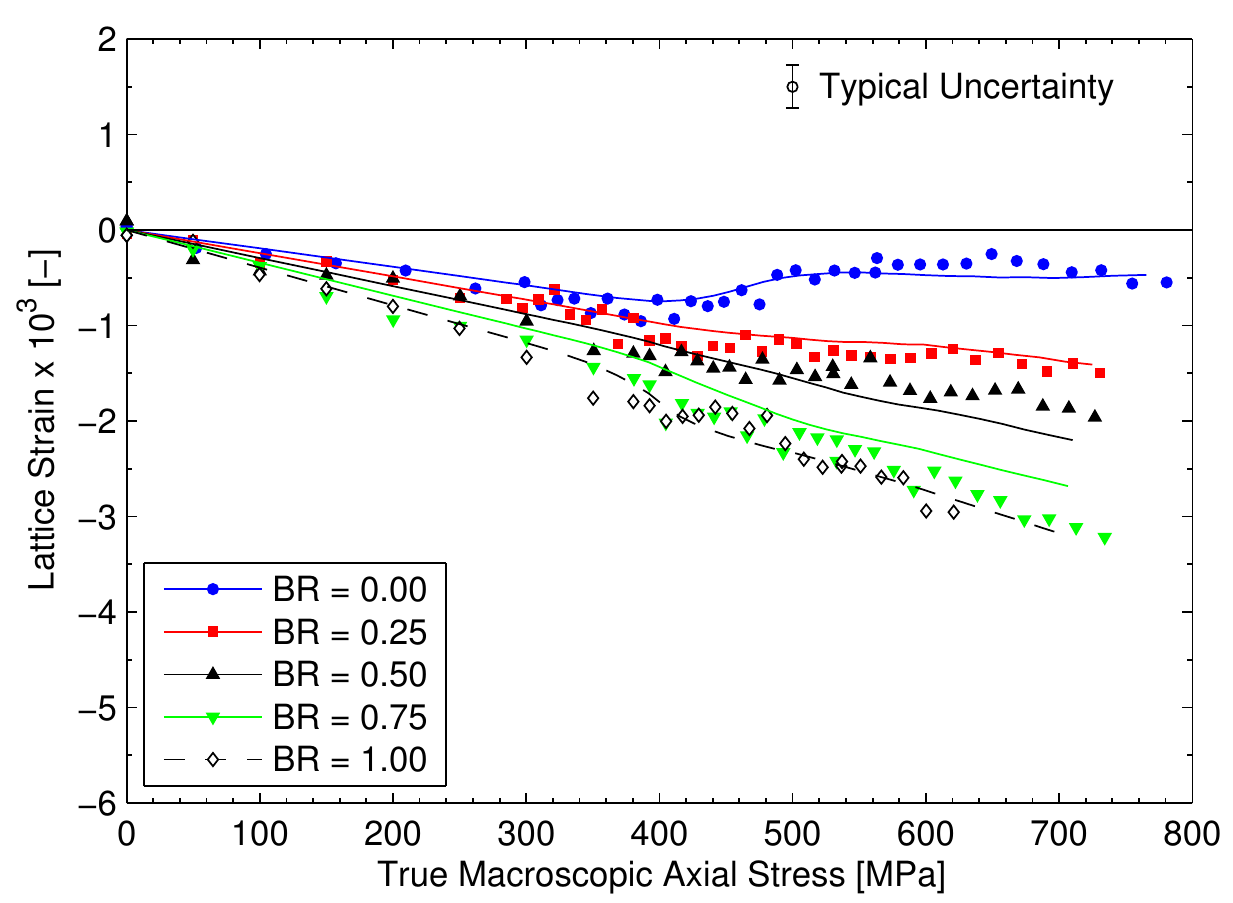}\label{fig:RadialBCC200}}
\subfigure[BCC \{110\}]{\includegraphics[width=0.49\linewidth]{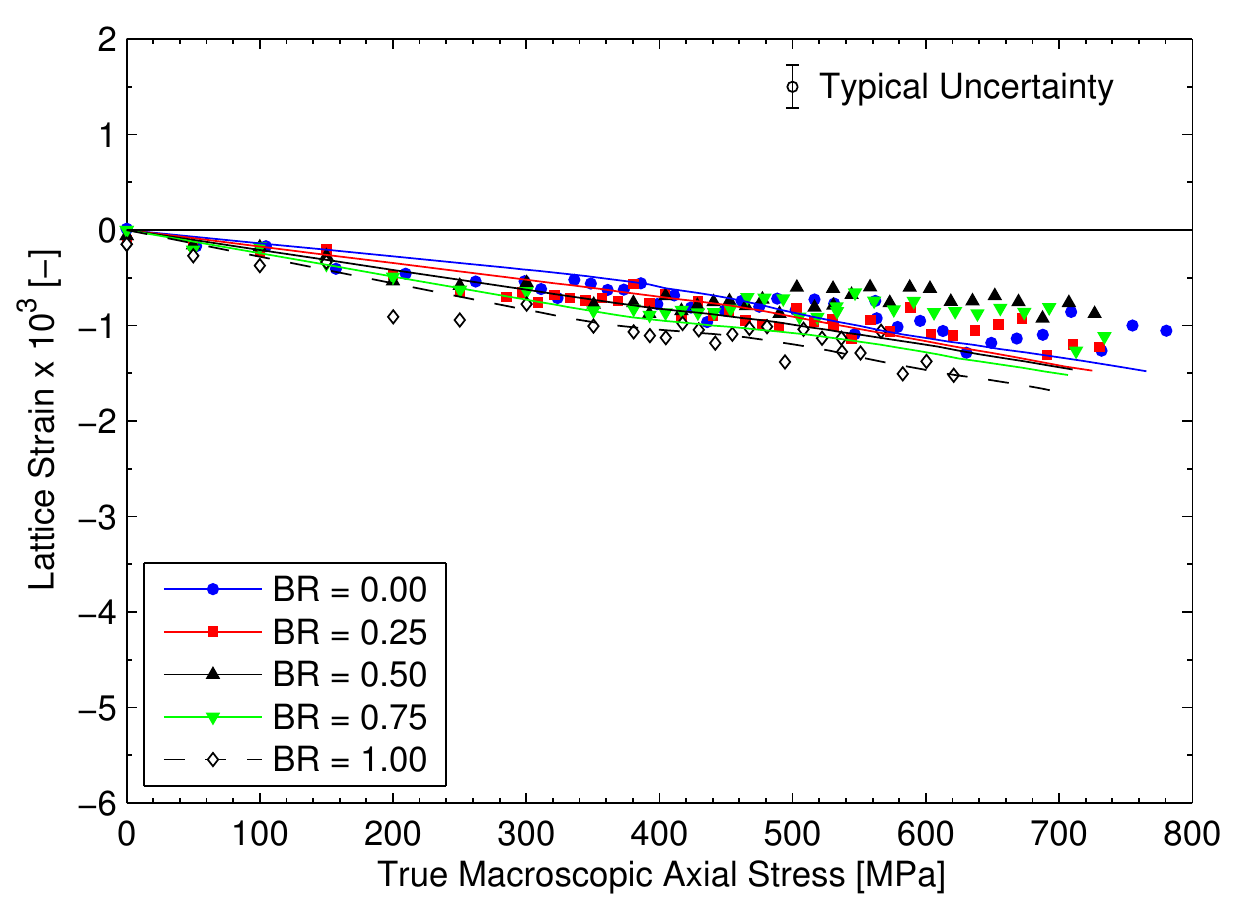}\label{fig:RadialBCC110}}
\subfigure[BCC \{211\}]{\includegraphics[width=0.49\linewidth]{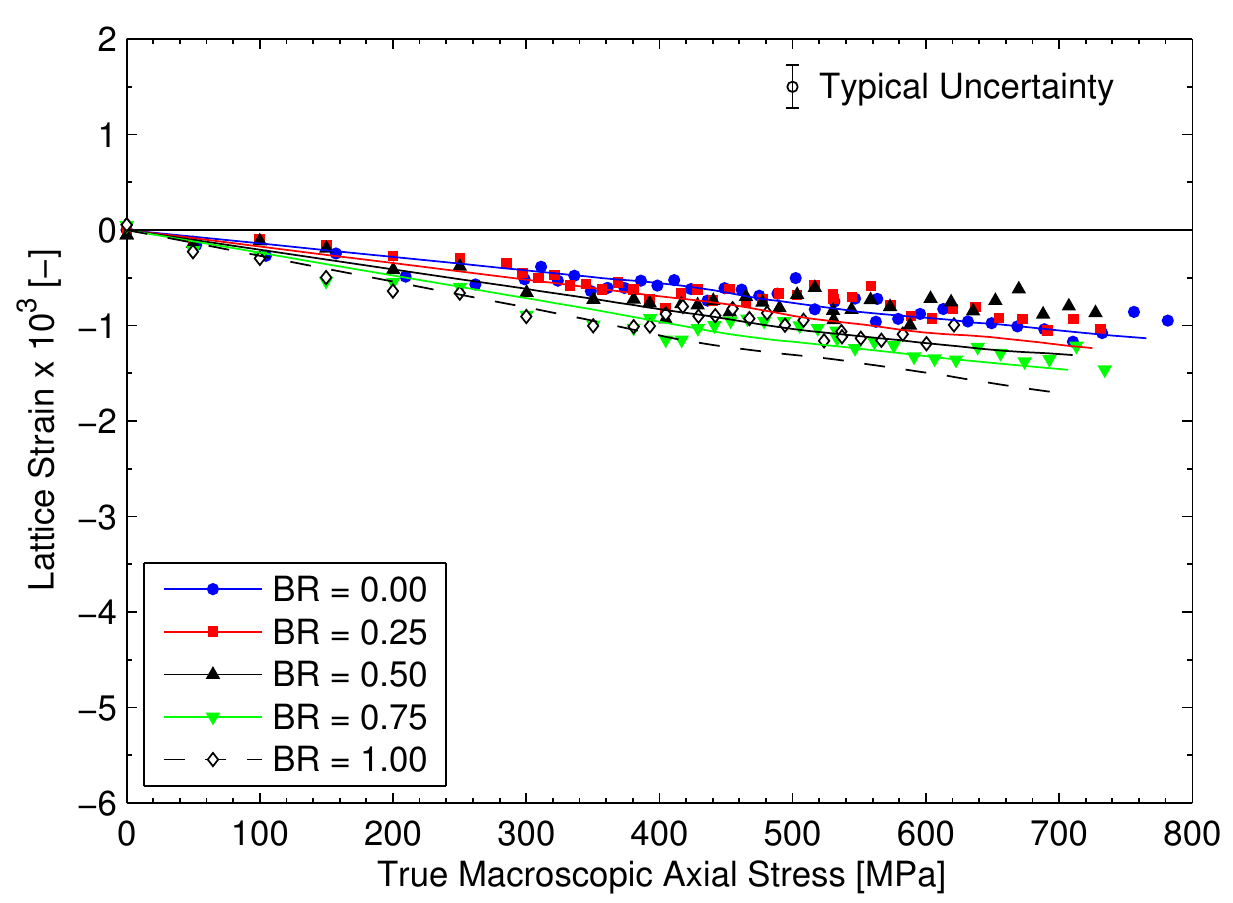}\label{fig:RadialBCC211}}
\caption{Radial lattice strains for both phases.}
\label{fig:radial_lattice_strains}
\end{figure}

Fiber averaged-lattice strain responses across the entire range of biaxial ratios are presented in Figures~\ref{fig:AxialFCC200}-\ref{fig:RadialBCC211}. Figures~\ref{fig:AxialFCC200}-\ref{fig:AxialBCC211} depict axial lattice strains, Figures~\ref{fig:HoopFCC200}-\ref{fig:HoopBCC211} depict hoop lattice strains, and Figures~\ref{fig:RadialFCC200}-\ref{fig:RadialBCC211} depict radial lattice strains. The lattice strain plots are grouped by crystallographic fiber and depict lattice strain for five levels of stress biaxiality. Overall, there is good general agreement between the simulations and experiments, which provides confidence in the model.

Of the axial lattice strains, the FCC \{200\} fiber (Figure~\ref{fig:AxialFCC200}) exhibits the greatest dependence on biaxial ratio, followed by the BCC \{200\} fiber (Figure~\ref{fig:AxialBCC200}). For both these fibers, there is a change in initial curvature from concave up to concave down as biaxial ratio increases. The other axial fibers are relatively insensitive to biaxial ratio. Hoop lattice strains exhibit the greatest dependence on biaxial ratio because the biaxial ratio is controlled by the hoop stress. For $BR=0$, the hoop lattice strain is negative due to Poisson contraction. At $BR=0.5$ and greater, hoop lattice strain is positive because it is dominated by extension due to hoop stress. Radial lattice strains are always negative due to Poisson contraction and become more negative with increasing biaxial ratio.

The simulation captures many of the key lattice strain trends, both in terms of magnitude and curvature. Not all the changes in curvature are matched in the plastic regime, for example the FCC\{200\} and FCC \{220\} axial lattice strains in Figures~\ref{fig:AxialFCC200} and~\ref{fig:AxialFCC220}. One might suppose that these discrepancies could be the result of a deformation-induced martensitic phase transformation that is not modeled in the simulation. However, such phase transformations for austenitic stainless steel typically occur below room temperature~\cite{De06a}. It is therefore unlikely that the material is undergoing phase transformation. Improvements to the fit in the plastic regime could potentially be made by further refinement of the hardening parameters. However, given the general agreement between the simulated and experimental lattice strain and macroscopic stress-strain responses, the current simulation does a good job of modeling the biaxial deformation of LDX-2101. The comparison between simulation and experiment builds confidence in the model, which is then used to investigate the deformation mechanics in ways that are complementary to the experiment. For example, in addition to fiber-averaged lattice strains, the model provides access to a range of complementary field data, including stress and plastic deformation rate. Simulation-based analysis of the initiation and propagation of yielding is presented in the subsequent chapters.


\pagebreak[4]
\clearpage

\section{Analysis of Yielding Based on Strength-to-Stiffness Ratios}
\label{sec:y2e}

In \cite{pos_daw_multiaxial-y2e}, the authors reported on a new strength-to-stiffness parameter formulated for multiaxial loading and demonstrated its application to the yielding of a single-phase stainless steel (AL6XN).   The  multiaxial strenth-to-stiffness parameter extended the development for uniaxial loading published earlier by Wong and Dawson~\cite{Wong10a}.   The authors also reported on a methodology for predicting the macroscopic stresses at which elements in a finite element mesh yield.  They demonstrated that it is possible to make this prediction using only the elastic strain data from a single increment of a purely elastic finite element simulation plus knowledge of the single-crystal yield surface.  In this section we apply this multiaxial strength-to-stiffness parameter to the two-phase LDX-2101 system, showing that the parameter is useful for predicting the initiation and propagation of yielding in two-phase systems as well.

\subsection{Strength-to-stiffness ratio for multiaxial stress states}\label{sec:Y2E_background}

A thorough development of the multiaxial strength-to-stiffness parameter is available in \cite{pos_daw_multiaxial-y2e}.   Here, we provide only a brief description of the parameter.  The strength-to-stiffness parameter, $r_{SE}$, draws the strength from the single crystal yield surface and the stiffness from Hooke's law.   For metals at low homologous temperatures, yielding corresponds to the onset of plastic deformation due to crystallographic slip. Crystallographic slip occurs on a restricted set of slip systems, where $\alpha$ denotes the slip system index. The resolved shear stress on the $\alpha$-slip system $\tau^\alpha$ is the projection of the local deviatoric stress onto the slip system
\begin{equation}\label{eqn:y2e_RSS}
\tau^\alpha = \boldsymbol{P}^\alpha:\boldsymbol{\sigma}^\prime
\end{equation}
where $\boldsymbol{P}^\alpha$ is the symmetric part of the Schmid tensor. In the rate-independent limit, yielding occurs when the magnitude of the resolved shear stress on any slip system is equal to the critical resolved shear stress $\tau^\alpha_{cr}$ for that slip system.   
This set of slip system constraints collectively define the single crystal yield surface.      The yield condition is given by 
\begin{equation}\label{eqn:y2e_YieldCondition}
\max_\alpha \left( \frac{\vert \tau^\alpha \vert}{\tau^\alpha_{cr}} \right) = 1
\end{equation}
Let $(\cdot)^*$ denote a slip system where 
\begin{equation}\label{eqn:y2e_TauStar}
\frac{\vert \tau^* \vert}{\tau^*_{cr}} \equiv \max_\alpha \left( \frac{\vert \tau^\alpha \vert}{\tau^\alpha_{cr}} \right)
\end{equation}
Equation~\ref{eqn:y2e_TauStar} embodies both the strength ($\tau^*_{cr}$) and the stress ($\tau^* $) acting on the crystal.  The stiffness can be inferred from the stress if the elastic strain is known.  To construct the strength-to-stiffness from Equation~\ref{eqn:y2e_TauStar}, the stiffness is introduced by dividing the stress by an appropriate estimate of the elastic strain.  Here,  $E_\mathit{eff}$, defined as the effective macroscopic elastic strain is used for this purpose.  This gives
\begin{equation}\label{eqn:y2e_Y2EEval}
r_{SE} = E_\mathit{eff} \frac{\tau^*_{cr}}{\vert \tau^* \vert}
\end{equation}
Equation~\ref{eqn:y2e_Y2EEval} may be evaluated a macroscopic stress in the elastic regime along the load path of interest. The crystal deviatoric stress needed to evaluate $\tau^*$ can either be evaluated from one load increment of a purely elastic finite element simulation or approximated with an isostrain assumption. 
$r_{SE}$ is a very simple, but quite general, multiaxial estimate of the strength-to-stiffness. 


\subsection{Analysis of initiation and propagation of yielding}\label{sec:Y2E_analysis}

\begin{figure}[h]
\centering
\subfigure[$BR = 0.00$]{\includegraphics[trim = 0in 0in 0.9in 0in, clip]{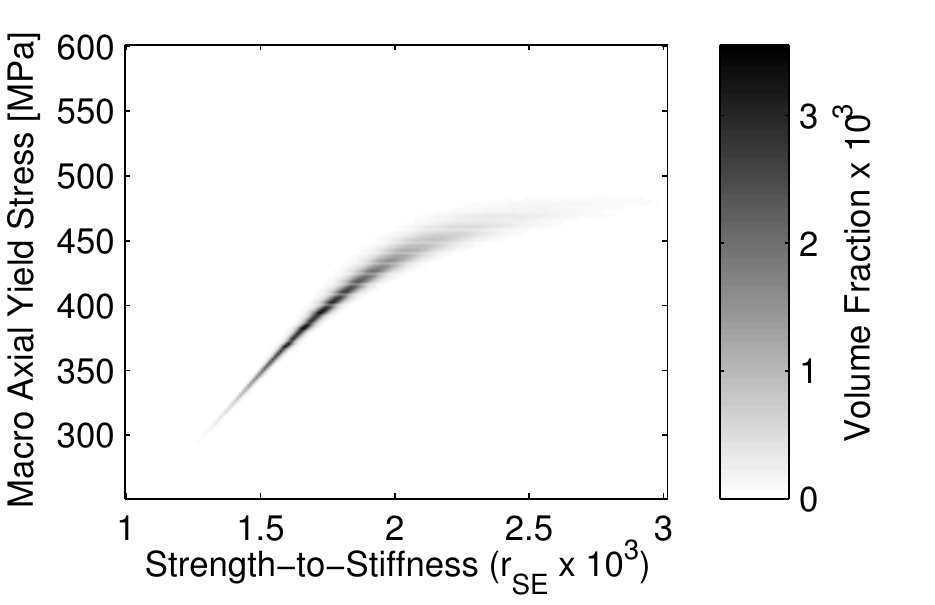}.pdf\label{fig:Y2E-BR000}} 
\subfigure[$BR = 0.25$]{\includegraphics[trim = 0in 0in 0.9in 0in, clip]{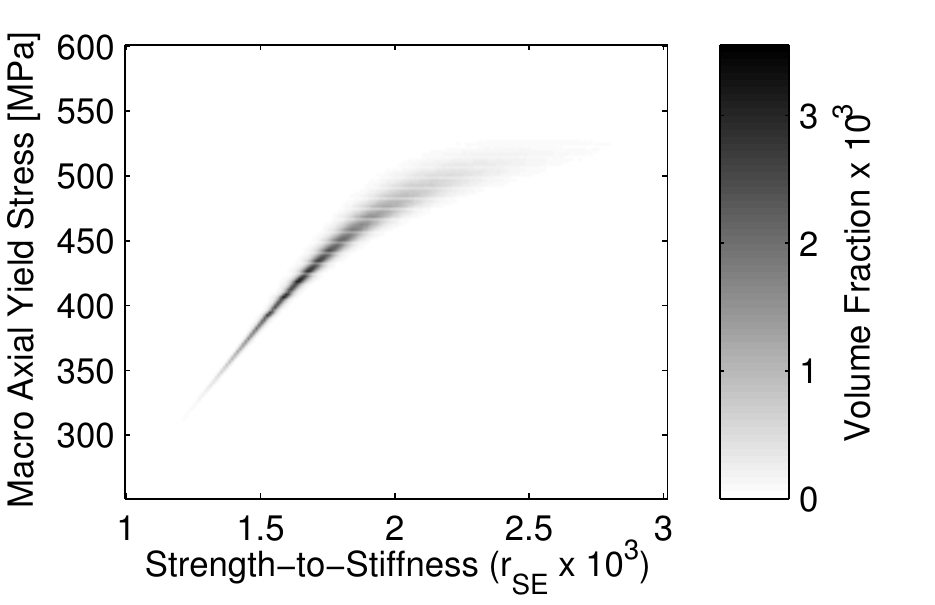}\label{fig:Y2E-BR025}}
\subfigure[$BR = 0.50$]{\includegraphics[trim = 0in 0in 0.9in 0in, clip]{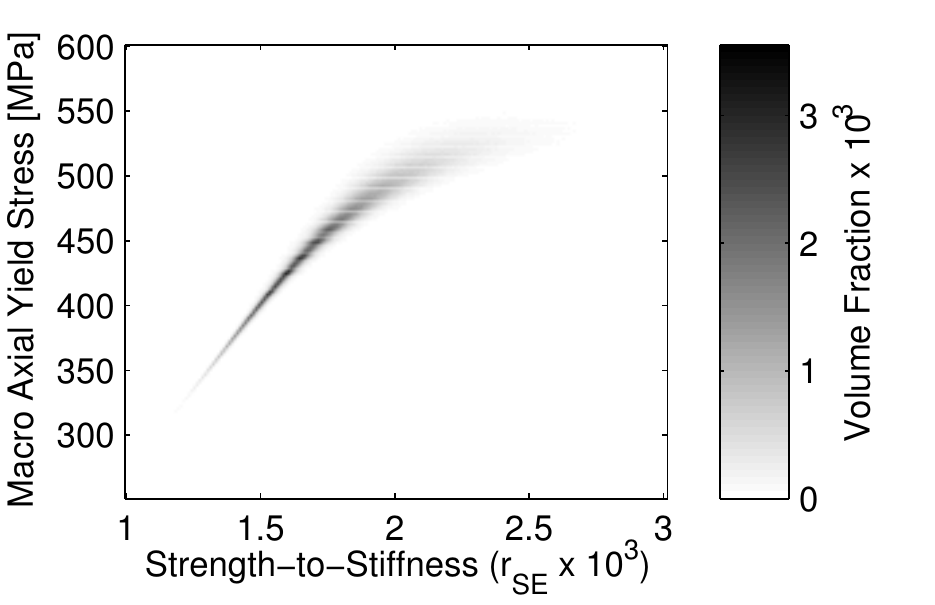}\label{fig:Y2E-BR050}}
\subfigure[$BR = 0.75$]{\includegraphics[trim = 0in 0in 0.9in 0in, clip]{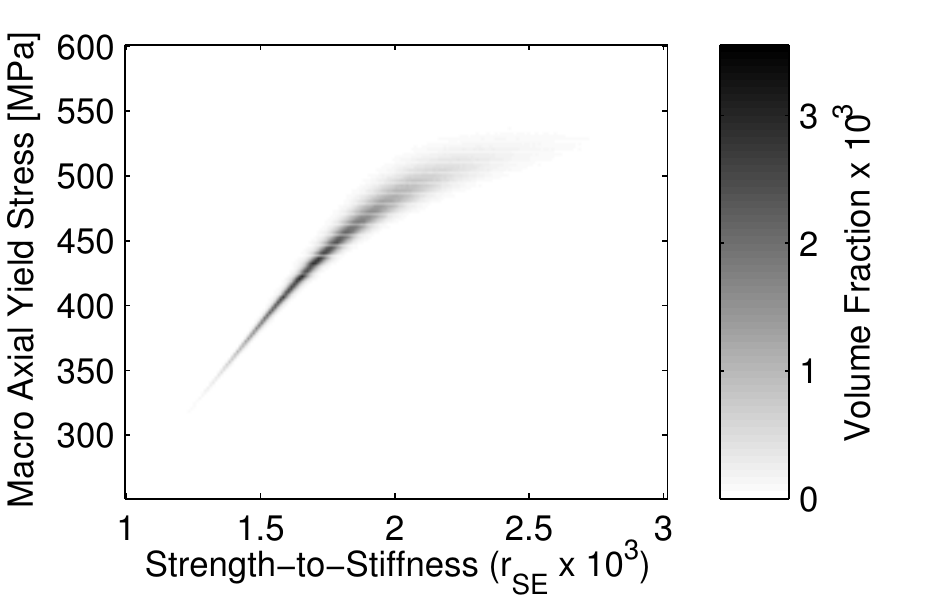}\label{fig:Y2E-BR075}}
\subfigure[$BR = 1.00$]{\includegraphics[trim = 0in 0in 0.9in 0in, clip]{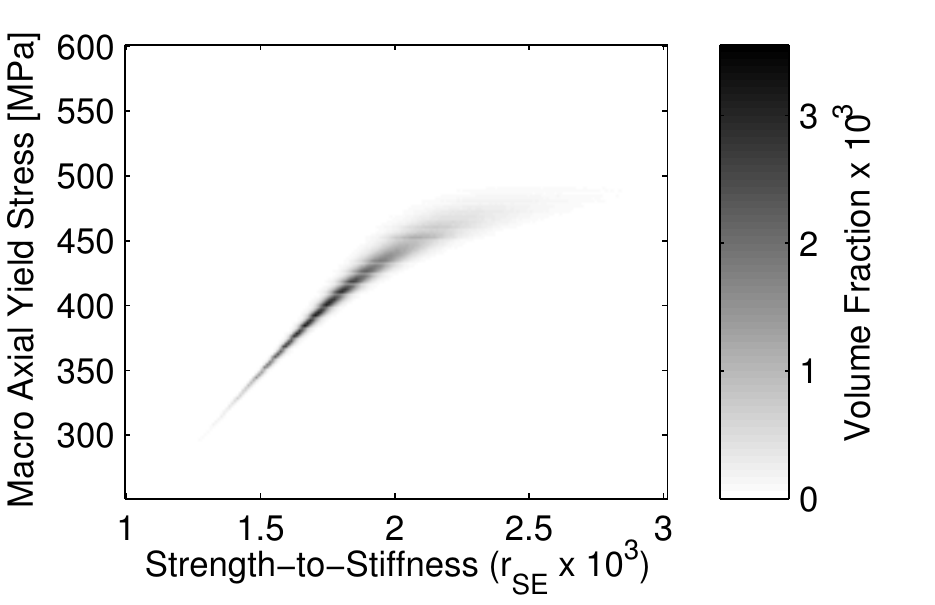}\label{fig:Y2E-BR100}}
\subfigure{\includegraphics[trim = 2.8in 0in 0in 0in, clip]{Y2E-BR000}}
\caption{The multiaxial strength-to-stiffness ratio governs the order in which elements yield. Elements with low strength-to-stiffness yield before elements with high strength-to-stiffness.}\label{fig:Y2E}
\end{figure}

Multiaxial strength-to-stiffness analysis is used to examine the initiation and propagation of yielding in LDX-2101. Binned scatter plots illustrating the correlation between strength-to-stiffness and the macroscopic stress at which elements yield are presented in Figure~\ref{fig:Y2E} for five levels of stress biaxiality.
In these plots, data are binned according to both strength-to-stiffness and macroscopic elemental yield stress. Intensity corresponds to the volume fraction of the aggregate contained in each bin. 
The behavior seen here for the two-phase system is similar to that observed 
for the single-phase system~\cite{pos_daw_multiaxial-y2e}.
As with the single-phase system, there is a strong, nonlinear correlation between strength-to-stiffness and macroscopic elemental yield stress for all biaxial stress states. The downward concavity of the curves is due to the increase in the local load increment, relative to the macroscopic load increment, that occurs to elastic elements when other elements yield. When an element yields, its ability to carry additional load is significantly reduced. Additional incremental load must be carry by the remaining elastic elements, and so the effective load increment for the elastic elements increases. As the local load increment increases relative to the macroscopic load increment, elements yield at lower macroscopic stresses than if the local load increment were constant, producing the downward curvature observed in the figures. The correlation between strength-to-stiffness and macroscopic elemental yield stress is stronger at low strength-to-stiffness than at high strength-to-stiffness. The correlation decreases over the course of the elasto-plastic transition because the analysis is based on a linearization of behavior in the elastic regime. As yielding progresses, local stresses evolve, deviating from the linearized values. The correlation is therefore stronger for elements that yield earlier in the elasto-plastic transition.

\begin{figure}[h]
\centering
\subfigure[Schmid factor]{\includegraphics[trim = 0in 0in 0.9in 0in, clip]{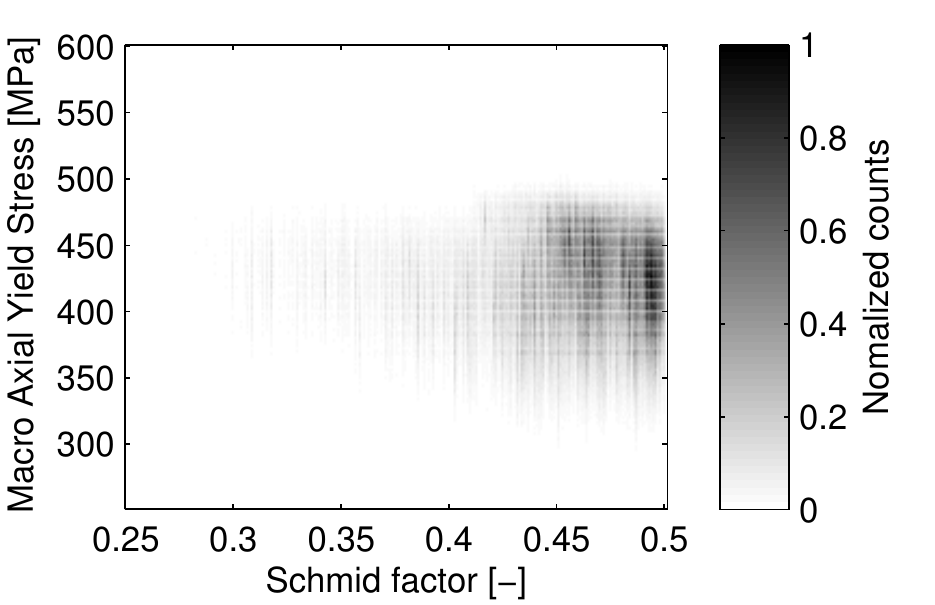}\label{fig:YieldStressSchmid}}
\subfigure[Taylor factor]{\includegraphics[trim = 0in 0in 0.9in 0in, clip]{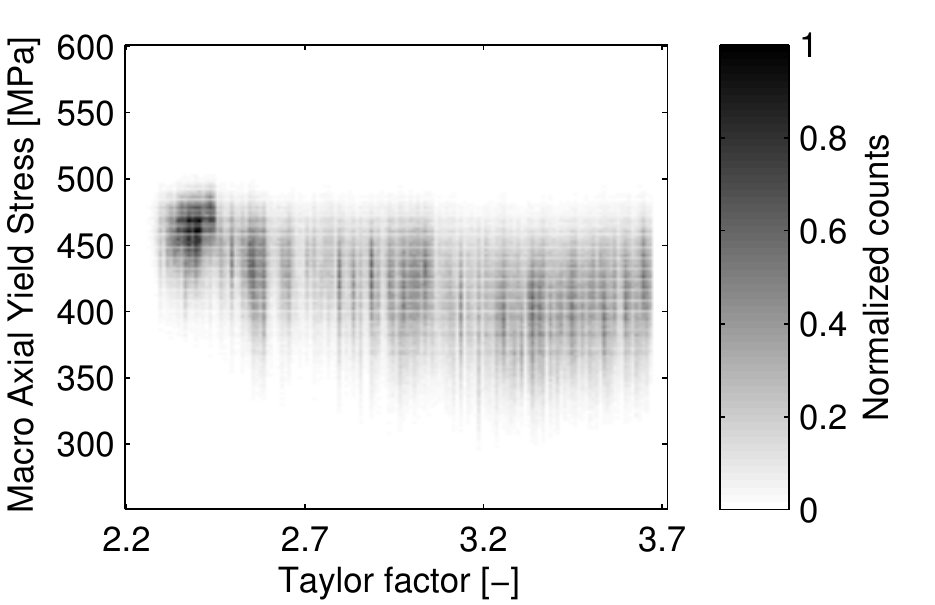}\label{fig:YieldStressTaylor}}
\subfigure[Strength-to-stiffness (isostrain)]{\includegraphics[trim = 0in 0in 0.9in 0in, clip]{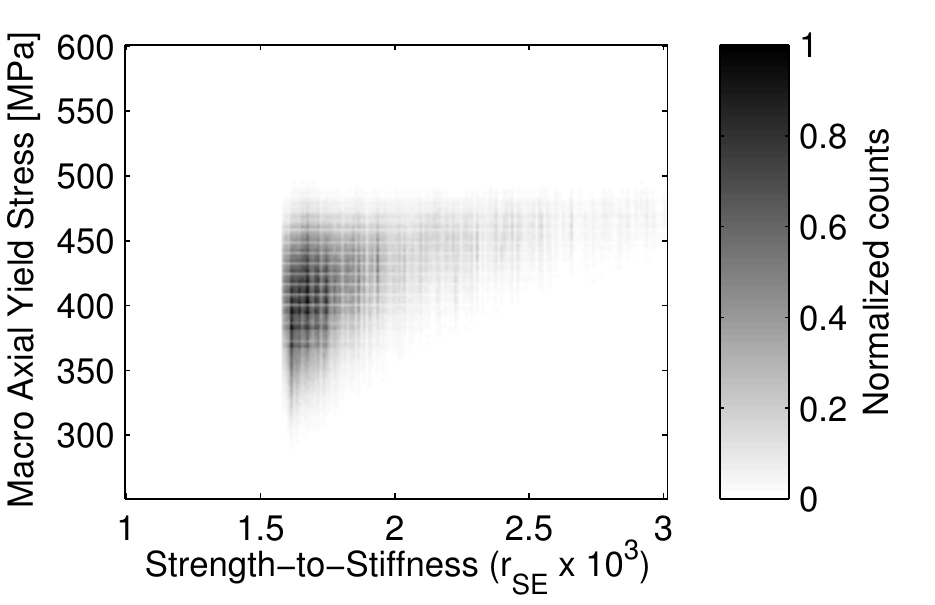}\label{fig:YieldStressY2EIso}}
\subfigure[Strength-to-stiffness]{\includegraphics[trim = 0in 0in 0.9in 0in, clip]{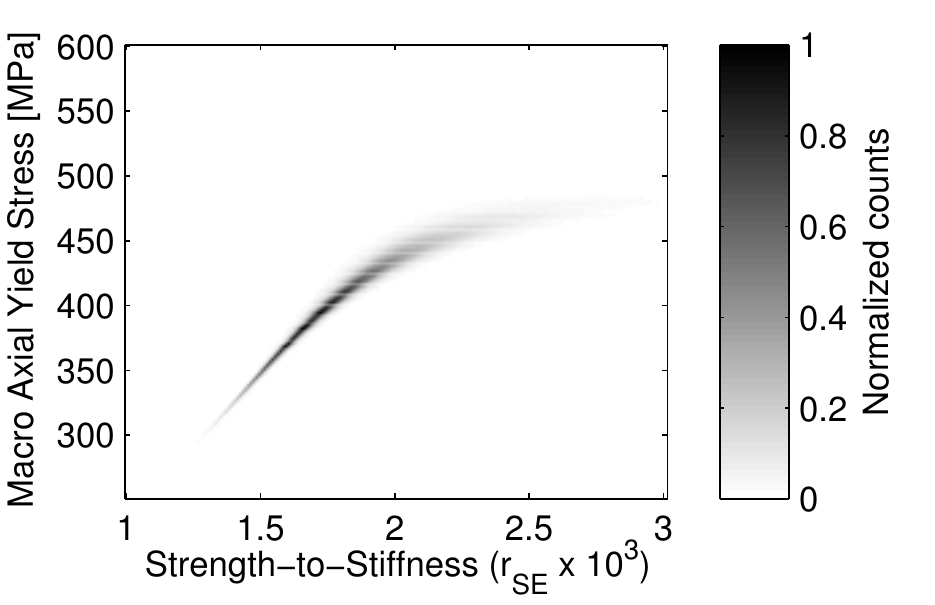}\label{fig:YieldStressY2E}}
\subfigure{\includegraphics[trim = 0.3in 0.2in 0.2in 1.4in, clip]{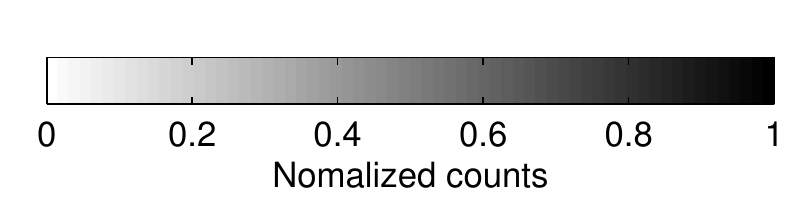}}
\caption{Correlation between elemental yield stress and strength and strength-to-stiffness measures for uniaxial loading. Both strength and stiffness are important in governing yield behavior. The strength-to-stiffness parameter, which incorporates neighborhood effects, captures the relative order in which elements yield.}\label{fig:ElemYieldStress}
\end{figure}

The correlation between macroscopic elemental yield stress and various strength and strength-to-stiffness parameters for uniaxial loading are presented in Figure~\ref{fig:ElemYieldStress}.
There is no correlation between the Schmid and Taylor factors and macroscopic elemental yield stress (Figures~\ref{fig:YieldStressSchmid} and~\subref{fig:YieldStressTaylor}). The vertical streaking is due to the fact that, although the strength parameter is constant over the grain, the entirety of a grain does not yield at the same macroscopic stress. The heavy streaks are likely produced by large BCC grains. It is also interesting to note that there is a pocket of elements with low Taylor factors that yield at a higher macroscopic stress than the other elements. This pocket corresponds to the $\{100\}||[001]$ crystallographic fiber, which has low strength, but also low directional stiffness. These crystals have relatively high strength-to-stiffness and therefore yield later than other crystals. Strength-to-stiffness formulated with an isostrain assumption exhibits some correlation with macroscopic elemental yield stress (Figure~\ref{fig:YieldStressY2EIso}). The elements that are first to yield all have low strength-to-stiffness, and elements with high strength-to-stiffness are among the last to yield. However, there is a large volume fraction of material with low strength-to-stiffness that yields over a range of 100~MPa. The isostrain approximation of strength-to-stiffness is not very predictive. It does not take into account that the local stress is different from the macroscopic stress due to intergranular interactions produced by compatibility constraints. Only when the local stress, or equivalently the local elastic strain, from the finite element simulation is incorporated into the strength-to-stiffness formulation is there a strong correlation between strength-to-stiffness and macroscopic elemental yield stress (Figure~\ref{fig:YieldStressY2E}). This comparison highlights the importance of neighborhood effects and illustrates that both strength and stiffness are important in governing the initiation and propagation of yielding.


\subsection{Methodology for predicting local yield}\label{sec:yield_prediction}

The multiaxial strength-to-stiffness parameter can be used to predict the macroscopic stress at which a region will yield. The same assumptions about the macroscopic and local load histories that were used for the strength-to-stiffness formulation are applied~\cite{pos_daw_multiaxial-y2e}. In this analysis, the incremental load is carried only by elastic regions as regions that have yielded as assumed to be unable to carry additional load (neglecting hardening over the elastic-plastic transition). Designating the local (crystal) stress as $\sigma$ and the macroscopic average stress as $\Sigma$, the ratio of local stress increment to macroscopic stress increment is $\Delta\sigma / \Delta\Sigma$  and is  zero for plastic regions. With the macroscopic stress increasing over the elastic-plastic transition, this ratio must increase for the remaining elastic regions as yielding progresses. As developed in \cite{pos_daw_multiaxial-y2e}, the local stress increment is approximated as being directly proportional to the applied macroscopic stress increment and inversely proportional to the elastic volume fraction raised to an empirical power $n$
\begin{equation}\label{eqn:local_stress_incr}
\Delta \sigma \varpropto \left( \frac{v}{v^e} \right)^n \Delta \Sigma
\end{equation}
where $v^e$ and $v$ are the elastic and total volumes, respectively. The volumetric scaling is an empirical factor that models the increase in local load increment, relative to the macroscopic load increment, for elastic regions that occurs as yielding propagates through an aggregate. The resolved shear stress is a projection of the local stress and therefore exhibits the same scaling

Introducing the constant of proportionality $\left( d \tau^* / d \Sigma \right)_0$ yields the equality
\begin{equation}\label{eqn:drss_dsigbar}
\frac{\Delta \tau^*}{\Delta \Sigma} = \left( \frac{d \tau^*}{d \Sigma} \right)_0 \left( \frac{v}{v^e} \right)^n
\end{equation}
where $\left( d \tau^* / d \Sigma \right)_0$ is the derivative of the resolved shear stress with respect to the macroscopic stress coefficient at zero load. This derivative can be evaluated from one load increment of a finite element simulation in the completely elastic regime. Equation~\ref{eqn:drss_dsigbar} can be integrated numerically by summing over the number of load steps ($N$) to calculate the macroscopic stress at which a region yields, corresponding to $\tau^* = \tau_{cr}^*$ in the rate-independent limit
\begin{equation}\label{eqn:rss_integrate}
\tau^* = \sum_{i=1}^N \frac{\Delta \tau^*}{\Delta \Sigma} \Delta\Sigma_i
\end{equation}
When evaluating the macroscopic stress at which elements in a finite element mesh yield, the elements are first binned according to strength-to-stiffness ratio to reduce numerical errors due to summing many small numbers. Each bin is associated with a range of macroscopic stresses over which all elements in the bin undergo yield. This range defines the macroscopic stress increment $\Delta\Sigma_i$ for the bin. The average elastic volume is used for the numerical integration.

\begin{figure}[h]
\centering
\subfigure[$n = 0$]{\includegraphics[trim = 0in 0in 0.9in 0in, clip]{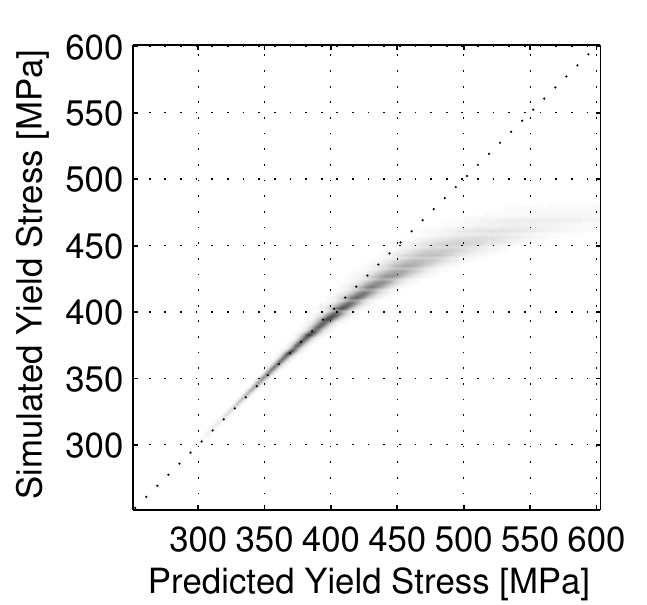}\label{fig:VolCorrExp-n000}} \qquad
\subfigure[$n = 2/3$]{\includegraphics[trim = 0in 0in 0.9in 0in, clip]{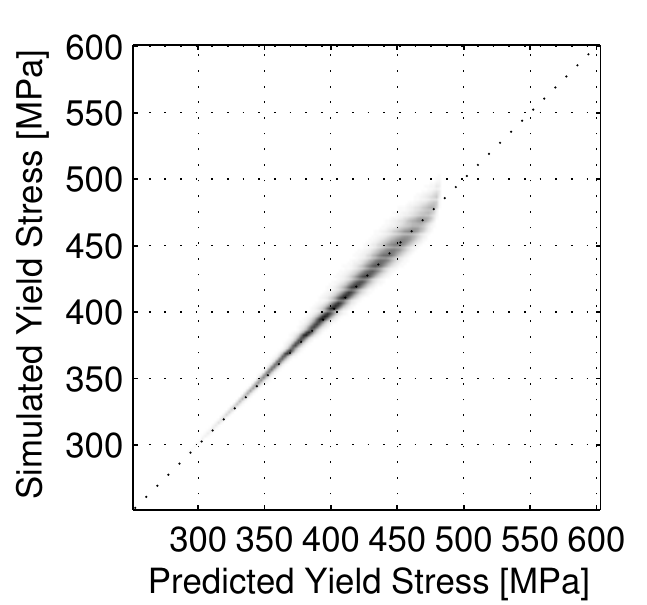}\label{fig:VolCorrExp-n067}} 
\subfigure[$n = 1$]{\includegraphics[trim = 0in 0in 0.9in 0in, clip]{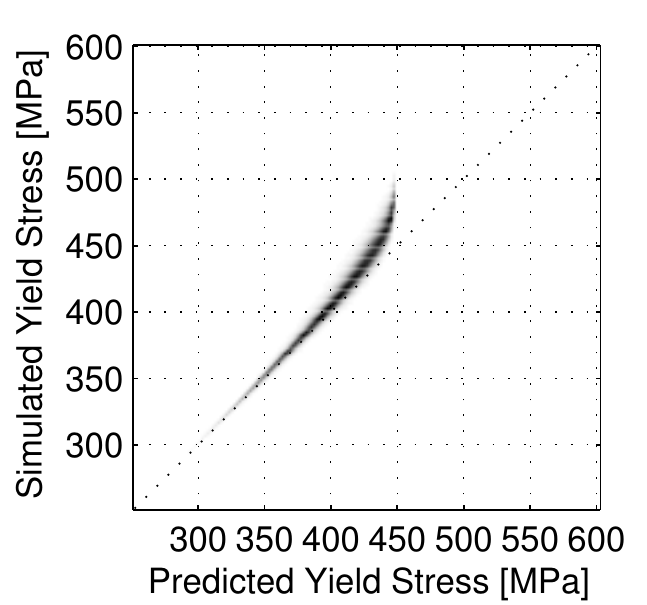}\label{fig:VolCorrExp-n100}} \qquad
\subfigure{\includegraphics[trim = 2.6in 0in 0in 0in, clip]{SimPredElemYldStress-BR000-n000.pdf}}
\caption{Effect of volume correction exponent on predicted elemental yield stress.}\label{fig:VolCorrExp}
\end{figure}

\begin{figure}[h]
\centering
\subfigure[$BR = 0.00$]{\includegraphics[trim = 0in 0in 0.9in 0in, clip]{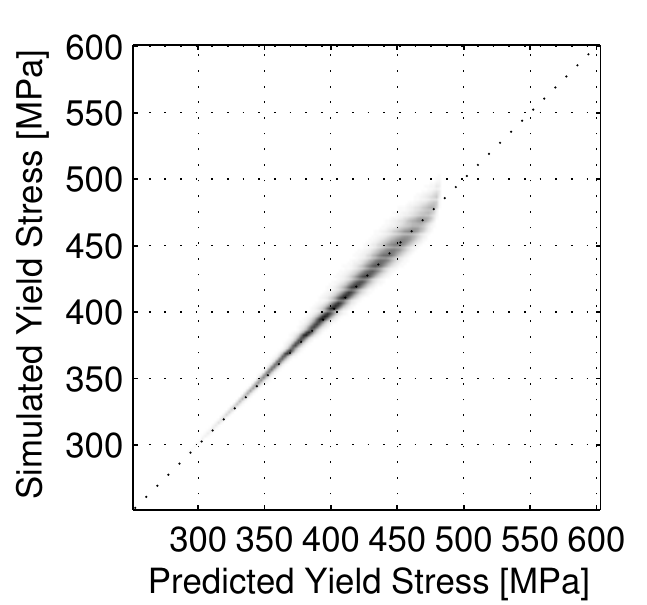}
\label{fig:ElemYldStress-BR000}} \qquad
\subfigure[$BR = 0.25$]{\includegraphics[trim = 0in 0in 0.9in 0in, clip]{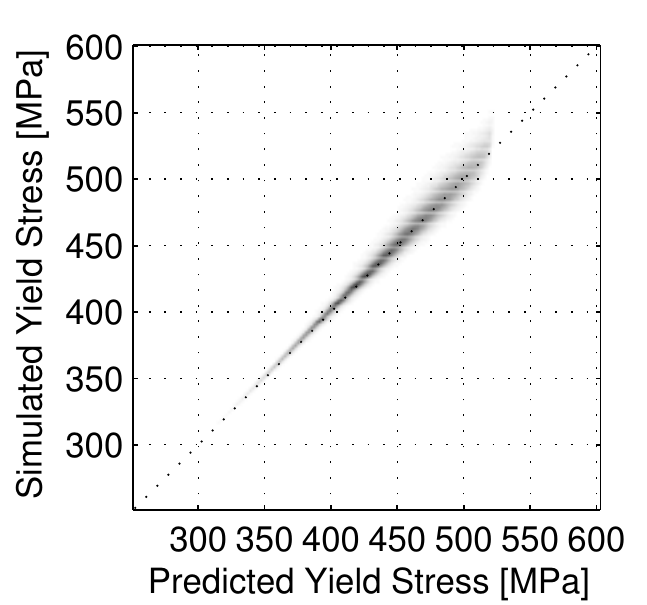}
\label{fig:ElemYldStress-BR025}}
\subfigure[$BR = 0.50$]{\includegraphics[trim = 0in 0in 0.9in 0in, clip]{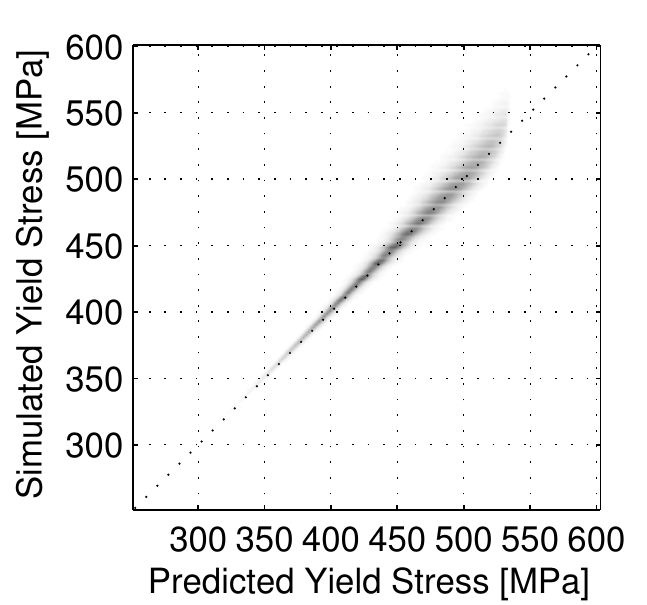}
\label{fig:ElemYldStress-BR050}} \qquad
\subfigure[$BR = 0.75$]{\includegraphics[trim = 0in 0in 0.9in 0in, clip]{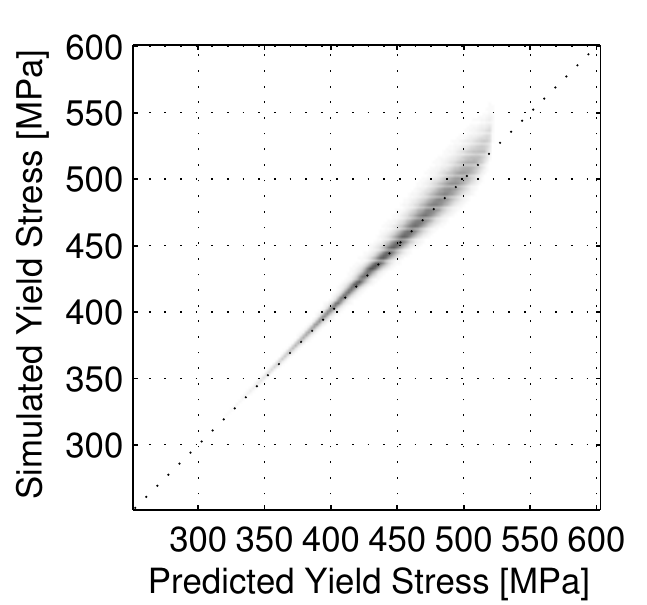}
\label{fig:ElemYldStress-BR075}}
\subfigure[$BR = 1.00$]{\includegraphics[trim = 0in 0in 0.9in 0in, clip]{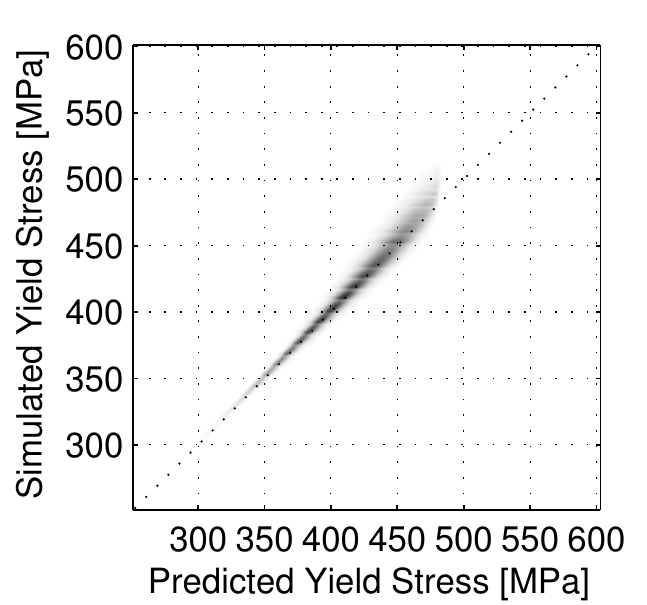}
\label{fig:ElemYldStress-BR100}} \qquad
\subfigure{\includegraphics[trim = 2.6in 0in 0in 0in, clip]{SimPredElemYldStress-BR000.pdf}}
\caption{Simulated and predicted elemental yield stresses.}\label{fig:SimPredElemYldStress}
\end{figure}

A volumetric scaling exponent $n$  of $2/3$  was demonstrated to give an excellent fit between the predicted value for yield from Equation~\ref{eqn:rss_integrate} and the yield extracted from simulation for single-phase stainless steel~\cite{pos_daw_multiaxial-y2e}. Here we examine the predictive capability of Equation~\ref{eqn:rss_integrate} for the two-phase, LDX-2101 system.   Binned scatter plots showing the relationship between predicted and simulated yield stresses for uniaxial loading are presented in Figure~\ref{fig:VolCorrExp} for three exponential values. 
As with the single-phase case, the correlation with an exponent of 2/3 provides excellent predictions of yielding, as indicated its nearness to the dotted line representing ideal agreement.
Figure~\ref{fig:SimPredElemYldStress} demonstrates the agreement between predicted and simulated macroscopic elemental yield stresses for five levels of stress biaxiality. As with the uniaxial case, there is good agreement between the prediction and simulation, with $R^2$ values greater than 0.93 for all five cases.

\begin{figure}[h]
	\centering
	\includegraphics{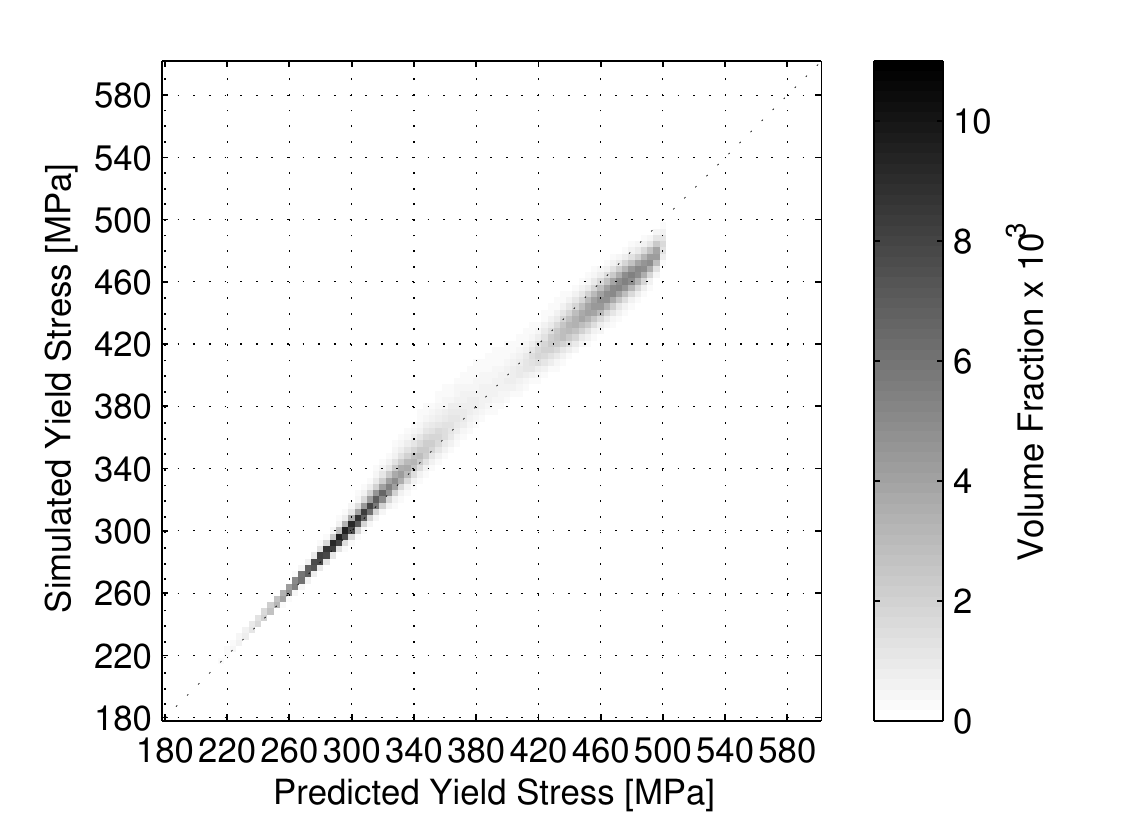}
	\caption{Simulated and predicted macroscopic elemental yield stresses for the case in which the initial slip system strength ($g_0$) of ferrite is twice that of austenite. The elemental yield prediction algorithm is valid for aggregates whose constitutive phases have different strengths.}
	\label{fig:ElemYldStressUnequalStrength}
\end{figure}

\begin{figure}[h]
	\centering
	\includegraphics{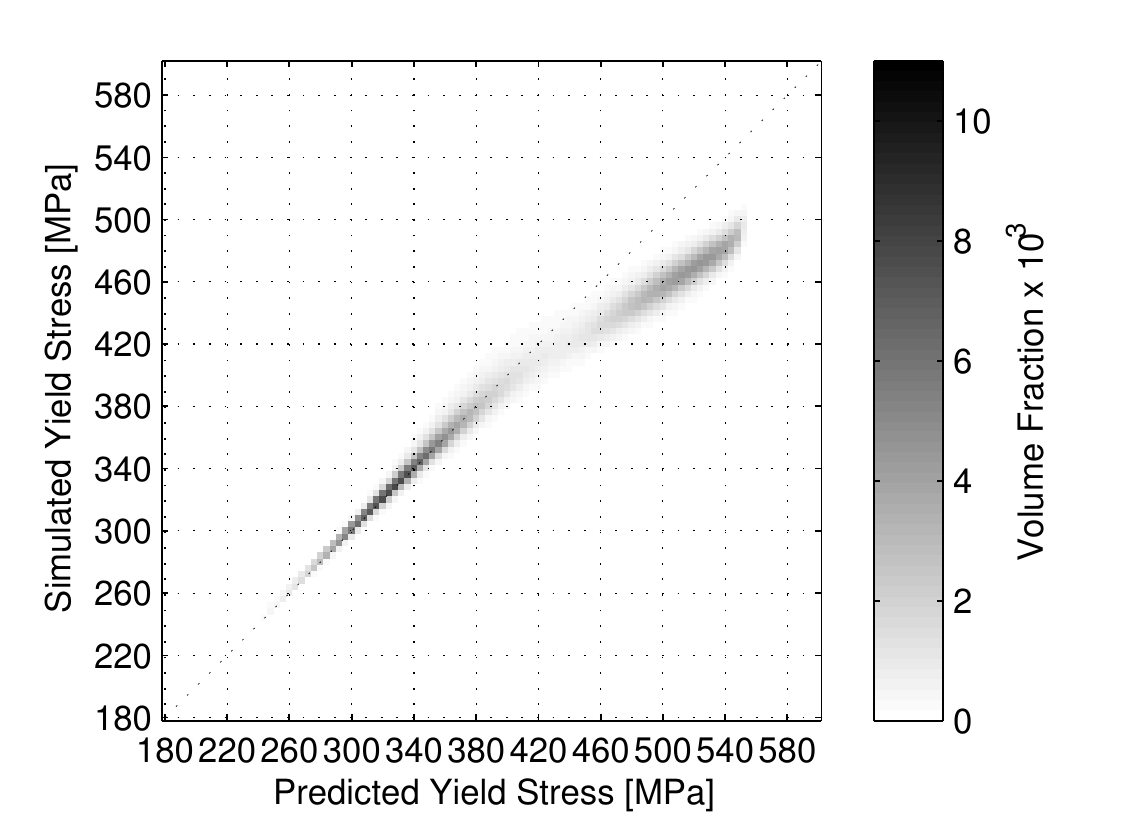}
	\caption{Simulated and predicted macroscopic elemental yield stresses for the case in which the stiffness of austenite is twice that of ferrite. The prediction is valid for the first stage of yielding, in which austenite yields, but breaks down for the second stage, in which ferrite yields. The prediction does not account for the jump in ratio of local to macroscopic stress increments that occurs once the majority of austenite has yielded. Multiphase systems comprised of phases with radically different stiffnesses, however, are rare among metals.}
	\label{fig:ElemYldStressUnequalStiff}
\end{figure}

To investigate the validity of the elemental yield prediction algorithm for dual phase systems in which the constituent phases have large differences in material properties, two additional simulations were conducted. For the first simulation, the initial slip system strength ($g_0$) of ferrite was twice that of austenite. The volume-averaged initial slip system strength was the same as that of LDX-2101. For the second simulation, austenite was twice as stiff as ferrite. The volume-averaged stiffness and single crystal elastic anisotropy ratios were the same as for LDX-2101. Binned scatter plots depicting the relationship between predicted and simulated macroscopic elemental yield stresses for the unequal phase strength and unequal phase stiffness cases are shown in Figures~\ref{fig:ElemYldStressUnequalStrength} and~\ref{fig:ElemYldStressUnequalStiff}, respectively. In both cases, the yielding is bimodal, with austenite yielding before ferrite. For the unequal phase strength case, the yield prediction is reasonably accurate, with an $R^2$ value of 0.978. At the end of the elasto-plastic transition, the algorithm overpredicts the elemental yield stresses by 20~MPa. However, for the unequal phase stiffness case, the prediction is not good, with an $R^2$ value of only 0.875. The prediction is valid for the first stage of yielding, in which the stiff austenite yields, but breaks down for the second stage, in which the compliant ferrite yields. The predicted macroscopic stresses at the end of the elasto-plastic transition are 60~MPa greater than the corresponding simulated values.

For the unequal phase strength case, the volume correction factor is able to account for the increase in the ratio of local to macroscopic stress increments that occurs in elastic regions as the aggregate yields. The correction works because the transition is relatively smooth. Considering the analogy of two phases loaded in parallel between rigid plates, both phases initially load at similar rates because they have similar stiffness. For the unequal phase stiffness case, however, the transition in load rate is not smooth. Stiffer austenite initially loads at a faster rate than the more compliant ferrite. Once austenite yields, the ratio of local to macroscopic stress increments jumps for ferrite. This jump is not accounted for in the present prediction. The jump in stress increment ratio produces the sudden change in slope observed in Figure~\ref{fig:ElemYldStressUnequalStiff}. As a result, the predicted macroscopic elemental yield stresses for ferrite are greater than the simulated values. The unequal phase stiffness example illustrates the limitations of the elemental yield prediction algorithm. It should be noted, however, that such cases of drastically unequal phase stiffness are rare among metallic alloys. The elemental yield prediction algorithm is valid for single-phase aggregates and dual phase aggregates with phases of roughly equal stiffness, even when the phases are of unequal strength.

\clearpage

\section{Conclusions}
\label{sec:conclusions}

The focus of this paper is on the initiation and propagation of plastic yielding in two-phase crystalline solids.  
In examining the initiation and propagation of yielding, the intent was to assess the ability of a recently developed metric for multiaxial strength-to-stiffness to rank crystals in terms of
the relative order in which they will yield as the intensity of the loading increases.  A range of biaxial stress states are considered from uniaxial tension to balanced biaxial tension.
The duplex stainless steel, LDX-2101, was chosen as a study material because it has relatively equal phase volume fractions and  comparable strengths and stiffnesses for the two phases. The phases differ in their crystal structures (FCC versus BCC), their morphologies and topologies, and in their single-crystal mechanical properties (elastic and plastic).     

The stress states at the crystal scale were simulated using an elasto-viscoplastic finite element formulation.  Based on microscopy of the LDX-2101, virtual samples were instantiated having columnar structure inherited from the processing history.   To verify that the computed stresses accurately estimated the real stresses, an extensive experimental program was conducted on tubular samples which were loaded with axial loads and internal pressures to produce a range of levels of stress biaxiality in the tube walls.  The samples were loaded to failure, pausing at numerous points along the loading curve to measure the lattice strains using neutron diffraction. Lattice strains were measured for scattering vectors in the axial, radial and hoop directions for several reflections in each of the two phases.   Material parameters were determined from one loading case (uniaxial tension), allowing comparisons across the remaining levels of stress biaxiality without further parameter adjustments.  Comparisons between the simulated and measured lattice strains 
indicated good overall ability of the model to detect inflection points in the strain histories.  
The inflection points are direct indicators of transitions associated with yielding and their relative 
behaviors provide evidence of the initiation and propagation of yielding.  
 
 Several conclusions  were drawn from the results: 
 \begin{enumerate}
\item The multidirectional strength-to-stiffness parameter, $r_{SE}$, correlates much better with  the macroscopic stress for yielding within crystals than either the Taylor factor or the Schmid factor.  
\item  The macroscopic strength prediction based on $r_{SE}$ works effectively  for both similar and dissimilar phase strengths with equal phase stiffnesses  across the full range of stress biaxiality from uniaxial tension to balanced biaxial tension.  The predictions for materials with highly contrasting phase stiffnesses (greater than a factor of two) is not as effective.
\item For LDX-2101, the difference in yield behavior between similarly oriented austenite and ferrite crystals is due to the difference in elastic anisotropy ratio. 
\end{enumerate}

\section{Acknowledgements}
\label{sec:acknowledgements}
Support was provided  by the US Office of Naval Research (ONR) under contract N00014-09-1-0447.
Neutron diffraction experiments were performed on a National Research Council (Canada) neu- tron diffractometer located at the NRU Reactor of AECL (Atomic Energy of Canada Limited).
\clearpage

\bibliographystyle{unsrt}
\bibliography{References}
\clearpage
\appendix
\section{Reduction of diffraction data}
\label{sec:exp_data_reduction}

\begin{figure}[h]
\centering
\subfigure[FCC \{111\}]{\includegraphics[scale = 0.85]{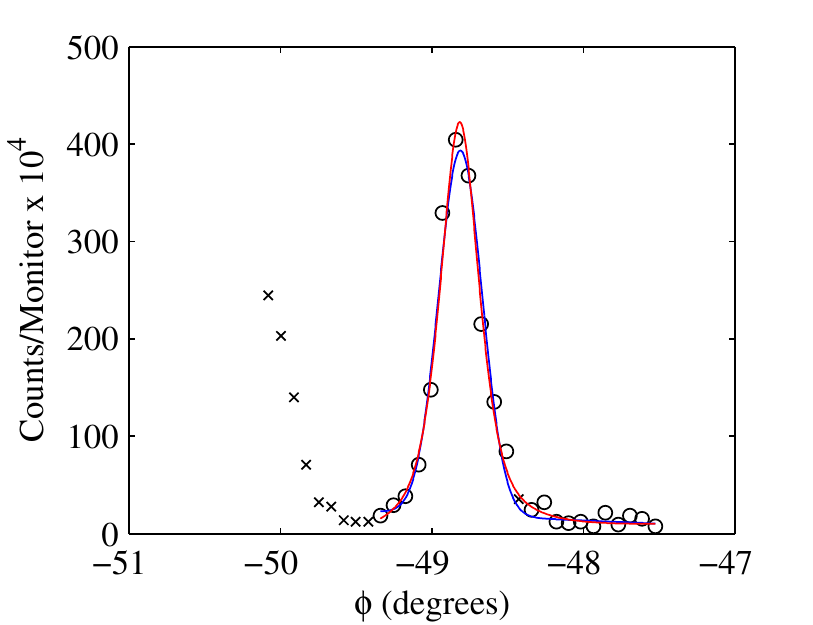}\label{fig:PeakFitFCC111}}
\subfigure[BCC \{110\}]{\includegraphics[scale = 0.85]{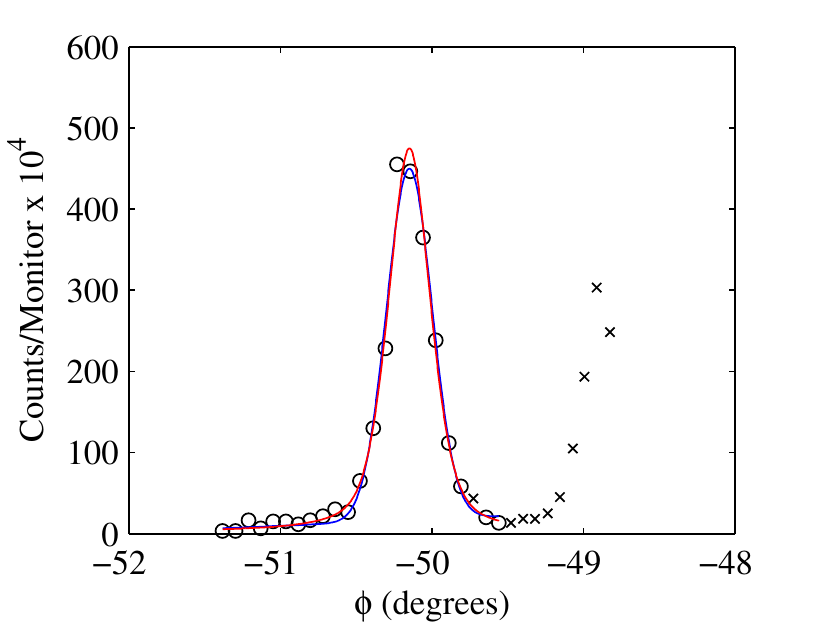}\label{fig:PeakFitBCC110}}
\subfigure[FCC \{200\}]{\includegraphics[scale = 0.85]{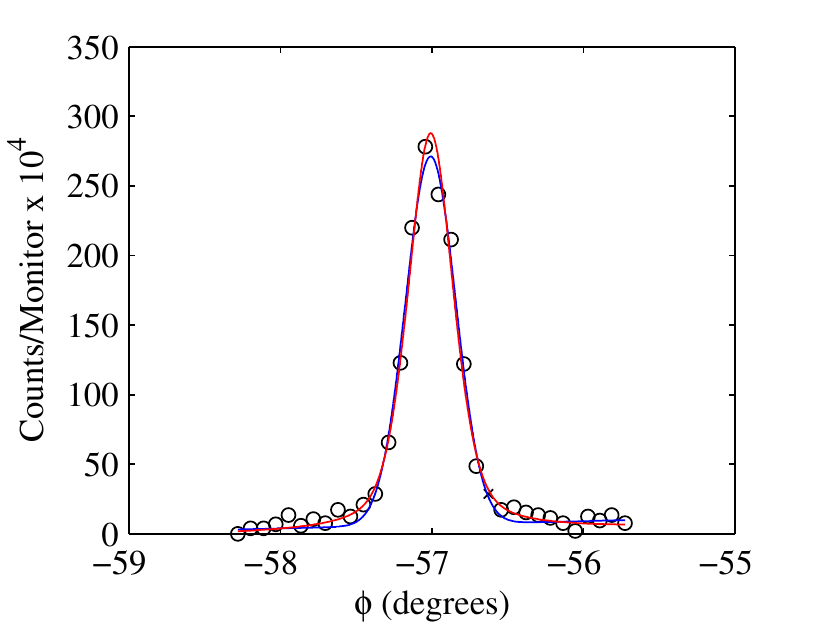}\label{fig:PeakFitFCC200}}
\subfigure[BCC \{200\}]{\includegraphics[scale = 0.85]{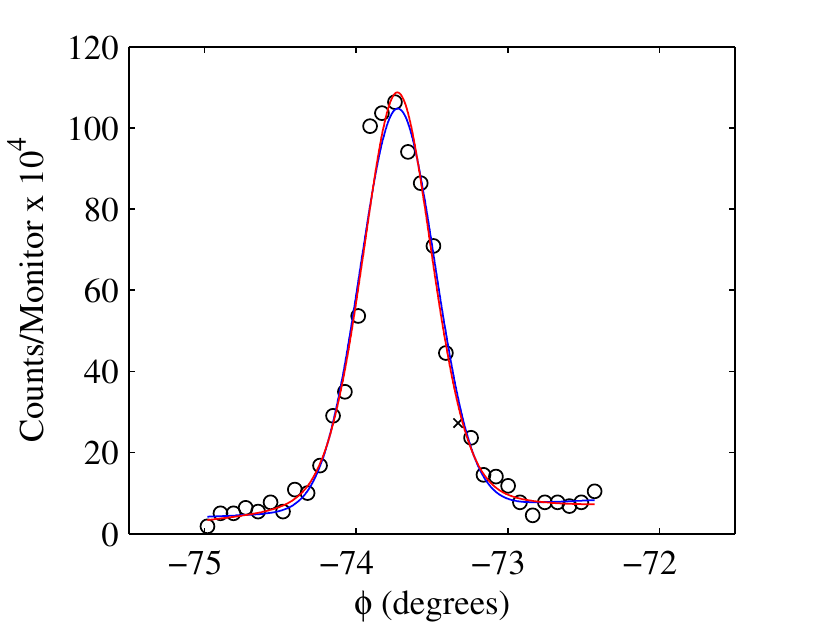}\label{fig:PeakFitBCC200}}
\subfigure[FCC \{220\}]{\includegraphics[scale = 0.85]{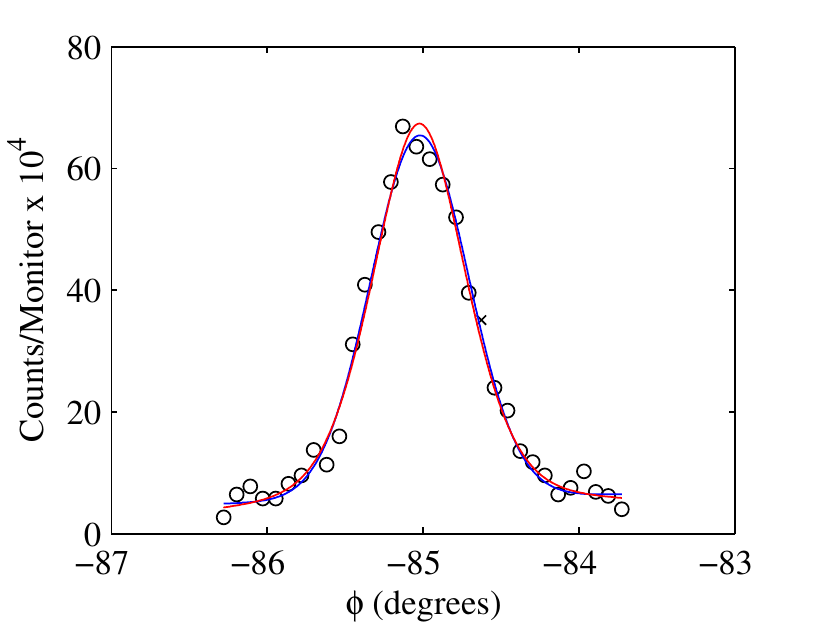}\label{fig:PeakFitFCC220}}
\subfigure[BCC \{211\}]{\includegraphics[scale = 0.85]{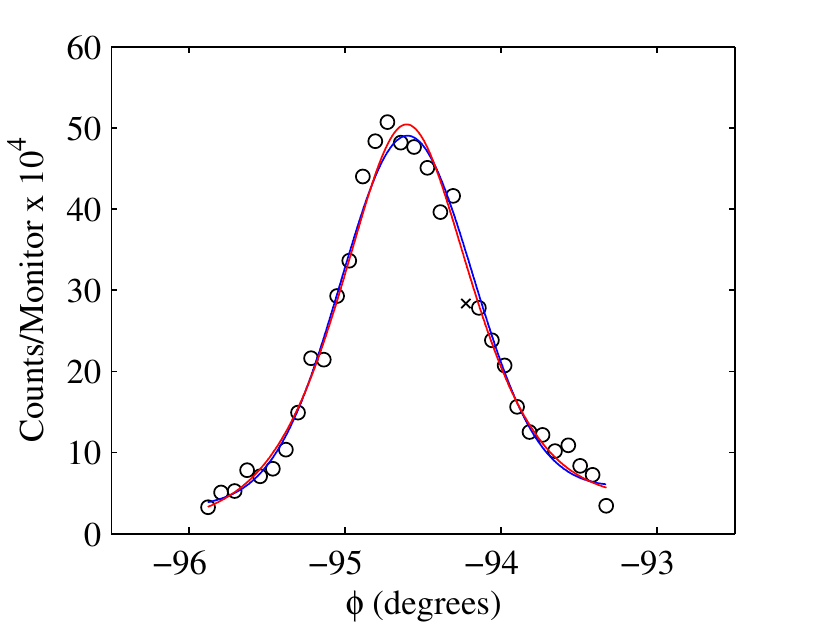}\label{fig:PeakFitBCC211}}
\subfigure{\includegraphics[scale = 0.4]{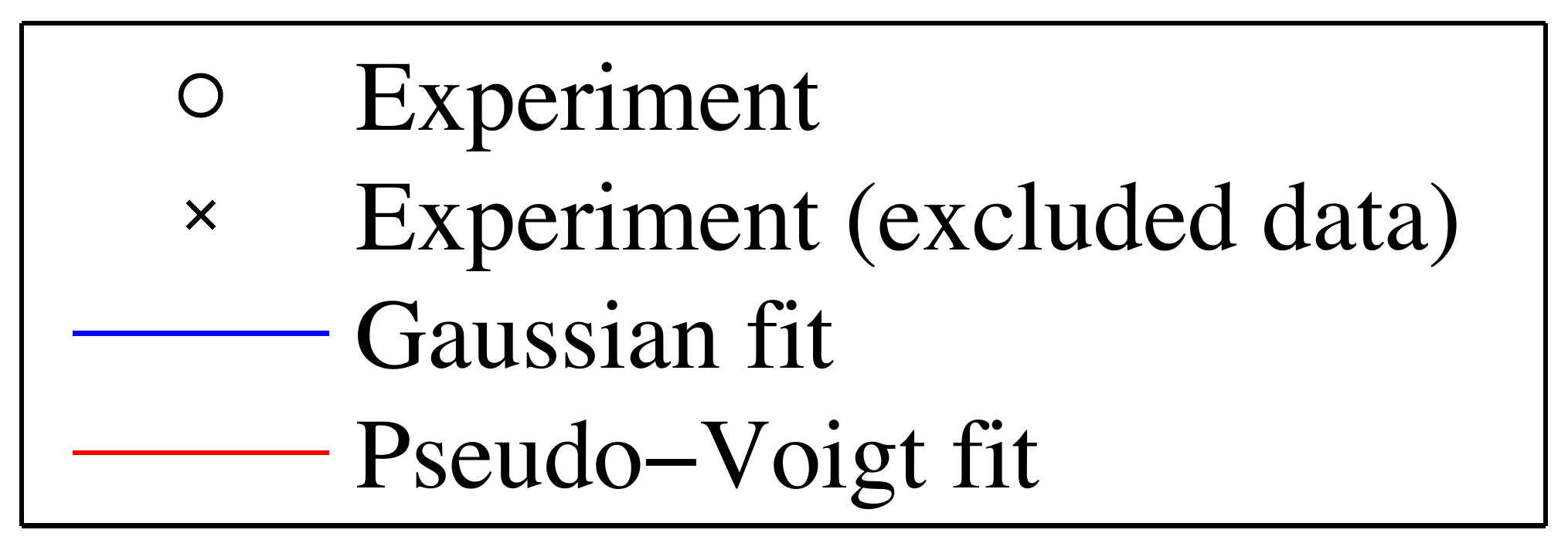}}
\caption{Representative diffraction peaks for the axial scattering vector at zero load.}\label{fig:PeakFit}
\end{figure}

The neutron detector registered one-dimensional profiles of diffracted neutron counts. Profiles were comprised of thirty-two points, corresponding to the thirty-two detector wires. Each profile was fit using an analytic function. The detector angle, $\phi$, was determined from the center of the analytic function. Two peak fit functions, Gaussian and pseudo-Voigt, were considered, both with a sloped background. The pseudo-Voigt is a linear combination of Gaussian and Lorentzian functions. The Gaussian and Lorentzian distributions are given by
\begin{equation}\label{eqn:gauss}
G(\phi) = \frac{2 \sqrt{\ln 2}}{H \sqrt{\pi}} \exp \left[ -4 \ln 2 \frac{(\phi - \mu)^2}{H^2} \right]
\end{equation}
\begin{equation}\label{eqn:lorentzian}
L(\phi) = \frac{2}{\pi H} \left[ 1 + \frac{4 (\phi - \mu)^2}{H^2} \right]^{-1}
\end{equation}
where $\mu$ is the mean and $H$ is the full-width at half-maximum. The analytic forms of the Gaussian and pseudo-Voigt peak fitting functions with sloped background are
\begin{equation}\label{eqn:gauss_fit}
y^{\mathrm{fit}} = I G(\phi) + m (\phi - \phi_1) + b
\end{equation}
\begin{equation}\label{eqn:pV_fit}
y^{\mathrm{fit}} = I \left[ \eta L(\phi) + (1-\eta) G(\phi) \right] + m (\phi - \phi_1) + b
\end{equation}
where $I$ is the intensity, $\eta$ is a mixing parameter between 0 and 1, $b$ is the constant background and $m$ is the background slope. The offset $\phi_1$ is the position of the first detector wire and is not a free parameter. The Gaussian function has five degrees of freedom, whereas the pseudo-Voigt has six. Figure~\ref{fig:PeakFit} shows representative diffraction peaks fit with both functions. The pseudo-Voigt function consistently performs better than the Gaussian function both qualitatively and in terms of the adjusted $R^2$ statistic. However, since the peaks are fairly symmetric, both functions perform about equally well at determining the center of the distribution, corresponding to the detector angle. Consequently, both functions perform about equally well in the strain calculation and yield similar results. The pseudo-Voigt function was used for all subsequent data reduction. 

There was overlap between the tails of the FCC \{111\} and BCC \{110\} peaks. The overlap region was excluded from the peak fit analysis and each peak was fit separately. Simultaneous fitting of the overlapping peaks was investigated, but did not produce a significant improvement. One of the detector wires (Wire 21) exhibited spurious behavior during some experiments. The twenty-first data point was therefore excluded from all analyses.

Diffraction peak fitting minimizes the weighted sum squared error $s$ between the measured data $y^{\mathrm{meas}}$ and fit data $y^{\mathrm{fit}}$
\begin{equation}\label{eqn:peakfit_resid}
s = \sum_{i=1}^n w_i \left(y_i^{\mathrm{meas}} - y_i^{\mathrm{fit}} \right)^2
\end{equation}
where $w$ represents the weight. The subscript $i$ indexes over the data points corresponding to each of the detector wires. The weighting function is related to the variance of the neutron counts that would arise if the peak were measured multiple times. Since neutron counting is a Poisson process, the variance of the counts is proportional to the number of counts. The weighting function is defined as the reciprocal of one plus the variance
\begin{equation}\label{eqn:peakfit_weight}
w_i =  \left( y_i^{\mathrm{meas}} + 1 \right)^{-1}
\end{equation}
The addition of one to the variance exists to avoid potential division by zero.

The Bragg angle, calculated from the diffraction peak fit, is used to calculate the lattice spacing. The Bragg angle is half the detector angle, $\phi$. Bragg's law describes the relationship between the lattice spacing, $d$, Bragg angle, $\theta$, and wavelength, $\lambda$.
\begin{equation}\label{eqn:Bragg}
\lambda = 2d \sin \theta
\end{equation}
The normal component of lattice strain, $\epsilon$, aligned with the scattering vector is the change in lattice spacing normalized by the initial lattice spacing
\begin{equation}\label{eqn:ls_def}
\epsilon = \frac{d - d_0}{d_0} = \frac{d}{d_0} - 1
\end{equation}
where the subscript $(\cdot)_0$ denotes a reference value. Combining Equations~\ref{eqn:Bragg} and~\ref{eqn:ls_def} yields
\begin{equation}\label{eqn:ls}
\epsilon = \frac{\sin \theta_0}{\sin \theta} - 1
\end{equation}
A relative, rather than absolute, reference Bragg angle was used to calculate strain. Since the material is assumed to be initially unstrained, the Bragg angle at zero load may be taken as the reference angle. To improve the precision of the reference Bragg angle, a linear fit was performed of the Bragg angle as a function of applied axial stress for the first six load steps in the elastic regime (0-250 MPa). The reference Bragg angle was then determined from the $y$-intercept of the fit line. This fitting procedure was used to determine a reference Bragg angle for each individual crystallographic fiber.

Bragg angle uncertainty, $\delta \theta$, is half the detector angle uncertainty, $\delta \phi$, calculated from the goodness of the diffraction peak fit. 
The relationship between lattice strain uncertainty, $\delta\epsilon$, and Bragg angle uncertainty, $\delta\theta$, is obtained by differentiating Equation~\ref{eqn:ls} with respect to $\theta$
\begin{equation}\label{eqn:ls_uncertainty}
\delta\epsilon = \frac{\sin\theta_0 \cos\theta}{\sin^2\theta} \delta\theta
\end{equation}
Bragg angle uncertainty is used with Equation~\ref{eqn:ls_uncertainty} to calculate the lattice strain uncertainty.   Uncertainty in the reference Bragg angle, $\theta_0$, is treated as bias, and is not included in this analysis.

\clearpage
\section{Boundary condition implementation}
\label{sec:sim-bc_implemetation}

\subsection{Constant load rate implementation}\label{sec:triaxclr_implement}

In constant load rate mode, the target load is known from the load rate and time increment. For each time increment, iterations are performed on the three surface velocities until the load recovered from the simulation is within tolerance of the target load. There are two rounds of surface velocity iterations. 

In the first round of surface velocity iterations, the ratios of the surface velocities are held constant. Iterations are performed on a single scale factor $\lambda$ applied to the surface velocities. Let the three surface velocities be expressed as an array $\{v\}$. A new iterate on the surface velocity is defined as
\begin{equation}\label{eqn:triaxclr_vel_update_1}
\{v\} = \lambda \{v_0\}\
\end{equation}
where $\{v_0\}$ is the current iterate on the surface velocities. The scale factor is computed from linear interpolation of the velocity-load response for the previous two surface velocity iterations. Iteration proceeds until the load in the primary control direction is within tolerance of the target load. 

In the second round of velocity iterations, the three surface velocities are treated independently. The surface velocities are perturbed sequentially and the corresponding equilibrium solutions calculated. A new iterate on the surface velocity can be represented as
\begin{equation}\label{eqn:triaxclr_vel_update_2}
\{v\} = \{v_0\} + \alpha \{\delta v_p\} + \beta \{\delta v_s\} + \gamma \{\delta v_t\} 
\end{equation}
where $\{v_0\}$ is the current iterate on the surface velocities, $\{\delta v_p\}$, $\{\delta v_s\}$, and $\{\delta v_t\}$ are perturbations to the surface velocity in the primary, secondary, and tertiary directions, and $\alpha$, $\beta$, and $\gamma$ are scalar coefficients. Let the load on each surface be expressed as an array $\{f\}$. The changes in load $\{\delta f_p\}$, $\{\delta f_s\}$, and $\{\delta f_t\}$, caused by each of the velocity perturbations, are calculated from the finite element solutions. Approximating the load-velocity relationship as locally linear in the vicinity of $\{f_0\}$, the load for the next velocity iteration may be approximated as
\begin{equation}\label{eqn:load_lin_approx_clr}
\{f\} = \{f_0\} + \alpha \{\delta f_p\} + \beta \{\delta f_s\} + \gamma \{\delta f_t\}
\end{equation}
where $\{f_0\}$ is the load for the current velocity iteration. Equating the load to the target load produces a system of three equations for the three unknown coefficients $\alpha$, $\beta$, and $\gamma$. The coefficients obtained from Equation~\ref{eqn:load_lin_approx_clr} are used with Equation~\ref{eqn:triaxclr_vel_update_2} to calculate new surface velocities. The equilibrium solution is then obtained for the new set of velocity boundary conditions. The second-round iteration proceeds until the recovered loads on all three surfaces are within tolerance of the target load.

\subsection{Constant strain rate implementation}\label{sec:triaxcsr_implement}

In constant strain rate mode, the strain rate in the primary loading direction is known. The velocity on the primary control surface is calculated directly from the strain rate. For each time increment, iterations are performed on the secondary and tertiary surface velocities until the ratios of the loads on the three surfaces are within tolerance.

For the first increment in a load or dwell episode, an initial guess of the secondary and tertiary surface velocities is obtained from isotropic elasticity. For all other increments, the surface velocities from the previous increment are used as the initial guess. The equilibrium solution is obtained for the initial set of boundary conditions. If the ratios of the recovered loads are within tolerance, then the simulation proceeds to the next time increment. If the load ratios are not within tolerance, iterations are performed on the secondary and tertiary surface velocities. Similar to the second-round iterations for constant load rate, perturbations are applied to the secondary and tertiary surface velocities, and the linearized response is used to calculate an update to the boundary conditions.

Let the three surface velocities be expressed as an array $\{v\}$. A new iterate on the surface velocity is defined as
\begin{equation}\label{eqn:triaxcsr_vel_update}
\{v\} = \{v_0\} + \alpha \{\delta v_s\} + \beta \{\delta v_t\} 
\end{equation}
where $\{v_0\}$ is the current iterate on the surface velocities, $\{\delta v_s\}$ and $\{\delta v_t\}$ are the perturbations to the surface velocity in the secondary and tertiary directions, respectively, and $\alpha$ and $\beta$ are scalar coefficients.

Similarly, the three principal loads are expressed as an array $\{f\}$. The load transitions linearly over each time step. The load at any time during the time step can be represented as the weighted sum of the initial load at the start of the time step $\{f\}^{i-1}$ and the final load at the end of the time step $\{f\}^i$
\begin{equation}\label{eqn:load_interpolate}
\{f\} = k \{f\}^i + (1-k) \{f\}^{i-1}
\end{equation}
where $k$ is the fraction of the time step that has been completed. Using a linear approximation, the load corresponding to the updated surface velocities can be represented as
\begin{equation}\label{eqn:load_lin_approx_csr}
\{f\} = \{f_0\} + \alpha \{\delta f_s\} + \beta \{\delta f_t\}
\end{equation}
where $\{f_0\}$ is the current load and $\{\delta f_s\}$ and $\{\delta f_t\}$ are the changes in load resulting from the perturbations to the surface velocities. Equating the loads in Equations~\ref{eqn:load_interpolate} and~\ref{eqn:load_lin_approx_csr} and rearranging terms produces a system of three equations for the three unknown coefficients $\alpha$, $\beta$, and $k$
\begin{equation}\label{eqn:triaxcsr_coeff}
\alpha \{\delta f_s\} + \beta \{\delta f_t\} + k \left( \{f\}^{i-1} - \{f\}^i \right) = \{f\}^{i-1} - \{f_0\}
\end{equation}
The coefficients obtained from Equation~\ref{eqn:triaxcsr_coeff} are then substituted into Equation~\ref{eqn:triaxcsr_vel_update} to form the next iterate on the surface velocities. The finite element solution is then computed using the new surface velocities. Iteration proceeds until the load ratios are within tolerance.

\end{document}